\documentclass[11pt]{article}
\usepackage{amsmath,amsthm,amssymb,epsfig,amsfonts,mathrsfs}
\usepackage{graphicx}
\usepackage{subfig}
\usepackage[usenames, dvipsnames]{color}
\usepackage{cite}
\usepackage{multirow}
\usepackage{hyperref}
\usepackage{verbatim} 
\usepackage{enumitem}
\usepackage{rotating}
\usepackage{xcolor}
\usepackage{multirow}
\usepackage{bm}
\usepackage[normalem]{ulem}
\usepackage{cleveref}
\usepackage{slashed}
\usepackage{enumitem}
\usepackage[fleqn,tbtags]{mathtools}
\usepackage{physics}
\usepackage{ytableau}
\usepackage{esint}
\usepackage{dsfont}
\usepackage{makecell}
\usepackage{stackengine, array}

\usepackage{tikz} 
\usepackage{tikz-cd}
\usepackage{extarrows}

\usepackage{booktabs}

\makeatletter

\usepackage{tikz}
\usetikzlibrary{arrows}
\usetikzlibrary{arrows.meta}
\usetikzlibrary{shapes.geometric,calc,arrows, positioning,shapes.misc,decorations.markings}
\tikzset{
  big arrow/.style={
    decoration={markings,mark=at position 1 with {\arrow[scale=2,#1]{>}}},
    postaction={decorate},
    shorten >=0.4pt},
  big arrow/.default=black}
\tikzset{every picture/.style={line width=0.75pt}}
\pgfdeclarelayer{edgelayer}
\pgfdeclarelayer{nodelayer}
\pgfsetlayers{edgelayer,nodelayer,main} 
\tikzstyle{none}=[inner sep=0pt] 

\tikzstyle{NodeCross}=[draw, shape=circle, cross out, inner sep=0pt, minimum size=6pt,line width=0.25mm]
\tikzstyle{Circle}=[draw, shape=circle, black,  fill=black, inner sep=0pt, minimum size=6pt]
\tikzstyle{Star}=[draw, shape=star, fill=black, star points=8, inner sep=0pt, minimum size=8pt]

\tikzstyle{DashedLine}=[-, densely dashed, line width=0.25mm]
\tikzstyle{DottedLine}=[-, dotted, line width=0.25mm]
\tikzstyle{ThickLine}=[-, line width=0.25mm]
\tikzstyle{ArrowLineRight}=[-, -{Stealth[scale=1.75]}, line width=0.1mm, scale=5]
\tikzstyle{RedLine}=[-, draw={rgb,255: red,191; green,0; blue,0}, fill=none, line width=0.25mm]
\tikzstyle{DottedRed}=[-, dotted, draw={rgb,255: red,191; green,0; blue,0}, fill=none, line width=0.25mm]
\tikzstyle{DashedLineThin}=[-, densely dashed, line width=0.125mm, fill=none, draw=black]
\tikzstyle{ArrowLineRed}=[-, -{Stealth[scale=1.75]}, draw={rgb,255: red,191; green,0; blue,0}, line width=0.25mm, scale=5]

\newtheorem{definition}{Definition}[section]
\newtheorem{theorem}{Theorem}[section]
\newtheorem{proposition}{Proposition}[section]
\newtheorem{example}{Example}[section]

\newcommand{\be}{\begin{equation}}
\newcommand{\ee}{\end{equation}}
\newcommand{\ba}{\begin{aligned}}
\newcommand{\ea}{\end{aligned}}

\newcommand{\bea}{\begin{eqnarray}}
\newcommand{\eea}{\end{eqnarray}}

\newcommand{\Z}{{\mathbb Z}}

\def\unit{{1\kern-.65ex {\rm l}}}
\def\1{{1\kern-.65ex {\rm l}}}

\def\bdm{\begin{displaymath}}
\def\edm{\end{displaymath}}
\def\ztwo{\mathbb{Z}_2}
\def\ringz{\mathbb{Z}}
\def\StSq{\textrm{Sq}}
\newcommand{\Bock}{\operatorname{Bock}_{\mathbb{Z}_2\to\mathbb{Z}}}

\def\CA{{\cal A}}

\def\CC{{\cal C}}
\def\CD{{\cal D}}

\def\CG{{\cal G}}

\def\CL{{\cal L}}
\def\CM{{\cal M}}

\def\CO{{\cal O}}

\def\CX{{\cal X}}

\def\CZ{{\cal Z}}

\def\bbC{{\mathbb{C}}}

\def\bbN{{\mathbb{N}}}

\def\bbR{{\mathbb{R}}}

\def\bbZ{{\mathbb{Z}}}

\newcount\hour \newcount\minute
\hour=\time \divide \hour by 60
\minute=\time
\def\now{%
\ifnum \hour<13
  \ifnum \hour=0 \advance \hour by 12 \number\hour:\else \number\hour:\fi%
     \ifnum \minute<10 0\fi%
     \number\minute%
\ A.M.%
\else \advance \hour by -12 \number\hour:%
  \ifnum \minute<10 0\fi%
  \number\minute%
  \ P.M.%
\fi%
}

\makeatother

\def\mb{\mathbb}
\def\mbf{\mathbf}
\def\mc{\mathcal}
\def\bp{\begin{pmatrix}}
\def\ep{\end{pmatrix}}

\def\ptl{\partial}

\def\ker{\mathrm{ker}}
\def\Aut{\mathrm{Aut}}

\def\mfr{\mathfrak}
\def\Fun{{\rm Fun}}
\def\Hom{{\rm Hom}}

\newcommand{\Rep}[1][]{\underline{#1}\mathrm{Rep}}

\newcommand{\Vect}{\mathbf{Vect}}
\newcommand{\sVect}{\mathbf{sVect}}
\newcommand{\Mod}{\operatorname{Mod}}
\newcommand{\Bimod}{\operatorname{Bimod}}
\newcommand{\TwoVect}{2\mathbf{Vect}}
\newcommand{\TwoRep}{2\mathbf{Rep}}
\newcommand{\SH}{\operatorname{SH}}
\newcommand{\br}{\mathrm{br}}

\usepackage{spectralsequences}
\begin{document}

\baselineskip=18pt  
\numberwithin{equation}{section}  
\allowdisplaybreaks  

\thispagestyle{empty}

\vspace*{0.8cm} 
\begin{center}
{\huge Understanding Non-Split 2-Group Symmetry: (3+1)D SymTFT, Anomaly and Bordism}\\

 \vspace*{1.5cm}

 {\large Zhenbang Gu$^1$, Ran Luo$^1$, Yi-Nan Wang$^{1,2}$, Yi Zhang$^3$}

\vspace*{1.0cm}


\smallskip
{\it $^1$ School of Physics, Peking University,\\
Beijing 100871, China\\ }

\smallskip

{\it  $^2$ Center for High Energy Physics, Peking University,\\
Beijing 100871, China}

\smallskip

{\it $^3$ Kavli IPMU (WPI), UTIAS, The University of Tokyo,\\
Kashiwa, Chiba 277-8583, Japan}

\vspace*{0.8cm}
\end{center}
\vspace*{.5cm}

\noindent
We present a systematic study of finite, non-split 2-group symmetries with a non-trivial Postnikov class, focusing on the simplest example with a $\mathbb{Z}_2$ 0-form symmetry and a $\mathbb{Z}_2$ 1-form symmetry, intertwined together by the non-trivial Postnikov class in $H^3(B\mathbb{Z}_2;\mathbb{Z}_2)\cong\mathbb{Z}_2$, denoted by $\mathcal{G}$. We classify the anomalies of this 2-group symmetry for physical theories in $d$-dimensional spacetime, by computing the oriented bordism groups $\Omega_{d+1}^{\rm SO}(B\mathcal{G})$ and the spin bordism groups $\Omega_{d+1}^{\rm Spin}(B\mathcal{G})$ for $d\leq 5$. For the case of $d=3$, we investigate the Symmetry TFT/TO of the 2-group symmetry $\mathcal{G}$ using the language of fusion 2-categories, as well as (3+1)D TQFT actions, for the cases without or with the 2-group anomaly classified by $\operatorname{Hom}\left(\widetilde{\Omega}_4^{\rm SO}(B\mathcal{G}),U(1)\right) \cong H^4(B\mathcal{G};U(1))\cong\mathbb{Z}_2$. We classify the minimal topological and physical boundary conditions of the Symmetry TFTs, and carry out the categorical Landau paradigm for such a non-split finite 2-group.
\newpage


\tableofcontents


\section{Introduction}

Despite the rapid development of generalized global symmetries~\cite{Gaiotto:2014kfa} in recent years (see e.g.~\cite{Sharpe:2015mja,Cordova:2022ruw,McGreevy:2022oyu,Gomes:2023ahz,Schafer-Nameki:2023jdn,Brennan:2023mmt,Bhardwaj:2023kri,Shao:2023gho,Luo:2023ive,Iqbal:2024pee,Costa:2024wks,Kaidi:2026urc} for reviews), the vast territory of higher-categorical symmetries remains largely unexplored. The simplest kind of higher-categorical symmetry is the 2-group symmetry~\cite{Baez:2003yaq,Baez:2004in,Baez:2010ya,Gukov:2013zka,Kapustin:2013uxa,Sharpe:2015mja,Tachikawa:2017gyf,Cordova:2018cvg,Delcamp:2018wlb,Benini:2018reh,Wan:2018djl,Hsin:2019fhf,Bhardwaj:2020phs,Chen:2020msl,Iqbal:2020lrt,Cordova:2020tij,DelZotto:2020sop,Yu:2020twi,DeWolfe:2020uzb,Gukov:2020btk,Brennan:2020ehu,Brauner:2020rtz,Hidaka:2020iaz,Hidaka:2020izy,Hsin:2021qiy,Lee:2021crt,Apruzzi:2021vcu,Apruzzi:2021mlh,Bhardwaj:2021wif,DelZotto:2022fnw,DelZotto:2022joo,Barkeshli:2022edm,Cvetic:2022imb,Cordova:2022qtz,Bhardwaj:2022scy,Bhardwaj:2022lsg,Barkeshli:2022wuz,Barkeshli:2023bta,Bhardwaj:2023wzd,Debray:2023rlx,Bartsch:2023pzl,Moradi:2023dan,Bartsch:2023wvv,Pace:2023kyi,Cvetic:2023pgm,Kang:2023uvm,Kan:2023yhz,Apruzzi:2024htg,Armas:2024caa,Nakajima:2024vgc,Bartsch:2024ech,Liu:2024znj,Tian:2025yrj,Bason:2026njw,Caro-Perez:2026esp}, which non-trivially mixes a 0-form symmetry group $\Pi_1$ and a 1-form symmetry group $\Pi_2$, with a homomorphism $\alpha:\Pi_1\rightarrow\mathrm{Aut}(\Pi_2)$ and a Postnikov class $\beta\in H^3_\alpha(B\Pi_1;\Pi_2)$ (for the case of finite $\Pi_1$ and $\Pi_2$)\footnote{$H^\bullet_\alpha(B\Pi_1;\Pi_2)$ is the (twisted) group cohomology of the $0$-form group $\Pi_1$ with coefficients in the $1$-form group $\Pi_2$~\cite{brown2012cohomology}. When $\Pi_1$ acts trivially on $\Pi_2$, it reduces to ordinary group cohomology $H^\bullet(B\Pi_1;\Pi_2)$.}. In particular, the 2-group is called split or non-split if the Postnikov class is trivial or non-trivial, respectively. To construct a simple example of a split 2-group, we can take the 1-form symmetry to be  $\Pi_2=\mb{Z}_N$ or $U(1)$ acting on the Wilson loops of a gauge theory, and the 0-form symmetry to be the charge conjugation $\Pi_1=\mb{Z}_2 \cong \{0,1\}$ acting on $\Pi_2$ as
\be
\alpha:1\rightarrow (x\rightarrow x^{-1})\quad (x\in\Pi_2)\,. 
\ee

On the other hand, the non-split 2-group symmetries are more involved, rich and interesting, which will be the main topic of this paper. Throughout most of the paper, we focus on the example of the simplest non-split 2-group $\mc{G}=(\mb{Z}_2,\mb{Z}_2, \alpha=\mathrm{triv.},\beta=1)$ as a demonstration. Here $\beta=1\in H^3(B\mb{Z}_2;\mb{Z}_2)$ denotes the non-trivial Postnikov class of the 2-group, represented by the function (in the additive group notation)\footnote{This case was studied in \cite{Liu:2024znj} in the strict 2-group formulation, but we will stick to the weak 2-group formulation in this paper.}
\be
\beta(g,h,k)= \begin{cases}
    \begin{aligned}
        1 \ &, \ g=h=k =1\in \bbZ_2  \\
        0 \ &, \ \text{otherwise} 
    \end{aligned}
    \ .
\end{cases} 
\ee 

An important physical aspect is the gauging of 2-group symmetries~\cite{Yetter:1993dh,Kapustin:2013uxa,Kapustin:2013qsa,Bullivant:2016clk,Delcamp:2019fdp,Carqueville:2025kqs,Huang:2020pki}. In particular, non-split 2-group symmetries naturally arise from gauging a subgroup of a 0-form symmetry group~\cite{Tachikawa:2017gyf}. Based on the formulation of weak 2-gauge theory, the $(d+1)$-dimensional on-shell gauge-invariant, closed topological action was discussed in \cite{Kapustin:2013uxa,Hsin:2019fhf}, and can be regarded as an anomaly theory for 2-group symmetry in $d$ spacetime dimensions. 

In this work we are going to scrutinize this point, based on the modern interpretation of anomaly in the bordism framework~\cite{Kapustin:2014tfa,Kapustin:2014dxa,Freed:2014iua,Freed:2016rqq,Yonekura:2018ufj,Garcia-Etxebarria:2018ajm}. Following such logic, we computed the cohomology of the classifying space $B\mc{G}$ of the weak 2-group~\cite{Baez0801}, and then the (reduced) oriented bordism groups $\widetilde{\Omega}_{d+1}^{\rm SO}(B\mc{G})$ as well as the (reduced) spin bordism groups $\widetilde{\Omega}_{d+1}^{\rm Spin}(B\mc{G})$ for $d\leq 5$, applicable to physical theories in $d$ spacetime dimensions, see Table~\ref{t:cobordism}. This is the first computation of the classifying space cohomology and the bordism groups for a finite, non-split 2-group, in addition to the cases of ``toric 2-groups''~\cite{Davighi:2023luh} and split 2-groups~\cite{Wan:2018bns,Wan:2019soo,Jia:2026jhj} in the literature. We present explicit descriptions of the anomaly as cochains for the oriented cases\footnote{The group cohomology cochains are also computed independently in an upcoming work by Yitao Feng and Yu-An Chen~\cite{Chen-Feng}.}.

\begin{table}
\begin{center}
\begin{tabular}{|c|c|c|c|c|c|c|c|}
\hline
\hline
$d+1$ & 1 & 2 & 3 & 4 & 5 & 6 \\
\hline
$H^{d+1}(B\mc{G};U(1))$ & $\mb{Z}_2$ & 0 & $\mb{Z}_2$ & $\mb{Z}_2$ & $\mb{Z}_2^2$ & $\mb{Z}_2$\\
\hline
$\widetilde{\Omega}_{d+1}^{\textrm{SO}}(B\mathcal{G})$ & $\ringz_2$ & $0$ & $\ringz_2$ & $\ringz_2$ & $\ringz_2^3$ & $\ringz_2^2$  \\
\hline
$\widetilde{\Omega}^{\textrm{Spin}}_{d+1}(B\mathcal{G})$& $\ztwo$&$\ztwo$&$\ringz_4\oplus\ztwo$&$\ztwo$&$\ztwo\oplus\ztwo$&0\\
\hline
\end{tabular}
\end{center}
\caption{Reduced oriented and spin bordism groups, characterizing anomalies for the non-split 2-group $\mc{G}=(\mb{Z}_2,\mb{Z}_2,\mathrm{triv.},1)$ in $d$-spacetime dimensions}\label{t:cobordism}
\end{table}

The next important physical perspective of non-split 2-group symmetry is its Symmetry Topological Field Theory (SymTFT/TO)~\cite{Witten:1998wy,Kong:2015flk,Kong:2017etd,Kong:2019byq,Kong:2019cuu,Kong:2020cie,Gaiotto:2020iye,Lichtman:2020nuw,Apruzzi:2021nmk,Apruzzi:2022dlm,DelZotto:2022ras,Moradi:2022lqp,Freed:2022qnc,Kaidi:2022cpf,vanBeest:2022fss,Kaidi:2023maf,Bhardwaj:2023ayw,Apruzzi:2023uma,Baume:2023kkf,Brennan:2024fgj,Argurio:2024oym,Bhardwaj:2024igy,Choi:2024tri,Tian:2024dgl,Najjar:2024vmm,Bonetti:2024etn,Antinucci:2024ltv,Antinucci:2024zjp,Bonetti:2024cjk,Gagliano:2024off,Tian:2025ooo,Putrov:2025xmw,DelZotto:2025yoy,Hung:2025gcp,Robbins:2025puq,Bonetti:2025dvm,Jia:2026tfh,Jia:2025vrj,Jia:2026vcr,Kong:2015flk,Ji:2019eqo,Ji:2019jhk,Ji:2021esj,Chatterjee:2022tyg,Chatterjee:2022kxb,Chatterjee:2022jll,Wen:2024udn,Wen:2023otf,Wen:2022tkg,Huang:2023pyk,Wen:2025thg,Luo:2025phx,Schafer-Nameki:2025fiy,Qi:2025tal}. Generally speaking, the SymTFT for a $d$-dimensional physical theory $\mc{T}_d$ is a $(d+1)$-dimensional TQFT in the bulk, with a topological boundary and a physical boundary. The assignment of different topological boundaries characterizes the generalized global symmetry of $\mc{T}_d$, while different physical boundary conditions classify the different phases of $\mc{T}_d$ such as the symmetric phase, spontaneous symmetry breaking (SSB) phase and symmetry protected topological (SPT) phases. This program is generally referred to as the categorical Landau paradigm~\cite{Kong:2015flk,Kong:2017etd,Kong:2019byq,Kong:2019cuu,Kong:2020cie,Bhardwaj:2023fca,Bhardwaj:2023idu,Bhardwaj:2023bbf,Hai:2023osv,Bhardwaj:2024qrf,Kong:2024ykr,Chen:2024ulc,Bhardwaj:2024wlr,Antinucci:2024bcm,Bhardwaj:2024qiv,Lootens:2024gfp,Bhardwaj:2025piv,Chen:2025uno,Ebisu:2026rnu}, generalizing the usual Landau paradigm of group symmetries. 

For the case of $d=2$, there is a successful equivalent description of the SymTFT in the language of category theory, already well-studied in the literature~\cite{Ostrik:2002ohv,Kitaev:2011dxc,Kong:2013aya,Lan:2014uaa,Cong:2017ffh,Bhardwaj:2017xup,Chang:2018iay,Thorngren:2019iar,Xu:2022rtj,Bhardwaj:2023idu,Hai:2023osv}. Namely, one assigns a 0-form symmetry in 2d with a symmetry category $\mc{S}$. For example for a finite symmetry group $G$ with anomaly $\omega\in H^3(BG;U(1))$, $\mc{S}=\mathbf{Vect}_G^\omega$, the fusion category of $G$-graded vector spaces with an associator characterized by $\omega$. Then the SymTFT corresponds to the Drinfeld center $\CZ(\mc{S})$. Different topological boundary conditions correspond to different Lagrangian algebra $\mc{L}$ of $\CZ(\mc{S})$, leading to different global forms of $\mc{T}_d$ that can be related by gauging of generalized global symmetries. For the physical boundary conditions, the Lagrangian algebra $\mc{L}$ of $\CZ(\mc{S})$ gives gapped boundary conditions while the non-maximal condensable algebras give gapless boundary conditions.

If one wants to generalize this program to 2-group symmetries $\mc{G}$, naturally one should define the symmetry category as a 2-fusion category $2\Vect_{\mc{G}}$~\cite{Douglas:2018qfz,Gaiotto:2019xmp,Johnson-Freyd:2020usu,Decoppet:2024htz} (in the absence of anomaly), and regard the SymTFT as its Drinfeld center $\CZ(2\Vect_{\mc{G}})$ (more precisely $\CZ_1(2\Vect_{\mc{G}})$\footnote{There are many constructions given the name of center for generic higher categories. We apply the definition of $E_k$-centers (the definition is reviewed in Appendix~\ref{app:classificationfusion2cat}) with the $E_1$ center (denoted as $\CZ_1(\cdot)$) being the Drinfeld center.})~\cite{Kong:2019brm,Decoppet:2023uoy,Decoppet2022Drinfeld,Xu:2024pwd}. Such approach was already proposed in \cite{Bhardwaj:2024qiv,Bhardwaj:2025piv,Bhardwaj:2025jtf}, and applied to the cases of finite split 2-groups. In this paper, we study the SymTFT of our example $\mc{G}=(\mb{Z}_2,\mb{Z}_2,\mathrm{triv.},1)$ for both the case without the 2-group anomaly and with 2-group anomaly (characterized by ${\rm Hom}(\widetilde{\Omega}_4^{\rm SO}(B\mc{G}),U(1))=H^4(B\mc{G};U(1))=\mb{Z}_2$ in this case).

For the case without the 2-group anomaly, the Drinfeld center is identified up to Morita equivalence with $\CZ_1(2\Vect_{\mb{Z}_2\times\mb{Z}_2}^{\pi_\beta})$, where $\pi_\beta\in H^4(B(\mb{Z}_2\times\mb{Z}_2);U(1))$ is a non-trivial group cohomology element. We identified seven minimal Lagrangian algebras~\cite{Zhao:2022yaw,Luo:2022krz,Xu2023,Chen2024} of $\CZ_1(2\Vect_{\mb{Z}_2\times\mb{Z}_2}^{\pi_\beta})$. In a parallel approach, we also found that the SymTFT has a topological action of
\be
S_{\rm SymTFT}=\frac{2}{2\pi}\int_{M_4}\left(B^1\wedge dA^1+B^2\wedge dA^2+\frac{1}{2\pi} A^1\wedge A^2\wedge dA^2\right)\,,
\ee
with $U(1)$-valued 1-form gauge fields $A^1,A^2$ and 2-form gauge fields $B^1,B^2$. In terms of $\mb{Z}_2$-valued 1-cochains $a^1,a^2$ and 2-cochains $b^1,b^2$, it reads
\be
S_{\rm SymTFT}=\pi\int_{M_4}\left(b^1\cup\delta a^1+b^2\cup\delta a^2+a^1\cup a^2\cup a^2\cup a^2\right)\,.
\ee
Such a 4d TQFT was already studied in~\cite{Putrov:2016qdo,Wang:2017loc,Zhang:2022rbg,Zhang:2023ynd,Zhang:2026uhx}. With the topological operator spectrum and correlation functions, we provide an analysis of seven different topological boundary conditions, which exactly one-to-one corresponds to the seven Lagrangian algebras of the Drinfeld center  $\CZ_1(2\Vect_{\mc{G}})$. They include one boundary condition $\mathcal{L}_1$ with $\mb{Z}_2^{(1)}\times\mb{Z}_2^{(2)}$ 0-form symmetry with mixed anomaly, two boundary conditions $\mc{L}_2^\pm$ with $\mc{G}=(\mb{Z}_2,\mb{Z}_2,\mathrm{triv.},1)$ 2-group symmetry, as well as four boundary conditions $\mc{L}_3^\pm$, $\mc{L}_4^\pm$ with the dual $2\mathbf{Rep}(\mc{G})$~\cite{elgueta2007representation,Huang:2024wdr} fusion 2-categorical symmetry. We also discussed the different physical boundary conditions, leading to symmetric, partial SSB phases, SSB phases and different types of $\mb{Z}_2$ and $\mc{G}$-SPT phases, summarized in Table~\ref{tab:phases-nonanomalous}. 

\begin{table}
\centering
\renewcommand{\arraystretch}{1.35}
\scriptsize
\setlength{\tabcolsep}{4pt}
\begin{tabular}{|>{\centering\arraybackslash}m{2.7cm}|>{\raggedright\arraybackslash}m{3.0cm}|>{\raggedright\arraybackslash}m{3.4cm}|>{\raggedright\arraybackslash}m{3.4cm}|>{\raggedright\arraybackslash}m{3.0cm}|}
\hline
\makecell{Top.\ b.c. $\downarrow$\\[-1pt] Phys.\ b.c. $\rightarrow$} & \centering\arraybackslash Physical $\mc{L}_1$ & \centering\arraybackslash Physical $\mc{L}_2^{\pm}$ & \centering\arraybackslash Physical $\mc{L}_3^{\pm}$ & \centering\arraybackslash Physical $\mc{L}_4^{\pm}$ \\
\hline
\makecell{$\mc{L}_1$\\[2pt] $\mb{Z}_2^{(1)}{\times}\mb{Z}_2^{(2)}$\\ 0-form\\ (mixed anomaly)}
 & full $\mb{Z}_2^{(1)}{\times}\mb{Z}_2^{(2)}$ SSB
 & $\mb{Z}_2^{(2)}$ SSB, $\mb{Z}_2^{(1)}$ preserved ($-$: stacks a $(a^1)^3$ SPT)
 & $\mb{Z}_2^{(1)}$ SSB, $\mb{Z}_2^{(2)}$ preserved, $+\,(a^2)^3$ SPT
 & diagonal $\mb{Z}_2$ SSB, off-diagonal $\mb{Z}_2$ preserved, $+$ SPT \\
\hline
\makecell{$\mc{L}_2^{+}$\\[2pt] 2-group $\CG$}
 & 0-form $\mb{Z}_2{\subset}\CG$ SSB, 1-form $\mb{Z}_2$ preserved
 & full $\CG$ SSB
 & $\CG$-symmetric ($-$: adds 2-group $\mc{G}$-SPT)
 & $\CG$-SPT phase \\
\hline
\makecell{$\mc{L}_3^{+}$\\[2pt] $2\mbf{Rep}(\CG)$}
 & non-invertible 0-form SSB, 1-form $\mb{Z}_2$ preserved
 & $2\mbf{Rep}(\CG)$-symmetric ($-$: adds $2\mbf{Rep}(\CG)$-SPT)
 & $2\mbf{Rep}(\CG)$ full SSB
 & $2\mbf{Rep}(\CG)$-SPT  \\
\hline
\makecell{$\mc{L}_4^{+}$\\[2pt] $2\mbf{Rep}(\CG)$}
 & non-invertible 0-form SSB, 1-form $\mb{Z}_2$ preserved
 & $2\mbf{Rep}(\CG)$-symmetric ($-$: adds $2\mbf{Rep}(\CG)$-SPT)
 & $2\mbf{Rep}(\CG)$-SPT
 & $2\mbf{Rep}(\CG)$ full SSB \\
\hline
\end{tabular}
\caption{The categorical Landau paradigm for $\CZ_1(2\Vect_\CG)$. The topological boundary conditions $\mc{L}_1$, $\mc{L}_2^+$, $\mc{L}_3^+$, $\mc{L}_4^+$ are listed, while the topological boundaries $\mc{L}_2^{-},\mc{L}_3^{-},\mc{L}_4^{-}$ follow from a $\pm$ interchange. No phase preserves only the $\mb{Z}_2$ 0-form symmetry while breaking the $1$-form symmetry, since the Postnikov class ties the 0-form anomaly to a $1$-form transformation. Here $a^1,a^2$ denote the $\mb{Z}_2^{(1)},\mb{Z}_2^{(2)}$ backgrounds.}
\label{tab:phases-nonanomalous}
\end{table}

The case with a 2-group anomaly is more intriguing. Building on \cite{Decoppet:2024moc} (see also \cite{Ambrosino:2024ggh} for the discussion of fermionic generalized symmetries), we identify the symmetry category in this case with a fermionic\footnote{In the fermionic description, the underlying manifold is not, in general, equipped with an ordinary spin structure, but with a twisted spin structure. This subtlety is discussed in Appendix~\ref{app:twistedspin}.} fusion 2-category through the monoidal equivalence $2\Vect_\CG^\omega \simeq 2\sVect_{(\bbZ_4,\underline{2})} \simeq \mfr{C}[(\bbZ_4,\underline{2}),\bbZ_2^f,\mathrm{triv.},\sVect,0] $. The SymTFT is its Drinfeld center, and we propose a Lagrangian realization of this (3+1)D TQFT using Stiefel--Whitney classes:
\be
S =\frac{2\pi}{4} \int_X B\cup \delta A + \pi \int_X (B\text{ mod } 2) \cup w_2(TX) \ , \ A\in C^1(X;\bbZ_4) \ ,\ B\in C^2(X;\bbZ_4) \ . 
\ee
There are two minimal Lagrangian algebras giving distinct gapped topological boundary conditions: one with the anomalous $\mc{G}$ 2-group symmetry, and the other one with a $\bbZ_4^f=(\bbZ_4,\underline{2})$ fermionic 1-form symmetry. We summarize the topological and physical boundary conditions in Table~\ref{tab:phases-anomalous}.

\begin{table}
\centering
\renewcommand{\arraystretch}{1.35}
\footnotesize
\setlength{\tabcolsep}{6pt}
\begin{tabular}{|>{\centering\arraybackslash}m{3.6cm}|>{\raggedright\arraybackslash}m{4.5cm}|>{\raggedright\arraybackslash}m{3.6cm}|}
\hline
\makecell{Top.\ b.c. $\downarrow$\\[-1pt] Phys.\ b.c. $\rightarrow$} & \centering\arraybackslash Physical $\mc{L}_1$ & \centering\arraybackslash Physical $\mc{L}_2$ \\
\hline
\makecell{$\mc{L}_1$\\[2pt] anomalous 2-group\\ $2\Vect_\CG^\omega\simeq 2\sVect_{(\bbZ_4,\underline{2})}$}
 & full $\CG$ SSB
 & $1$-form $\mb{Z}_2$  SSB, 0-form $\mb{Z}_2$ preserved \\
\hline
\makecell{$\mc{L}_2$\\[2pt] fermionic $1$-form\\ $\bbZ_4^f=(\bbZ_4,\underline{2})$, $2\mbf{sRep}(\bbZ_4,\underline{2})$}
 & $1$-form $\bbZ_4^f$ SSB to $\bbZ_2^f=(\bbZ_2,\underline{1})$
 & full SSB \\
\hline
\end{tabular}
\caption{Gapped phases of the non-split 2-group in the \emph{anomalous} case, from the categorical Landau paradigm for $\CZ_1(2\Vect_\CG^\omega)\simeq\CZ_1(2\sVect_{(\bbZ_4,\underline{2})})$. }
\label{tab:phases-anomalous}
\end{table}

The structure of this paper is as follows: in Section~\ref{sec:2-group} we review the notion of finite weak 2-group symmetry, its background gauge fields and the modified flatness condition $\delta b=\beta(a)$. In Section~\ref{sec:anomaly} we classify the 't Hooft anomalies of the non-split 2-group $\mc{G}$ via bordism: we compute the cohomology of the classifying space $B\mc{G}$ and the oriented and spin bordism groups summarized in Table~\ref{t:cobordism}, and present explicit cochain-level anomaly actions in each spacetime dimension $d\leq 5$. In Section~\ref{sec:category} we develop the fusion 2-categorical description of the symmetry: Section~\ref{sec:cat-no-anomaly} identifies the classifying data of $2\Vect_\mc{G}$ and the Lagrangian algebras of its Drinfeld center; Section~\ref{sec:cat-anomaly} establishes the equivalence of the anomalous symmetry category $2\Vect_\mc{G}^\omega$ with the fermionic fusion 2-category $2\sVect_{(\bbZ_4,\underline{2})}$; and Section~\ref{sec:SymTFT-anomaly} presents the categorical data of the corresponding SymTFT. In Section~\ref{sec:TQFT} we study the $(3+1)$D SymTFT from the Lagrangian point of view, for the non-anomalous case (Section~\ref{sec:without2-groupanomaly}) and the anomalous case (Section~\ref{sec:with2-groupanomaly}). We classify the different topological and physical boundary conditions and discuss the physical interpretations. In Section~\ref{sec:physical} we briefly discuss the realizations of the (anomalous) non-split 2-group symmetry in physical systems. Appendix~\ref{app:fusion2cat} reviews the classification of fusion 2-categories and the examples used in the main text. Appendix~\ref{app:spin-bordism-dual} reviews the three-layer approximation to spin bordism and (twisted) supercohomology with explicit cochain models. Appendix~\ref{app:spectral} collects the spectral sequence computations of the cohomology and bordism groups of $B\mc{G}$.

\section{Finite 2-group symmetry}
\label{sec:2-group}

\begin{figure}
\centering
\begin{tikzpicture}
    \draw [yslant = -0.5, fill = cyan] (0,0) rectangle (1,2) (1,2.5)node[above , cyan] {$g\in\Pi_1 $};
    \draw[densely dotted, thick, orange] (-1,0.8) node[anchor = south] {$\alpha_{g}(h)$} -- (0.5,0.8)  ;
    \draw[ultra thick, orange] (0.5,0.8) -- (1.7,0.8) node[anchor = south]{$h\in \Pi_2$};
\end{tikzpicture}
\caption{The action of the codimension-one defect labeled by a group element $g\in\Pi_1 $ on the codimension-two defect labeled by a group element $h\in\Pi_2$.}\label{2-group-rho}
\end{figure}
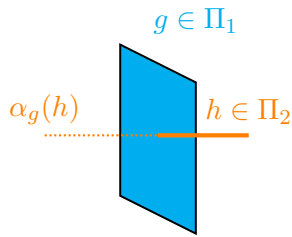

We consider a physical theory in $d$-spacetime dimensions, with a weak 2-group symmetry. A weak 2-group is labeled by a quadruple $(\Pi_1,\Pi_2,\alpha,\beta)$. $\Pi_1$ denotes a 0-form symmetry group generated by codimension-one topological defects and $\Pi_2$ denotes an abelian, 1-form symmetry group generated by codimension-two topological defects in spacetime. $\alpha:\Pi_1\rightarrow\mathrm{Aut}(\Pi_2)$ is a group homomorphism from $\Pi_1$ to the automorphism group of $\Pi_2$. The physical effect of $\alpha$ is depicted in Figure~\ref{2-group-rho}. A codimension-two defect labeled by $h\in \Pi_2$ passing through a codimension-one defect labeled by $g\in \Pi_1$ is twisted by the action $\alpha$ and becomes that labeled by $\alpha_g(h)$.

On the other hand, the element $\beta\in H^3_\alpha(B\Pi_1;\Pi_2)\cong H^3_{\text{grp},\alpha}(\Pi_1;\Pi_2)$ is called the Postnikov class of the weak 2-group. If the group action $\alpha$ is trivial, $\beta$ is an element of the group cohomology, otherwise it is an element of the twisted group cohomology with action $\alpha$. 

\begin{figure}
    \centering
\definecolor{darkred}{rgb}{0.76,0.13,0.08}

\begin{tikzpicture}[line width=0.8pt]

\coordinate (A)  at (0.68,7.63);   
\coordinate (Ab) at (0.68,3.92);   
\coordinate (B)  at (7.96,7.65);   
\coordinate (Bb) at (7.96,3.90);   
\coordinate (P)  at (4.79,8.25);   
\coordinate (L)  at (3.44,6.47);   
\coordinate (R)  at (5.77,6.54);   
\coordinate (T)  at (4.32,6.385);  
\coordinate (C)  at (4.32,4.50);   
\coordinate (D)  at (4.32,2.575);  
\coordinate (LL) at (1.97,4.45);   
\coordinate (E1) at (2.90,2.85);   
\coordinate (F)  at (7.12,4.06);   
\coordinate (Bv) at (7.14,0.26);   
\coordinate (Bt) at (0,1);   

\draw (A) -- (Ab);                                          
\draw (B) -- (Bb);                                          
\draw (A) .. controls (2.70,6.00) and (6.20,5.98) .. (B);   
\draw          (Ab) .. controls (2.20,2.95) and (3.30,2.60) .. (D);
\draw          (D)  .. controls (4.90,2.56) and (5.30,2.64) .. (5.63,2.78);
\draw[dashed]  (5.63,2.78) .. controls (6.10,2.98) and (6.70,3.20) .. (7.12,3.36);
\draw          (7.12,3.36) .. controls (7.50,3.53) and (7.80,3.72) .. (Bb);

\draw (L) -- (P) -- (R);
\draw[dashed] (L) -- (LL) -- (E1);     

\draw (T) -- (F);            
\draw (F) -- (Bv) -- (D);    

\draw[darkred,line width=1.2pt] (T) -- (D);
\draw[darkred,line width=1.2pt]        (E1) .. controls (3.35,3.60) and (3.85,4.15) .. (C);
\draw[darkred,line width=1.2pt]        (C)  .. controls (4.62,4.72) and (4.95,5.02) .. (5.12,5.26);
\draw[darkred,line width=1.2pt,dashed] (5.12,5.26) .. controls (5.24,5.43) and (5.38,5.64) .. (5.48,5.85);
\draw[darkred,line width=1.2pt]        (5.48,5.85) .. controls (5.58,6.06) and (5.70,6.38) .. (R);

\draw[blue,line width=1.1pt,-{Stealth[length=3.4mm]}]
      (C) .. controls (3.95,4.55) and (3.62,4.30) .. (Bt);

\fill (C) circle (2.4pt);

\node at (1.05,6.87) {$g$};
\node at (4.87,7.52) {$h$};
\node at (7.60,6.82) {$k$};
\node at (5.32,6.15) {$hk$};
\node at (3.70,3.05) {$gh$};
\node at (6.66,1.32) {$ghk$};
\node at (3.10,4.66) {$\beta(g,h,k)$};

\end{tikzpicture}
    \caption{The physical meaning of the Postnikov class $\beta(g,h,k)\in \Pi_2$ as an emergent 1-form symmetry generator in the presence of three codimension-one defects labeled by $g,h,k\in\Pi_1$. The figure shows the case of (2+1)D.}\label{2-group-beta}
\end{figure}
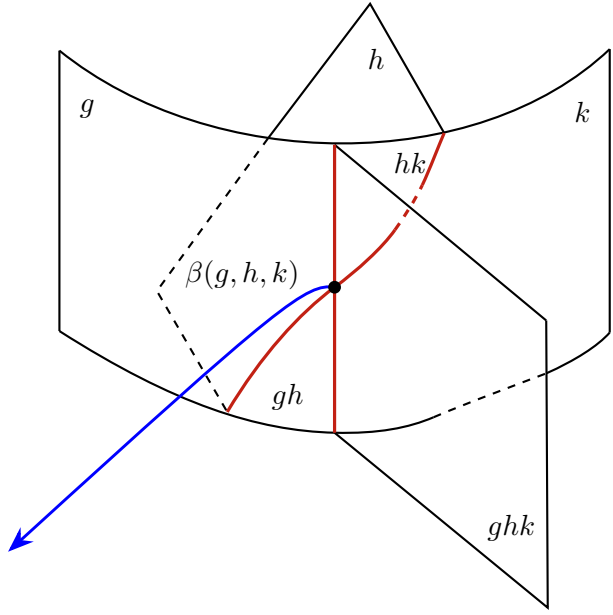

The information of $\beta$ is encoded in the fusion of 0-form symmetry generators, see Figure~\ref{2-group-beta} for an illustration of the (2+1)D case. At the junction point a new generator of the 1-form symmetry $\beta(g,h,k)\in \Pi_2$ emerges. It can also be viewed as a manifestation of the F-move, when one looks at the 1+1D slices of the picture from top to the bottom.  

When $\beta\in H^3_\alpha(B\Pi_1;\Pi_2)$ is a trivial group element, the 2-group is called a split 2-group, otherwise it is non-split. 

We will discuss the gauging of a finite 2-group symmetry, and we review the weak formulation for a finite weak 2-group $(\Pi_1,\Pi_2,\alpha,\beta)$~\cite{Kapustin:2013uxa}. For simplicity we assume that $\Pi_1$ is also abelian. The weak 2-gauge fields are $a\in C^1(M;\Pi_1)$, $b\in C^2(M;\Pi_2)$, and we use the additive notation for the abelian groups $\Pi_1$ and $\Pi_2$. The flatness conditions for $a$ and $b$ are (up to mod $N$ for $\mb{Z}_N$ groups)
\be
\label{weak-flat}
\delta a=0\ ,\ \delta_a b=\beta(a)\,,
\ee
where $\delta_a b$ is the twisted coboundary
\be
\label{weak-flatness}
(\delta_a b)_{ijkl}=\alpha(a_{ij})b_{jkl}-b_{ikl}+b_{ijl}-b_{ijk}\,,
\ee
and $\beta(a)$ is 
\be
\beta(a)_{ijkl}=\beta(a_{ij},a_{jk},a_{kl})\,.
\ee
The gauge transformation rules include the 0-form $g$-gauge transformation
\be
a\rightarrow a^g=a+\delta g\,,
\ee
\be
b\rightarrow \alpha_g b+\zeta(a,g)\,.
\ee
$\zeta(a,g)$ is a compensator function satisfying
\be
\delta_{a^g} \zeta(a,g)=\beta(a^g)-\alpha_g (\beta(a))\,,
\ee
to guarantee that the gauge-transformed $b^g$ still satisfies the flatness conditions (\ref{weak-flatness}) on-shell.

There is also a 1-form $\lambda$-gauge transformation with $\lambda\in C^1(M;\Pi_2)$:
\be
a\rightarrow a\ ,\ b\rightarrow b^\lambda=b+\delta_a\lambda\,.
\ee

The central example in this paper would be the simplest non-split 2-group, with $\Pi_1=\Pi_2=\mb{Z}_2$, $\alpha=\mathrm{triv.}$ (as $\mb{Z}_2$ has a trivial automorphism group), and $\beta=1\in H^3(B\mb{Z}_2;\mb{Z}_2)=\mb{Z}_2$ is the non-trivial Postnikov class. The function $\beta(g,h,k)$ can be represented as
\be
\beta(g,h,k)=ghk
\ee
in the additive group notation. We denote this 2-group as $\mc{G}=(\mb{Z}_2,\mb{Z}_2, \mathrm{triv.},1)$.

For the gauge transformation of such weak 2-group gauge fields $a$ and $b$, we can write explicitly $\beta(a)=a\cup a\cup a\equiv a^3$. The flatness conditions are
\be
\label{Z2Z2-flatness}
\delta a=0 \ ,\ \delta b=a\cup a\cup a\,.
\ee
The $g$-gauge transformation rules are
\be
a\rightarrow a+\delta g\ ,\ b\rightarrow b+\zeta(a,g)\,,
\ee
where 
\be
\delta\zeta(a,g)=(a+\delta g)^3-a^3\,.
\ee

If the flatness condition $\delta a=0$ is imposed, both $a$ and $h\equiv\delta g$ are cocycles. However, closed cochains do not commute under the cup product at the cochain level, so all the seven ordered terms remain:
\be
\ba
\delta\zeta(a,g)=\,&h\cup a\cup a+a\cup h\cup a+a\cup a\cup h\cr
&+h\cup h\cup a+h\cup a\cup h+a\cup h\cup h+h\cup h\cup h\,.
\ea
\ee
With $\mb{Z}_2$ coefficients, the cup-1 product satisfies $\delta(a\cup_1 h)=a\cup h+h\cup a$ for the closed 1-cochains $a$ and $h$. Using this identity, one solution for $\zeta$ is
\be
\label{zeta-canonical}
\zeta(a,g)=g\cup(a\cup a+h\cup a+h\cup h)+a\cup(a\cup_1 h)+(a\cup_1 h)\cup h\,.
\ee
Different solutions differ by closed 2-cochains; this ambiguity will be
relevant in Section~\ref{sec:with2-groupanomaly}.
Hence the full gauge transformation rules on-shell are
\be
\ba
&a\rightarrow a+\delta g\ ,\ b\rightarrow b+g\cup(a\cup a+\delta g\cup a+\delta g\cup \delta g)\cr
&\hphantom{a\rightarrow a+\delta g\ ,\ b\rightarrow b}+a\cup(a\cup_1\delta g)+(a\cup_1\delta g)\cup\delta g\cr
&a\rightarrow a\ ,\ b\rightarrow b+\delta\lambda\,.
\ea
\ee
\section{Anomalies of non-split 2-group symmetry}
\label{sec:anomaly}

\subsection{Classification of 't Hooft anomalies via bordism}\label{sec:Anomalyclassification}

From the modern viewpoint, a 't Hooft anomaly of a quantum field theory\footnote{The existence of a Dijkgraaf--Witten phase in $6+1$ dimensions that is nontrivial on general simplicial complexes but trivial on manifolds~\cite{SaitoTachikawaZhang2026} shows that anomalies of QFTs defined on smooth manifolds and those of lattice Hamiltonian models are genuinely different notions.} in $d$ spacetime dimensions is encoded, through anomaly inflow, by an invertible QFT in $d+1$ dimensions. A QFT is called invertible~\cite{Freed:2014iua} if it assigns a one-dimensional Hilbert space to every closed spatial manifold and its partition function on spacetime is a pure phase. Hence, the anomaly theories are also known as invertible phases. It was conjectured~\cite{Kapustin:2014dxa,Kapustin:2014tfa} that these anomaly theories can be classified by generalized cohomology theories. Freed and Hopkins~\cite{Freed:2016rqq} formulated the deformation classes of invertible, reflection-positive field theories in terms of the Anderson dual $I_{\mathbb{Z}}\Omega$ of the relevant bordism theory $\Omega$. See also a more accessible and explicit approach for physicists~\cite{Yonekura:2018ufj}.

To formulate the classification problem, one fixes the spacetime dimension $d$, a tangential structure $\mathcal{S}$ on spacetime, and a target space $X$ classifying the internal symmetry background. A background on a spacetime $M$ is represented topologically by a map $M\to X$. The deformation classes of the corresponding $(d+1)$-dimensional anomaly theories form the group
\bdm
\mathrm{Inv}^{d+1}_{\mathcal{S}}(X)
:=(I_{\mathbb{Z}}\Omega^{\mathcal{S}})^{d+2}(X),
\edm
where $\Omega^{\mathcal{S}}$ is the bordism theory associated with $\mathcal{S}$ and $I_{\mathbb{Z}}\Omega^{\mathcal{S}}$ denotes its Anderson dual.

The Anderson dual is related to the bordism groups by the universal coefficient exact sequence
\bdm
0\longrightarrow
\operatorname{Ext}^1_{\mathbb{Z}}\!\left(\Omega^{\mathcal{S}}_{d+1}(X),\mathbb{Z}\right)
\longrightarrow
(I_{\mathbb{Z}}\Omega^{\mathcal{S}})^{d+2}(X)
\longrightarrow
\operatorname{Hom}_{\mathbb{Z}}\!\left(\Omega^{\mathcal{S}}_{d+2}(X),\mathbb{Z}\right)
\longrightarrow 0.
\edm
For the finitely generated bordism groups considered here,
\bdm
\begin{aligned}
\operatorname{Ext}^1_{\mathbb{Z}}\!\left(\Omega^{\mathcal{S}}_{d+1}(X),\mathbb{Z}\right)
&\simeq
\operatorname{Hom}_{\mathbb{Z}}\!\left(\operatorname{Tor}\Omega^{\mathcal{S}}_{d+1}(X),\mathbb{R}/\mathbb{Z}\right),
\\
\operatorname{Hom}_{\mathbb{Z}}\!\left(\Omega^{\mathcal{S}}_{d+2}(X),\mathbb{Z}\right)
&\simeq
\operatorname{Hom}_{\mathbb{Z}}\!\left(\operatorname{Free}\Omega^{\mathcal{S}}_{d+2}(X),\mathbb{Z}\right).
\end{aligned}
\edm
The first term is the Pontryagin dual of the torsion subgroup and captures global anomalies, while the second detects the free part and captures perturbative anomalies described by anomaly polynomials. 

For an ordinary internal 0-form symmetry group $G$, the target space is $X=BG$. A 2-group $\mathcal{G}$, including a non-split one, likewise has a classifying space $X=B\mathcal{G}$~\cite{Baez0801}. The corresponding bordism problem is therefore obtained by replacing $BG$ with $B\mathcal{G}$.

For the theories considered here, the spacetime structure and internal target are
\bdm
\mathcal{S}=\left\{\begin{array}{ll}
\textrm{Spin}&\quad \textrm{for fermionic theories},\\
\textrm{SO}&\quad \textrm{for bosonic theories},
\end{array}
\right.
\qquad X=B\mathcal{G}.
\edm
The anomaly classification therefore reduces to computing the corresponding spin or oriented bordism groups of $B\mathcal{G}$, from which the Anderson-dual group above is obtained. Applications of this general framework to 2-group symmetries include the works of~\cite{Wan:2018bns,Lee:2020ewl,Davighi:2023luh}; here we apply it to finite non-split 2-groups.
An important feature of bordism groups is that it decomposes canonically 
\begin{equation}
    \Omega^{\mathcal{S}}_{\bullet}(X) \simeq \Omega^{\mathcal{S}}_{\bullet}(\rm pt) \oplus \tilde \Omega^{\mathcal{S}}_{\bullet}(X) \,,
\end{equation}
where $\tilde \Omega^{\mathcal{S}}_{\bullet}(X)$ is called the reduced bordism group. For our interests of finite 2-groups, $\tilde \Omega^{\mathcal{S}}_{\bullet}(B\CG)$ consists only of torsion elements. Then the Anderson dual of degree $d+2$ reduces to the $\operatorname{Ext}$-factor, which is just the Pontryagin dual
\begin{equation}
    \operatorname{Hom}_{\mathbb{Z}}\!\left(\tilde \Omega^{\mathcal{S}}_{d+1}(B\CG),\mathbb{R}/\mathbb{Z}\right).
\end{equation}

For computing $\Omega^{\mathcal{S}}_{\bullet}(B\CG)$, we will use the Atiyah--Hirzebruch spectral sequence (AHSS) and the Adams spectral sequence (ASS). Since the Pontryagin dual $\operatorname{Hom}_{\mathbb{Z}}\!\left(\tilde \Omega^{\mathcal{S}}_{\bullet}(B\CG),\mathbb{R}/\mathbb{Z}\right)$ also defines a generalized cohomology theory, we can use the cohomology version of AHSS for a direct computation. The details of calculations for $\mc{G}=(\mb{Z}_2,\mb{Z}_2, \alpha=\mathrm{triv.},\beta =1)$ are presented in the Appendix~\ref{app:spectral}.

\paragraph{The classifying space of a finite 2-group.}  For a finite 2-group $\CG=(\Pi_1,\Pi_2, \alpha=\mathrm{triv.},\beta)$ whose 0-form symmetry acts trivially on its 1-form symmetry, the classifying space is described by the following fibration:
\begin{equation}
\begin{tikzcd}\label{eq:2groupfibration}
K(\Pi_2,2) \simeq B^2\Pi_2  \arrow[r, hook] & B\mathcal{G} \arrow[d]   \\
                             &  \quad \qquad \quad \quad B\Pi_1 \simeq K(\Pi_1,1) \,.
\end{tikzcd}
\end{equation}
where $B\Pi_1$ and $B^2\Pi_2$ are the classifying spaces for the $\Pi_1$ 0-form symmetry and for the $\Pi_2$ 1-form symmetry. They are just the Eilenberg--MacLane spaces $K(\Pi_1,1)$ and $K(\Pi_2,2)$, respectively. One can take the homotopy fiber to get another fibration 
\begin{equation}
    B\mathcal{G} \longrightarrow K(\Pi_1,1) \xlongrightarrow{\beta} K(\Pi_2,3)\,,
\end{equation}
by which we have the class $\beta\in [K(\Pi_1,1),K(\Pi_2,3)]\simeq H^3(B\Pi_1;\Pi_2)$ characterizing the fibration~\eqref{eq:2groupfibration}. It is called the Postnikov class or Postnikov $k$-invariant.\\

When the 0-form symmetry acts non-trivially on the 1-form symmetry, $\alpha:\Pi_1\to\operatorname{Aut}(\Pi_2)$ is a non-trivial map. The classifying space $B\mathcal{G}$ still sits as the middle space in a fibration like~\eqref{eq:2groupfibration}, but the action $\alpha$ controls the possibly non-trivial monodromy as the following. 

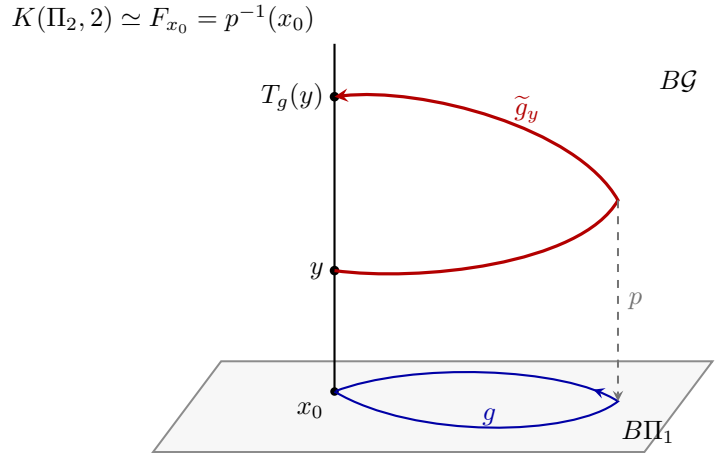
\begin{figure}[ht]
\centering
\begin{tikzpicture}[>=stealth, every node/.style={font=\small}]
    \filldraw[fill=black!3,draw=black!45]
        (-1.5,-0.65) -- (5.0,-0.65) -- (4.1,-1.85) -- (-2.4,-1.85) -- cycle;
    \node at (4.15,-1.58) {$B\Pi_1$};

    \coordinate (x0) at (0,-1.05);
    \fill (x0) circle (1.6pt);
    \node[below left] at (x0) {$x_0$};

    \draw[thick,blue!65!black,
        decoration={markings,mark=at position 0.55 with {\arrow{>}}},
        postaction={decorate}]
        (x0) .. controls (1.05,-1.68) and (3.15,-1.65) .. (3.75,-1.18)
             .. controls (3.05,-0.72) and (1.00,-0.66) .. (x0);
    \node[blue!65!black] at (2.05,-1.38) {$g$};

    \draw[thick] (x0) -- (0,3.55);
    \node[above left,align=right] at (0,3.55)
        {$ K(\Pi_2,2)\simeq  F_{x_0}=p^{-1}(x_0)$ };

    \coordinate (y) at (0,0.55);
    \coordinate (Tgy) at (0,2.85);
    \fill (y) circle (1.8pt);
    \fill (Tgy) circle (1.8pt);
    \node[left] at (y) {$y$};
    \node[left] at (Tgy) {$T_g(y)$};

    \draw[->,very thick,red!70!black]
        (y) .. controls (1.10,0.40) and (3.25,0.62) .. (3.75,1.48)
            .. controls (3.20,2.45) and (1.15,2.98) .. (Tgy);
    \node[red!70!black] at (2.55,2.68) {$\widetilde g_y$};
    \node at (4.55,3.05) {$B\mathcal{G}$};

    \draw[->,dashed,black!55] (3.75,1.48) -- (3.75,-1.18)
        node[midway,right] {$p$};
\end{tikzpicture}
\caption{The vertical line is the single fiber $F_{x_0}$ over $x_0$. The closed loop $g$ in the base lifts to a path $\widetilde g_y$ in $B\mathcal{G}$ from $y$ to $T_g(y)$; the lifted path need not be closed.}
\label{fig:2group-monodromy}
\end{figure}

Denote the fibration map by $p:B\mathcal{G}\to B\Pi_1$ and write $F_{x_0}=p^{-1}(x_0)\simeq K(\Pi_2,2)$ for the fiber over $x_0\in B\Pi_1$. Given a loop $g$ in $B\Pi_1$ based at $x_0$ and a point $y\in F_{x_0}$, let $\widetilde g_y$ be the lift of $g$ beginning at $y$. Its endpoint, denoted by $T_g(y)$, lies again in $F_{x_0}$. As $y$ varies, this gives, up to homotopy, a self-homotopy equivalence $T_g:F_{x_0}\to F_{x_0}$. The lifted path need not be closed: it joins $y$ to $T_g(y)$ in the same fiber, as illustrated in Figure~\ref{fig:2group-monodromy}. Under the identification $F_{x_0}\simeq K(\Pi_2,2)$, the self-equivalence $T_g$ is induced by the automorphism $\alpha_g:\Pi_2\to\Pi_2$.

The homotopy class of $T_g$ is the monodromy around $g$. For fixed $\Pi_1$, $\Pi_2$, and $\alpha$, the Postnikov class now takes value in the 
\begin{equation}
    \beta\in H^3_\alpha(B\Pi_1;\Pi_2).
\end{equation} specifies the homotopy type of the middle space. When $\alpha={\rm triv.}$, this loop action is trivial and one recovers the homotopy-fiber description above. For non-trivial $\alpha$, there is in general no ordinary map $B\Pi_1\to K(\Pi_2,3)$ whose homotopy fiber is $B\mathcal{G}$.

Having specified the homotopy type of $B\mathcal{G}$, its integral cohomology can be computed using the Serre spectral sequence
\begin{equation}
    E_2^{p,q}=H^p\!\left(B\Pi_1;H^q_\alpha(B^2\Pi_2;\mathbb{Z})\right)\Longrightarrow H^{p+q}(B\mathcal{G};\mathbb{Z}),
\end{equation}
where the $\Pi_1$-action on $H^q(B^2\Pi_2;\mathbb{Z})$ is induced by $\alpha$. Hence $\alpha$ enters already on the $E_2$-page, while $\beta$ affects the differentials~\cite{BROWN1994235, DATUASHVILI2001352,may2009spectral}.  For the purpose of our article, we are looking at the case $\alpha=\mathrm{triv.}$.
The ordinary (co)homology of the classifying space $B\mathcal{G}$ itself is the key ingredient of the aforementioned bordism computations. In the pioneering works on finite 2-groups~\cite{Kapustin:2013uxa,Tachikawa:2017gyf}, the authors adopted the Serre spectral sequence (SSS) to compute these (co)homological data. The result for $\mc{G}=(\mb{Z}_2,\mb{Z}_2, \alpha=\mathrm{triv.},\beta =1)$ can be found in Appendix~\ref{app:spectral}.

\subsection{Weak 2-group gauge fields and anomalies in lower dimensions}
\label{sec:anomaly-pol}

We now translate the bordism classes into explicit topological actions
constructed from the weak $2$-group gauge fields.\footnote{We thank Yunqin
Zheng for comments and suggestions on the presentation of the identification
between bordism invariants and cochain expressions constructed from the
$2$-group gauge fields.}

Let $X=B\CG$ for
$\CG=(\mb{Z}_2,\mb{Z}_2, \mathrm{triv.},1)$.  A background weak $2$-group gauge field on a manifold $M$
is a map $f:M\to X$, represented at the cochain level by
\begin{equation}
    a\in Z^1(M;\mb{Z}_2),
    \qquad
    b\in C^2(M;\mb{Z}_2),
    \qquad
    \delta b=a^3.
\end{equation}
Here $a$ is the $\Pi_1$ gauge field associated with the base
$B\Pi_1$, while $b$ is the $\Pi_2$ gauge field associated with the
fiber $B^2\Pi_2$.  Let $\mathfrak{a}\in H^1(X;\mb{Z}_2)$ be the class pulled
back from the base; then $[a]=f^*\mathfrak{a}$.  Although $b$ is not itself
closed, combinations of $a$, $b$, and the higher cup products can
define cocycles on $X$.
 
Such cocycles provide useful candidates for bosonic anomalies, but
closedness alone does not imply that the corresponding topological
action is nontrivial.  A class $z\in H^{d+1}(X;\mb{Z}_2)$ defines the
$U(1)$-valued phase
\begin{equation}
    \exp\left(\pi i\int_{M_{d+1}}f^*z\right),
\end{equation}
whose anomaly class is the image
$\frac12z\in H^{d+1}(X;U(1))$.  The oriented bordism
calculation~\eqref{eq:reduced-oriented-result} gives
\begin{equation}\label{eq:oriented-anomaly-groups}
\operatorname{Hom}\!\left(
    \widetilde{\Omega}_{d+1}^{\mathrm{SO}}(X),U(1)
\right)
\cong
\begin{cases}
0,&d=1,\\
\mb{Z}_2,&d=2,\\
\mb{Z}_2,&d=3,\\
\mb{Z}_2^3,&d=4,\\
\mb{Z}_2^2,&d=5.
\end{cases}
\end{equation}
In particular, a nonzero mod-$2$ cohomology class may still give a
trivial phase on every closed oriented manifold.

The Serre spectral sequence in Appendix~\ref{app:spectral} gives a
systematic starting point for the cochain representatives.  The relevant
degree-$3$ and degree-$4$ classes restrict on the fiber to
$u_3=\mathrm{Sq}^1u_2$ and $u_2^2$, respectively.  Since $a$ vanishes on the
fiber and $b$ restricts to $u_2$, their natural initial lifts are
$b\cup_1b$ and $b\cup b$.  They are not closed when $\delta b=a^3$:
\begin{equation*}
\delta(b\cup_1b)=a^3\cup_1b+b\cup_1a^3,
\qquad
\delta(b\cup b)=a^3\cup b+b\cup a^3.
\end{equation*}
The higher-cup-product identities show that the mixed terms in the first
expression are cancelled by $b\cup_2a^3$, leaving
$\mathrm{Sq}^1(a^3)=a^4$, which is cancelled by $a\cup b$.  Similarly,
$b\cup_1a^3$ cancels the mixed terms in the second expression, leaving
$\mathrm{Sq}^2(a^3)=a^5$, which is cancelled by $a^2\cup b$.  This cochain
descent constructs the two basic cocycles needed below:
\begin{align}
v_3&=
a\cup b+b\cup_1b+b\cup_2a^3,
\label{eq:v3-weak-gauge}\\
v_4&=
b\cup b+b\cup_1a^3+a^2\cup b.
\label{eq:v4-weak-gauge}
\end{align}
Using $\delta b=a^3$ and the higher-cup-product identities, their
coboundaries reduce to
\begin{align}
\delta v_3
&=a^4+a^3\cup_2a^3
=a^4+\mathrm{Sq}^1(a^3)=0,\\
\delta v_4
&=a^3\cup_1a^3+a^5
=\mathrm{Sq}^2(a^3)+a^5=0,
\end{align}
where the last equalities follow from the Cartan formula.
We use the same symbols $v_3$ and $v_4$ for the resulting cohomology classes
on $X$.
 
\paragraph{$d=1$.}
The only candidate from the base field is $\mathfrak{a}^2\in
H^2(X;\mb{Z}_2)$.  Let $\Bock$ denote the Bockstein associated with
$0\to\mb{Z}\xrightarrow{2}\mb{Z}\to\mb{Z}_2\to0$.  Then
\begin{equation}
    \mathfrak{a}^2=\rho_2\!\left(\Bock(\mathfrak{a})\right),
\end{equation}
so $\frac12\mathfrak{a}^2=0$ in $H^2(X;U(1))$.  This agrees with
$\widetilde{\Omega}_2^{\mathrm{SO}}(X)=0$: there is no nontrivial
oriented anomaly in this degree.
 
\paragraph{$d=2$.}
The class $v_3$ gives the topological action
\begin{equation}
    S_{\mathrm{3d}}
    =\pi\int_{M_3}v_3.
\end{equation}
Its Bockstein
$\Bock(v_3)$ is the nonzero class in
$H^4(X;\mb{Z})\cong\mb{Z}_2$.  Hence $\frac12v_3$ generates
\begin{equation}
    \operatorname{Hom}\!\left(
        \widetilde{\Omega}_3^{\mathrm{SO}}(X),U(1)
    \right)\cong\mb{Z}_2.
\end{equation}
 
\paragraph{$d=3$.}
The mod-$2$ cohomology has the two candidate classes
\begin{equation}
    H^4(X;\mb{Z}_2)
    =\mb{Z}_2\langle \mathfrak{a}v_3,v_4\rangle.
\end{equation}
The first does not define a nontrivial oriented anomaly, since
\begin{equation}
    \mathfrak{a}v_3
    =\mathrm{Sq}^1(v_3)
    =\rho_2\!\left(\Bock(v_3)\right).
\end{equation}
By contrast,
$\Bock(v_4)$ generates
$H^5(X;\mb{Z})\cong\mb{Z}_2$.  Therefore the unique nonzero anomaly is
\begin{equation}\label{4-cocycle}
\begin{aligned}
S_{\mathrm{4d}}
&=\pi\int_{M_4}v_4\\
&=\pi\int_{M_4}
\left(b\cup b+b\cup_1a^3+a^2\cup b\right).
\end{aligned}
\end{equation}
Equivalently,
\begin{equation}
    \operatorname{Hom}\!\left(
        \widetilde{\Omega}_4^{\mathrm{SO}}(X),U(1)
    \right)
    \cong\mb{Z}_2\langle\frac12v_4\rangle.
\end{equation}
 
\paragraph{$d=4$.}
The reduced bordism computation gives
\begin{equation}
\widetilde{\Omega}_5^{\mathrm{SO}}(X)
\cong H_5(X;\mb{Z})\oplus H_1(X;\mb{Z})
\cong\mb{Z}_2^2\oplus\mb{Z}_2.
\end{equation}
The three gauge-field cocycles found in mod-$2$ cohomology are
\begin{equation}
H^5(X;\mb{Z}_2)
=\mb{Z}_2\langle \mathfrak{a}^2v_3,\mathfrak{a}v_4,v_5\rangle,
\qquad
v_5:=\mathrm{Sq}^2(v_3)=[v_3\cup_1v_3].
\end{equation}
They do not give three independent $U(1)$-valued actions.  Indeed,
\begin{equation}
    \mathfrak{a}^2v_3
    =\mathrm{Sq}^1(v_4)
    =\rho_2\!\left(\Bock(v_4)\right),
\end{equation}
so $\frac12\mathfrak{a}^2v_3$ is trivial.  Moreover,
$\Bock(\mathfrak{a}v_4)$ and
$\Bock(v_5)$ form a basis of
$H^6(X;\mb{Z})\cong\mb{Z}_2^2$.  The two independent gauge-field
anomalies can therefore be represented by
\begin{equation}
    S_{\mathrm{5d},1}
    =\pi\int_{M_5}a\cup v_4,
    \qquad
    S_{\mathrm{5d},2}
    =\pi\int_{M_5}v_3\cup_1v_3.
\end{equation}
These two characters are dual to the $H_5(X;\mb{Z})$ summand.  The
remaining $H_1(X;\mb{Z})$ summand of the reduced bordism group is
represented by
\begin{equation}
    S_{\mathrm{5d},3}
    =\pi\int_{M_5}a\cup w_2(TM_5)^2.
\end{equation}
Here and below, $w_i(TM)$ denotes the $i$-th Stiefel--Whitney class
of the tangent bundle.
Together, these three characters generate
\begin{equation}
\operatorname{Hom}\!\left(
    \widetilde{\Omega}_5^{\mathrm{SO}}(X),U(1)
\right)\cong\mb{Z}_2^3,
\end{equation}
in agreement with Eq.~\eqref{eq:oriented-anomaly-groups}.
 
\paragraph{$d=5$.}
In total degree $6$, the reduced oriented bordism group is
\begin{equation}
\widetilde{\Omega}_6^{\mathrm{SO}}(X)
\cong H_6(X;\mb{Z})\oplus H_1(X;\mb{Z}_2)
\cong\mb{Z}_2\oplus\mb{Z}_2.
\end{equation}
The degree-$6$ mod-$2$ cohomology is
\begin{equation}
H^6(X;\mb{Z}_2)
=\mb{Z}_2\langle v_3^2,\mathfrak{a}^2v_4,\mathfrak{a}v_5\rangle.
\end{equation}
The coefficient Bockstein vanishes on $v_3^2$ and $\mathfrak{a}^2v_4$, whereas
\begin{equation}
\Bock(\mathfrak{a}v_5)
\end{equation}
generates $H^7(X;\mb{Z})\cong\mb{Z}_2$.  Thus the character dual to
$H_6(X;\mb{Z})$ is represented by
\begin{equation}
S_{\mathrm{6d},1}
=\pi\int_{M_6}a\cup v_5
=\pi\int_{M_6}a\cup(v_3\cup_1v_3).
\end{equation}
The other character comes from the $H_1(X;\mb{Z}_2)$ summand.
The generator of $\Omega_5^{\mathrm{SO}}(\mathrm{pt})\cong\mb{Z}_2$
is detected by the Stiefel--Whitney number $w_2w_3$
~\cite{MilnorStasheff1974}; hence this character is represented by
\begin{equation}
S_{\mathrm{6d},2}
=\pi\int_{M_6}
a\cup w_2(TM_6)\cup w_3(TM_6).
\end{equation}
Consequently,
\begin{equation}
\operatorname{Hom}\!\left(
\widetilde{\Omega}_6^{\mathrm{SO}}(X),U(1)
\right)\cong\mb{Z}_2^2,
\end{equation}
as in Eq.~\eqref{eq:oriented-anomaly-groups}.

\section{Symmetry category of a non-split 2-group}
\label{sec:category}

In this section, we give a categorical description of the non-split 2-group symmetry $\CG=(\bbZ_2,\bbZ_2,\mathrm{triv.},\beta=1)$
in $(2+1)$ dimensions. In the absence of an additional 't Hooft anomaly, its symmetry category is $2\Vect_\CG$, the fusion 2-category of finite semisimple 2-vector spaces graded by $\CG$\footnote{See Example~\ref{exm:2-groupsym} for an explicit definition}. We first determine its classifying data and then study the $(3+1)$D SymTFT with topological operators described by the Drinfeld center $\CZ_1(2\Vect_\CG)$ and its Lagrangian condensable algebras. We subsequently turn to the anomalous case and its Drinfeld center.

\subsection{2-Group without Anomaly}
\label{sec:cat-no-anomaly}

In addition to group-theoretical and cohomological data, the general classification of fusion 2-categories involves a non-degenerate braided fusion category $\CA$ equipped with an action of the relevant subgroup~\cite{Decoppet:2024htz}. Physically, $\CA$ describes intrinsic topological order on the $(2+1)$-dimensional boundary; a non-trivial $\CA$ therefore corresponds to non-minimal boundary conditions~\cite{Bhardwaj:2024qiv}. This datum is not part of the 2-group $\CG=(\Pi_1,\Pi_2,\alpha,\beta)$ and does not arise from the gauging construction of~\cite{Tachikawa:2017gyf}. For $2\Vect_\CG$, one has $\CA=\Vect$, so this additional sector is absent~\cite{Decoppet:2024htz,Decoppet:2023uoy,Wen:2025thg,Xi:2023djc}.

To identify the remaining classifying data, let $\widehat{\Pi}_2=\Hom(\Pi_2,U(1))$ denote the Pontryagin dual of $\Pi_2$. The action $\alpha$ of $\Pi_1$ on $\Pi_2$ induces the dual action $\widehat{\alpha}$ on $\widehat{\Pi}_2$,
\be
\big(\widehat{\alpha}_g\chi\big)(h)
=\chi\big(\alpha_{g^{-1}}(h)\big),
\qquad
g\in\Pi_1,\quad \chi\in\widehat{\Pi}_2,\quad h\in\Pi_2.
\ee
The notation $\Pi_1\ltimes_{\widehat{\alpha}}\widehat{\Pi}_2$ denotes the corresponding semidirect product, with multiplication
\be
(g,\chi)(g',\chi')
=\big(gg',\,\chi\,\widehat{\alpha}_g(\chi')\big).
\ee

The gauging construction of~\cite{Tachikawa:2017gyf} explains the remaining data. In three spacetime dimensions, the parent theory has ordinary 0-form symmetry $\Pi_1\ltimes_{\widehat{\alpha}}\widehat{\Pi}_2$, reducing to $\Pi_1\times\widehat{\Pi}_2$ when $\alpha$ is trivial. Let $a$ be the $\Pi_1$ background introduced previously and $c$ the degree-one background for $\widehat{\Pi}_2$. The parent theory has the mixed anomaly
\be
[\pi_\beta]
=\big[c\cup\beta(a)\big]
\in H^4\big(B(\Pi_1\ltimes_{\widehat{\alpha}}\widehat{\Pi}_2);U(1)\big),
\ee
where the evaluation pairing between $\widehat{\Pi}_2$ and $\Pi_2$ is understood. Gauging the normal Abelian subgroup $\widehat{\Pi}_2$ produces the dual 1-form symmetry $\Pi_2$ with the degree-two background $b$ introduced previously. The mixed anomaly above becomes the modified flatness condition
\be
\delta_a b=\beta(a)\,,
\ee
the second condition in~\eqref{weak-flat}. The remaining 0-form symmetry $\Pi_1$ and emergent 1-form symmetry $\Pi_2$ therefore form the 2-group $\CG$ with Postnikov class $\beta$. At the cochain level,
\be
\pi_\beta\big((g_1,\chi_1),(g_2,\chi_2),(g_3,\chi_3),(g_4,\chi_4)\big)
=\chi_1\big(\beta(g_2,g_3,g_4)\big).
\ee
The restriction to $\widehat{\Pi}_2$ is trivial. Indeed, its elements have the form $(e,\chi)$, where $e$ is the identity of $\Pi_1$, and the normalization of $\beta$ gives
\be
\pi_\beta\big((e,\chi_1),(e,\chi_2),(e,\chi_3),(e,\chi_4)\big)
=\chi_1\big(\beta(e,e,e)\big)=1.
\ee
Hence $\widehat{\Pi}_2$ can be gauged without a counterterm. The final entry in the general classification records a trivialization of the restricted anomaly, which is trivial here and is written as $\mathrm{triv.}$; we obtain
\be
2\Vect_\CG\simeq
\mfr{C}\big[
\Pi_1\ltimes_{\widehat{\alpha}}\widehat{\Pi}_2,\,
\widehat{\Pi}_2,\,
\pi_\beta,\,
\CA =\Vect,\,
\mathrm{triv.}
\big].
\ee

Specializing to $\Pi_1\cong\Pi_2\cong\bbZ_2$ and trivial $\alpha$ gives $\widehat{\Pi}_2\cong\bbZ_2$ and
\be
\Pi_1\ltimes_{\widehat{\alpha}}\widehat{\Pi}_2
=\Pi_1\times\widehat{\Pi}_2
\cong\bbZ_2\times\bbZ_2.
\ee
The non-trivial class $[\beta]\in H^3(B\bbZ_2;\bbZ_2)$ is represented by the normalized cocycle
\be
\beta(a,b,c)=abc,
\qquad
a,b,c\in\bbZ_2.
\ee

Gauging in the 2-group symmetry is equivalent to choosing a condensable $E_1$ algebra $A$ of $2\Vect_\CG$. Here we take another approach: gauge the 1-form symmetry in $2\Vect_\CG$ to switch back to $2\Vect_{\bbZ_2\times \bbZ_2}^{\pi_\beta}$, then consider the condensable $E_1$ algebra in $2\Vect_{\bbZ_2\times \bbZ_2}^{\pi_\beta}$\footnote{The notion of fusion 2-categories $2\Vect_G^\pi$ (0-form symmetry $G$ with anomaly $\pi\in H^4(BG;U(1))$) and $2\Vect_\CG$ (2-group symmetry) are introduced in detail in Example~\ref{exm:0-formGsym} and Example~\ref{exm:2-groupsym} respectively.}. Generically the condensable $E_1$ algebra takes a complicated form, but here we only consider the ones corresponding to the minimal gapped boundary conditions of the SymTFT, whose corresponding fusion 2-categories take the form of
\be  \mfr{C}[G,H,\pi,\CA = \Vect , \psi] \ . \ee
In the following, we will abbreviate the notation as $\mfr{C}[G,H,\pi,\psi]$. Once we have a symmetry category, an immediate question is its relation with other symmetry categories through gauging. The mathematical notion for this relation between two symmetry categories is Morita equivalence, see Appendix~\ref{app:fusion2cat} for elaborations. 
The Morita equivalent symmetry categories are given by condensable $E_1$ algebras of $2\Vect_{\bbZ_2\times \bbZ_2}^{\pi_\beta}$ (which is in one-to-one correspondence with $E_1$ algebras in $2\Vect_{\CG}$) that take the form of
\be\ba
&\Vect_H^\psi \\ 
&H\le \bbZ_2\times \bbZ_2 \text{ up to conjugation} \\
&\psi\in C^3(H;U(1)) \ \text{such that} \  d\psi = \pi\big|_H \ .
\ea\ee
The fusion 2-category corresponding to the $E_1$ algebra $\Vect_H^\psi$ is~\cite{Wen:2025thg,Decoppet:2023uoy,Decoppet:2024htz}
\be
\mfr{C}[\bbZ_2\times \bbZ_2,H,\pi_\beta,\psi] \simeq \mathbf{Bimod}_{\Vect_H^\psi-\Vect_H^\psi}(2\Vect_{\bbZ_2\times \bbZ_2}^{\pi_\beta}) \ .
\ee
To describe the relative SPT decoration $\psi$, let us choose a reference
trivialization $\psi_0$ satisfying
\be
\delta\psi_0=\pi_\beta|_H.
\ee
Any other trivialization can be written as
\be
\psi=\psi_0+\eta,
\qquad
\delta\eta=0,
\qquad
[\eta]\in H^3(BH;U(1)).
\ee
The class $[\eta]$ labels a $(2+1)$-dimensional $H$-SPT.
Accordingly, the fusion $2$-category
$\mfr{C}[\bbZ_2\times\bbZ_2,H,\pi_\beta,\psi]$ can be physically
understood through the following procedure:
\begin{enumerate}
    \item Start with the anomalous $0$-form symmetry fusion
    $2$-category $2\Vect_{\bbZ_2\times\bbZ_2}^{\pi_\beta}$.
    \item Choose a subgroup $H$ for which
    $[\pi_\beta|_H]=0$, together with a reference trivialization
    $\psi_0$ of the restricted anomaly.
    \item Stack the condensation interface with the
    $(2+1)$-dimensional $H$-SPT labeled by
    $[\eta]=[\psi-\psi_0]\in H^3(BH;U(1))$.
    \item Gauge the subgroup $H$ using the resulting trivialization
    $\psi=\psi_0+\eta$.
\end{enumerate}

We can also study the gauging from the SymTFT. The SymTFT is a (3+1)D TQFT whose defects are given by the Drinfeld center $\CZ_1(2\Vect_\CG)$, which is equivalent to the Drinfeld center of the anomalous 0-form symmetry $2\Vect_{\bbZ_2\times \bbZ_2}^{\pi_\beta}$
\be
\CZ_1(2\Vect_\CG) \simeq \CZ_1(2\Vect_{\bbZ_2\times \bbZ_2}^{\pi_\beta}) \ ,
\ee
which can further be decomposed into components
\be\ba  \CZ_1(2\Vect_{\bbZ_2\times \bbZ_2}^{\pi_\beta}) = &\boxplus_{h\in {\rm Cl}(G)} 2\mathbf{Rep}(C_G(h),\tau_h(\pi_\beta)) \\ 
= &2\mathbf{Rep}\Big(\bbZ_2\times \bbZ_2,\tau_{(0,0)}(\pi_\beta)\Big)  \boxplus 2\mathbf{Rep}\Big(\bbZ_2\times \bbZ_2,\tau_{(0,1)}(\pi_\beta)\Big) \\ &\boxplus 2\mathbf{Rep}\Big(\bbZ_2\times \bbZ_2,\tau_{(1,0)}(\pi_\beta)\Big) \boxplus 2\mathbf{Rep}\Big(\bbZ_2\times \bbZ_2,\tau_{(1,1)}(\pi_\beta)\Big) \ . 
\ea\ee
where the transgression $\tau_h : H^{k+1}(C_G(h);U(1))\to H^k(C_G(h);U(1))$ is given by
\be  \tau_h(\pi)(g_1,\dots,g_k) = \prod_{0\le i \le k} \pi(g_1,\dots,g_i,h,g_{i+1},\dots,g_k)^{(-1)^i} \ .  \ee

The topological boundaries of the SymTFT are classified by the Lagrangian $E_2$ algebras of $\CZ_1(2\Vect_\CG)$. However, not all these topological conditions of the SymTFT give symmetry categories that are Morita equivalent to $2\Vect_\CG$, we intend to consider the topological boundary conditions that come from gauging in $2\Vect_\CG$, i.e. the minimal boundary conditions~\cite{Bhardwaj:2024qiv}. For each $E_1$ algebra $\mathtt{A}$ in fusion 2-category $\mfr{C}$, the Lagrangian algebra in $\CZ_1(\mfr{C})$ corresponding to the boundary condition $\mathbf{Bimod}_{\mathtt{A}-\mathtt{A}}(\mfr{C})$ is the full center\footnote{See Appendix~\ref{app:gaugingfullcenter} for explicit derivations of this result.},
\be  \mathbf{Z}_1^\mfr{C}(\mathtt{A}) \in {\rm Alg}_{E_2}\big(\CZ_1(\mfr{C})\big) . \ee
The possible choices are presented in Table~\ref{tab:E1algebra}. We will discuss the realization of these topological boundary conditions of SymTFT through the TQFT action in Section~\ref{sec:without2-groupanomaly}.
\begin{table}[htbp]
\begin{center}
\renewcommand{\arraystretch}{1.5}
\begin{tabular}{|c|c|c|c|c|}
\hline
Subgroup $H$ & & \makecell{Condensable $E_1$  algebra\\ of $2\Vect_{\bbZ_2\times \bbZ_2}^{\pi_\beta}$} & Fusion 2-category   & Representative of $\psi$  \\ 
\hline
$\{e \}$ & $\mc{L}_1$ & $\Vect$  & $2\Vect_{\bbZ_2\times \bbZ_2}^{\pi_\beta}$ & 0 \\ \hline
\multirow{2}{*}{$\bbZ_2\cong \widehat{\Pi}_2$} & $\mc{L}_2^+$ & $\Vect_{\bbZ_2}$          & $2\Vect_\CG$                                            & 0                            \\ \cline{2-5}  & $\mc{L}_2^-$ & $\Vect_{\bbZ_2}^{\psi_1}$ & $\mfr{C}[\bbZ_2\times \bbZ_2, \bbZ_2,\pi_\beta,\psi_1]$ & $\psi_1(a,b,c) = (-1)^{abc}$ \\ \hline
\multirow{2}{*}{$ \bbZ_2\cong \Pi_1 $}         & $\mc{L}_3^+$ & $\Vect_{\bbZ_2}$          & $2\mathbf{Rep}(\CG)$                                    & 0                            \\ \cline{2-5}  & $\mc{L}_3^-$ & $\Vect_{\bbZ_2}^{\psi_2}$ & $\mfr{C}[\bbZ_2\times \bbZ_2, \bbZ_2,\pi_\beta,\psi_2]$ & $\psi_2(a,b,c) = (-1)^{abc}$ \\ \hline
\multirow{2}{*}{$\bbZ_2^{\rm diagonal}$}       & $\mc{L}_4^+$ & $\Vect_{\bbZ_2}^{\psi_3}$ & $\mfr{C}[\bbZ_2\times \bbZ_2, \bbZ_2,\pi_\beta,\psi_3]$ & $\psi_3(a,b,c) = i^{abc}$    \\ \cline{2-5}  & $\mc{L}_4^-$ & $\Vect_{\bbZ_2}^{\psi_4}$ & $\mfr{C}[\bbZ_2\times \bbZ_2, \bbZ_2,\pi_\beta,\psi_4]$ & $\psi_4(a,b,c) = (-i)^{abc}$ \\ \hline
\end{tabular}
\caption{Classification of minimal group-theoretical $E_1$ condensable algebras in $2\Vect_{\bbZ_2\times \bbZ_2}^{\pi_\beta}$ and their corresponding fusion 2-categories. For each condensable $E_1$ algebra $A$, the corresponding Lagrangian algebra of $\CZ_1(2\Vect_{\bbZ_2\times \bbZ_2}^{\pi_\beta})$ is the full center $\mbf{Z}_1^{2\Vect_{\bbZ_2\times \bbZ_2}^{\pi_\beta}}(\mathtt{A})$.} \label{tab:E1algebra}
\end{center}
\end{table}

\subsection{Symmetry Category of 2-Group with Anomaly}
\label{sec:cat-anomaly}

In Section~\ref{sec:Anomalyclassification}, we found that the group of anomalies in the bosonic case is
\be
\begin{aligned}
\operatorname{Hom}\bigl(\widetilde\Omega_4^{\mathrm{SO}}(B\CG),U(1)\bigr)
&\cong \operatorname{Hom}\bigl(H_4(B\CG;\bbZ),U(1)\bigr)\\
&\cong H^4(B\CG;U(1))
\cong\bbZ_2.
\end{aligned}
\ee
Let $\omega$ denote its non-trivial element. To determine its restriction to the 1-form symmetry, consider the fiber inclusion in the Postnikov fibration
\be
B^2\Pi_2=B^2\bbZ_2
\xrightarrow{\ \mathbf{i}\ }
B\CG
\longrightarrow
B\Pi_1=B\bbZ_2.
\ee
Let $\mathfrak{b}\in H^2(B^2\bbZ_2;\bbZ_2)$ be the universal class. On the fiber, $a$ vanishes and $b$ restricts to $\mathfrak{b}$. Since $\omega$ is represented by $\frac12v_4$, where $v_4$ was defined in~\eqref{eq:v4-weak-gauge},
\be\label{eq:anomalyon1-form}
\mathbf{i}^*v_4=\mathfrak{b}\cup \mathfrak{b}=\operatorname{Sq}^2\mathfrak{b},
\qquad
\mathbf{i}^*\omega=\frac12\operatorname{Sq}^2\mathfrak{b}.
\ee
This is precisely the Postnikov invariant
$\frac12\operatorname{Sq}^2(\bar\iota_2)$ governing the degree-two and degree-three layers of the three-layer cochain model reviewed in Appendix~\ref{app:spin-bordism-dual}, with $\bar\iota_2=b$.
Thus $\mathbf{i}^*\omega$ is the unique element of order two in
$H^4(B^2\bbZ_2;U(1))\cong\bbZ_4$. For a 1-form background $b$ on a closed four-manifold, the corresponding phase is
\be
(-1)^{\int_{M_4}\operatorname{Sq}^2\mathfrak{b}}.
\ee

This non-trivial restriction implies (cf. Example 4.12 of~\cite{Decoppet:2024moc} and Construction 2.1.16 of~\cite{Douglas:2018qfz}) that the looping of $2\Vect_\CG^\omega$ is $\sVect$,
\be\ba
&\Omega(\mfr{C}):= {\rm End}_{\mfr{C}}(\mathbf{1}) \ ,
&\Omega 2\Vect_\CG^\omega\simeq\sVect \ .
\ea\ee
Thus, the category $2\Vect_\CG^\omega$ fits into the definition of a fermionic strongly fusion 2-category:
\begin{definition}
    A fusion 2-category $\mfr{C}$ is called bosonic strongly fusion if $\Omega\mfr{C}\simeq\Vect$. A fusion 2-category $\mfr{C}$ is called fermionic strongly fusion if $\Omega\mfr{C}\simeq\sVect$.
\end{definition}

Bosonic/fermionic strongly fusion 2-categories are group-like:
\begin{theorem}[\cite{Xu:2026EtaleAlgebras,Johnson-Freyd:2020ivj}]
    Every strongly fusion 2-category $\mfr{C}$ satisfies one of the following:
    \begin{enumerate}
        \item If $\Omega \mfr{C} \simeq \Vect$, then $\mfr{C}$ is monoidally equivalent to $2\Vect^\pi_G$ for some finite group $G$ and cohomology class $\pi\in H^4(BG;U(1))$.
        \item If $\Omega \mfr{C} \simeq \sVect$, then $\mfr{C}$ is monoidally equivalent to $2\sVect^\varpi_{(G,z)}$ for some finite supergroup $(G,z)$ and supercohomology class $\varpi\in SH^4(B(G,z))$.
    \end{enumerate}
\end{theorem}

Our category $2\Vect_\CG^\omega$ falls into the latter description. The remaining job is to determine the supergroup $(G,z)$ and the anomaly $\varpi\in SH^4(B(G,z))$.

\paragraph{Determining the group theoretical data.}
Let $\mfr{C}$ be a fermionic strongly fusion 2-category, and let 
\be \pi_0(\mfr{C}):=
\frac{\{\text{equivalence classes of simple objects of }\mathfrak C\}}
{\text{existence of a nonzero $1$-morphism}}.
\ee
be its group of connected components whose multiplication is induced by the monoidal structure of $\mfr{C}$\footnote{See Definition~\ref{def:connectedcomponents} and Proposition~\ref{prop:pi0group} for an explicit definition.}. In the case of $\mfr C = 2\sVect_{(G,z)}^\varpi$, it is demonstrated that $\pi_0(\mfr{C}) = G/\expval{z}$~\cite{Decoppet:2024moc}. The physical intuition of the group $\pi_0(\mfr C)$ is the group of codimension-1 defects modulo condensation of line defects~\cite{Gaiotto:2019xmp}. Since the 't Hooft anomaly does not change the labeling of symmetry defects, the 2-group symmetry categories with or without anomaly shall have the same $\pi_0$,
\be  \pi_0\left(2\Vect_\CG \right)  =\pi_0\left(2\Vect_\CG ^\omega\right) = \pi_0\left(2\sVect_{(G,z)}^\varpi\right) \cong \Pi_1 \ . \ee

Thus the data we need to determine are
\be
\upsilon\in H^2(B\Pi_1;\bbZ_2) \ \text{and} \
\varpi\in SH^{4}_{\upsilon}(B\Pi_1).
\ee
The first class determines the supergroup $(G,z)$ appearing in the fermionic
description through the central extension
\be\label{eq:general-fermionic-extension}
1\longrightarrow \expval{z}\cong\bbZ_2
\longrightarrow G_\upsilon
\longrightarrow \Pi_1
\longrightarrow 1.
\ee

A class in $SH^{4}_{\upsilon}(B\Pi_1)$ may be represented schematically by
three layers
\begin{equation}
    (\nu_4,n_3,n_2)
    \in
    \frac{C^4(B\Pi_1;\bbR/\bbZ)}
    {\delta C^3(B\Pi_1;\bbR/\bbZ)}
    \times
    C^3(B\Pi_1;\bbZ_2)\times
    Z^2(B\Pi_1;\bbZ_2).
\end{equation}
subject to the coupled cocycle conditions and equivalence relations
reviewed in Appendix~\ref{app:spin-bordism-dual}. Thus these three
entries are not independent cohomology classes.  In particular, when a
fermionic strongly fusion 2-category is presented as
$2\Vect_{\mathcal G}^{\omega}$ with $\Omega2\Vect_{\mathcal G}^{\omega}\simeq\sVect$, Example~4.12
of~\cite{Decoppet:2024moc} identifies
\be
n_2=0,
\qquad
n_3=\beta,
\ee
where $\beta$ is the Postnikov class of $\mathcal G$.  Determining the
fermionic data therefore requires two logically distinct steps:
\begin{enumerate}
    \item choose the extension class $\upsilon$ so that $\beta$ lies
    in the kernel of the outgoing twisted AHSS differential,
    \be
    d_2^\upsilon(\beta)=0.
    \ee
    \item after imposing
    this condition, determine whether $\beta$ survives to
    $E_\infty$ or is in the image of an incoming differential.  Only
    its class modulo such images contributes to $\varpi$.
\end{enumerate}

\paragraph{Application to the non-split $\bbZ_2$ 2-group.}
For the category $2\Vect_\CG^\omega$ under consideration,
\be
\pi_0\bigl(2\Vect_\CG^\omega\bigr)
\cong \Pi_1
\cong \bbZ_2.
\ee
Consequently, the bosonic quotient of the desired supergroup must be $G/\expval{z}\cong\Pi_1\cong\bbZ_2$. Thus $G$ fits into the central extension
\be\label{eq:fermionicextension}
1\rightarrow \expval{z}\cong\bbZ_2
\rightarrow G
\rightarrow \Pi_1\cong\bbZ_2
\rightarrow 1.
\ee
Such extensions are classified by $\upsilon\in H^2(B\bbZ_2;\bbZ_2)\cong\bbZ_2$.

Let $\mathfrak{a}\in H^1(B\bbZ_2;\bbZ_2)$ denote the non-trivial class and write $\mathfrak{a}^k:=\mathfrak{a}\cup\cdots\cup \mathfrak{a}$.  The Postnikov class of the non-split
2-group is $\beta=\mathfrak{a}^3$. There are two possible extension classes, 
\be\ba
&\upsilon= 0 \ ,\ G = \bbZ_2\times \bbZ_2 \\
&\upsilon= \mathfrak{a}^2 \ , \ G = \bbZ_4 \ .
\ea\ee
On the relevant associated-graded pieces of the twisted AHSS, the
differential is~\cite{Debray:2025kfg}
\be
d_2^\upsilon(x)
=\operatorname{Sq}^2(x)+\upsilon\cup x \ .
\ee
Since $\operatorname{Sq}^2(\mathfrak{a}^3)=\mathfrak{a}^5$, the condition
$d_2^\upsilon(\beta)=0$ becomes
\be
d_2^\upsilon(\beta)=\mathfrak{a}^5+\upsilon\cup \mathfrak{a}^3=0.
\ee
It is solved uniquely by $\upsilon=\mathfrak{a}^2$.  Thus we have obtained the data of
extension $\upsilon = \mathfrak{a}^2$.
The extension~\eqref{eq:fermionicextension} is therefore a
non-split extension
\be
1\rightarrow\bbZ_2^f
\rightarrow\bbZ_4^f
\rightarrow\bbZ_2
\rightarrow 1,
\ee
so that the supergroup data is
\be
(G,z)=(\bbZ_4,\underline{2})\equiv \mb{Z}_4^f\  .
\ee

It remains to determine whether the middle-layer datum
$n_3=\beta$ represents a non-trivial element of
$SH^{4}_\upsilon(B\bbZ_2)$.  Since $n_2=0$, the relation
\be
\delta n_3=(\operatorname{Sq}^2+\upsilon\cup)n_2
\ee
from \eqref{eq:twisted-degree-four-equations} reduces to the cocycle
condition $\delta n_3=0$.  However, closure is not sufficient: one must
still quotient by incoming AHSS differentials.  For $\upsilon=\mathfrak{a}^2$, the generator
$\mathfrak{a}\in H^1(B\bbZ_2;\bbZ_2)$ satisfies~\cite{Debray:2025kfg}
\be
\begin{aligned}
d_2^\upsilon(\mathfrak{a})
&=\operatorname{Sq}^2(\mathfrak{a})+\mathfrak{a}^2\cup \mathfrak{a}\\
&=\mathfrak{a}^3
=\beta,
\end{aligned}
\ee
where we have $\operatorname{Sq}^2(\mathfrak{a})=0$. Thus
$\beta$ is $d_2^\upsilon$-exact and does not survive to the
$E_\infty$ page.  Equivalently, the three-layer representative
$(0,\beta,0)$ represents the trivial total supercohomology class.
This gives the explicit mechanism behind Example~4.13
of~\cite{Decoppet:2024moc}.

Indeed, Example~4.13 of~\cite{Decoppet:2024moc} finds
\be
SH^{4}_{\upsilon}(B\bbZ_2)\cong\bbZ_2.
\ee
The non-trivial element of this group has degree-two layer $n_2=\mathfrak{a}^2$,
whereas the 2-group-graded category under consideration has $n_2=0$.
Moreover,
\be
H^4(B\bbZ_2;U(1))=0,
\ee
so there is no independent degree-four group-cohomology contribution.
We therefore conclude that
\be
\varpi=0\in SH^{4}_{\upsilon}(B\bbZ_2).
\ee
This is a statement about the total supercohomology class; it does not
require every representative to have a pointwise vanishing
degree-four cochain $\nu_4$.

Combining the determination of the supergroup and of the total
supercohomology class, we obtain the monoidal equivalence
\be\label{eq:anomalous-2group-fermionic-equivalence}
2\Vect_\CG^\omega
\simeq
2\sVect_{(\bbZ_4,\underline{2})}.
\ee
Equivalently, in the notation of
Def.~\ref{def:fermionicfusion2cat}, we have
\be
2\Vect_\CG^\omega
\simeq
2\sVect_{(\bbZ_4,\underline{2})}
\simeq
\mfr{C}\bigl[
(\bbZ_4,\underline{2}),
\bbZ_2^f,
0,
\sVect,
0
\bigr].
\ee
We emphasize that \eqref{eq:anomalous-2group-fermionic-equivalence} is a monoidal equivalence of symmetry fusion 2-categories within the classification of fermionic strongly fusion 2-categories, see Examples~4.11--4.13 of~\cite{Decoppet:2024moc}. Here $(\bbZ_4,\underline{2})$ is the finite supergroup given by $\bbZ_4$ with its distinguished central element of order two, and $2\sVect_{(\bbZ_4,\underline{2})}$ is not the bosonic symmetry category $2\Vect_{\bbZ_4}$. In particular, \eqref{eq:anomalous-2group-fermionic-equivalence} is not obtained by gauging the order-two subgroup of an ordinary bosonic $\bbZ_4$ symmetry. The essential fermionic input is the restriction $\mathbf{i}^*\omega=\frac12\operatorname{Sq}^2\mathfrak{b}$ in \eqref{eq:anomalyon1-form}, which makes $\Omega 2\Vect_\CG^\omega\simeq\sVect$. The adjective ``fermionic'' refers to the symmetry category and its lines, and does not by itself mean that the theory has local fermions. The equivalence implies the equivalence of the Drinfeld centers, which determines the SymTFT in the next subsection.

\subsection{SymTFT of Anomalous 2-Group Symmetry}
\label{sec:SymTFT-anomaly}

With the above categorical identification, we are able to construct the categorical data of the SymTFT of the anomalous 2-group symmetry $2\Vect_\CG^\omega$.

The SymTFT of $2\Vect_\CG^\omega$ is given by the Drinfeld center
\be
\CZ_1(2\Vect_\CG^\omega) = \CZ_1(2\sVect_{(\bbZ_4,\underline{2})}) \ .
\ee
The data of the Drinfeld center has been previously computed by~\cite{Decoppet:2024htz,Xu:2026EtaleAlgebras}, for the untwisted cases
\be
\CZ_1(2\sVect_{(\bbZ_4,\underline{2})}) \simeq \boxplus_{[g]\in {\rm Cl}(G)} 2\mathbf{sRep}(C_G(g),z)  =  2\mathbf{sRep}(\bbZ_4,\underline{2})^{\boxplus 4} \ .
\ee
This calculation reveals that the SymTFT possesses 4 connected flux sectors (codimension-2 defects modulo condensation of line operators) and 4 simple codimension-3 defects organized by $\Omega2\mathbf{sRep}(\bbZ_4,\underline{2}) = \mathbf{sRep}(\bbZ_4,\underline{2})$.

The natural candidate for a Lagrangian description of the SymTFT is thus a fermionic $\bbZ_4$ BF-theory. We will propose a Lagrangian form of this SymTFT and unravel its boundary conditions in Section~\ref{sec:with2-groupanomaly}.

Now we can give the classification of the minimal topological boundary conditions of $\CZ_1(2\sVect_{(\bbZ_4,\underline{2})})$. Each minimal topological boundary condition of $\CZ_1(2\sVect_{(G,z)})$ is given by~\cite{Xu:2026EtaleAlgebras}
\begin{enumerate}
    \item an inclusion of finite supergroup $(H,z)\hookrightarrow(G,z)$ up to conjugation;
    \item a class $\mu \in SH^3(B(H,z))$.
\end{enumerate}
Therefore, there are in total 2 minimal topological boundary conditions of this SymTFT:
\begin{itemize}
    \item The minimal supersubgroup is $H=\langle z\rangle\cong\bbZ_2^f$. Its bosonic quotient $H_{\mathrm b}=H/\langle z\rangle$ is a point. Here $SH^3(B\bbZ_2^f)$ is not the supercohomology of $B\bbZ_2$ with an ordinary internal symmetry: the fermion-parity subgroup is absorbed into the tangential structure. Thus
    \be
    \mu\in SH^3(B\bbZ_2^f)
    \equiv SH^3(\mathrm{pt})=0.
    \ee
    The corresponding fermionic fusion 2-category is
    \be
    \mfr{C}[(\bbZ_4,\underline{2}),\bbZ_2^f,0,\sVect,0]
    \simeq 2\Vect_\CG^\omega \ .
    \ee We denote this boundary condition by $\mc{L}_1$.
    \item The sub-supergroup $(H,z) = (\bbZ_4,\underline{2})$ with $\mu  \in SH^3_\upsilon(B\bbZ_2)= 0$ where $\upsilon$ is the non-trivial element of $H^2(B\bbZ_2;\bbZ_2)\cong \bbZ_2$~\cite{Wang:2018pdc}. The corresponding fermionic fusion 2-category is 
    \be\mfr{C}[(\bbZ_4,\underline{2}),(\bbZ_4,\underline{2}),0,\sVect,0]\simeq 2\mathbf{sRep}(\bbZ_4,\underline{2}) \ .\ee We denote this boundary condition by $\mc{L}_2$.
\end{itemize}

\section{(3+1)D SymTFT of non-split 2-group symmetry}
\label{sec:TQFT}

In this section, we discuss the (3+1)D SymTFT (topological order) of the non-split 2-group $\mc{G}=(\mb{Z}_2,\mb{Z}_2,\mathrm{triv.},1)$ for both the non-anomalous and the anomalous cases. In Section~\ref{sec:without2-groupanomaly}, we write the TQFT both in terms of $U(1)$-valued gauge fields $A^i$, $B^i$ $(i=1,2)$ and in terms of $\mb{Z}_2$-valued simplicial cochains $a^i$, $b^i$ on a triangulated spacetime with a branching structure. In Section~\ref{sec:with2-groupanomaly}, $A$ and $B$ without superscripts denote $\mb{Z}_4$-valued cochains.

\subsection{Without the 2-group anomaly}
\label{sec:without2-groupanomaly}

\subsubsection{TQFT action}

For the case without the 2-group anomaly, i.e. corresponding to the trivial element in $\Hom(\widetilde{\Omega}_4^{\rm SO}(B\mc{G}),U(1))=\mb{Z}_2$, we claim that the (3+1)D SymTFT can be modeled by the following topological action with $U(1)$-valued 1-form gauge fields $A^1,A^2$ and 2-form gauge fields $B^1,B^2$, constructed and studied in e.g.~\cite{Putrov:2016qdo,Wang:2017loc,Chan:2017eov,Wang:2018iwz,Zhang:2020kgc,Zhang:2022rbg}:
\be
\label{AAdA}
S_{\rm SymTFT}=\frac{2}{2\pi}\int_{M_4}\left(B^1\wedge dA^1+B^2\wedge dA^2+\frac{1}{2\pi} A^1\wedge A^2\wedge dA^2\right)\,.
\ee
The gauge fields are quantized such that the magnetic fluxes over closed submanifolds are integer multiples of $2\pi$. In terms of $\mb{Z}_2$-valued 1-cochains $a^1,a^2\in C^1(M_4;\mb{Z}_2)$ and 2-cochains $b^1,b^2\in C^2(M_4;\mb{Z}_2)$, the action reads
\be
\label{AAdA-Z2}
S_{\rm SymTFT}=\pi\int_{M_4}\left(b^1\cup\delta a^1+b^2\cup\delta a^2+a^1\cup a^2\cup a^2\cup a^2\right)\,.
\ee
The two formulations are related by $A^i\rightarrow\pi\tilde a^i$, $B^i\rightarrow\pi\tilde b^i$, $d\rightarrow\delta$ and $\wedge\rightarrow\cup$, where $\tilde x$ denotes the lift of a $\mb{Z}_2$-valued cochain $x$ to an integer-valued cochain with values in $\{0,1\}$. In particular, $\frac{1}{2\pi}dA^2\rightarrow\frac12\delta\tilde a^2=\tilde a^2\cup\tilde a^2$ for $\delta a^2=0$, so that the twist term in (\ref{AAdA}) becomes $\pi\, a^1\cup a^2\cup a^2\cup a^2$ mod $2\pi$. Summing over $b^1$ and $b^2$ imposes the flatness conditions $\delta a^1=\delta a^2=0$. Hence (\ref{AAdA-Z2}) is the $\mb{Z}_2\times\mb{Z}_2$ Dijkgraaf-Witten theory twisted by the 4-cocycle $a^1\cup a^2\cup a^2\cup a^2$, which represents $\pi_\beta$ in Section~\ref{sec:cat-no-anomaly}. 

The gauge transformations are
\be
\ba
\label{SymTFT-gauge-trans}
&A^1\rightarrow A^1+dg^1\cr
&A^2\rightarrow A^2+dg^2\cr
&B^1\rightarrow B^1+d\lambda^1-\frac{1}{2\pi} g^2\wedge dA^2\cr
&B^2\rightarrow B^2+d\lambda^2+\frac{1}{2\pi} g^1\wedge dA^2\,.\cr
\ea
\ee
$g^1$, $g^2$ are 0-form gauge parameters and $\lambda^1$, $\lambda^2$ are 1-form gauge parameters. In the cochain formulation, with $g^1,g^2\in C^0(M_4;\mb{Z}_2)$ and $\lambda^1,\lambda^2\in C^1(M_4;\mb{Z}_2)$, they read
\be
\ba
\label{SymTFT-gauge-trans-Z2}
&a^1\rightarrow a^1+\delta g^1\cr
&a^2\rightarrow a^2+\delta g^2\cr
&b^1\rightarrow b^1+\delta\lambda^1+\zeta(a^2,g^2)\cr
&b^2\rightarrow b^2+\delta\lambda^2+g^1\cup(a^2+\delta g^2)\cup(a^2+\delta g^2)+a^1\cup\big(a^2\cup_1\delta g^2+g^2\cup\delta g^2\big)\,,\cr
\ea
\ee
where $\zeta$ is the compensator in (\ref{zeta-canonical}). The terms $-\frac{1}{2\pi}g^2\wedge dA^2$ and $\frac{1}{2\pi}g^1\wedge dA^2$ in (\ref{SymTFT-gauge-trans}) correspond to $g^2\cup a^2\cup a^2$ and $g^1\cup a^2\cup a^2$ in (\ref{SymTFT-gauge-trans-Z2}). The remaining terms in (\ref{SymTFT-gauge-trans-Z2}) contain $\delta g^2$. They are needed since $a^2\cup a^2$, unlike $dA^2$, is not invariant under $a^2\rightarrow a^2+\delta g^2$ at the cochain level. On shell, the twist term in (\ref{AAdA-Z2}) changes by the coboundary $\delta\big(g^1\cup(a^2+\delta g^2)^3+a^1\cup\zeta(a^2,g^2)\big)$, and the shifts of $b^1$ and $b^2$ make the surface operators (\ref{Lmn-Z2}) below gauge invariant.

The equations of motion are
\be
\ba
dA^1&=0\cr
dA^2&=0\cr
dB^1+\frac{1}{2\pi}A^2\wedge dA^2&=0\cr
dB^2-\frac{1}{2\pi}A^1\wedge dA^2&=0\,,
\ea
\ee
and in the cochain formulation
\be
\ba
\delta a^1&=0\cr
\delta a^2&=0\cr
\delta b^1+a^2\cup a^2\cup a^2&=0\cr
\delta b^2+a^1\cup a^2\cup a^2&=0\,,
\ea
\ee
where the first two equations are imposed by summing over $b^1$ and $b^2$. 

The gauge-invariant topological operators are line operators $(m,n=0,1)$
\be
\label{Pmn}
P_{mn}=\exp\left(im\oint_C A^1+in\oint_C A^2\right)=\exp\left(\pi im\oint_C a^1+\pi in\oint_C a^2\right)\,,
\ee
and surface operators $(m,n=0,1)$
\be
\ba
\label{Lmn}
L_{10}&=\exp\left(i\int_{S}B^1+\frac{i}{2\pi}\int_{\Omega}A^2 \wedge dA^2\right)\,,\cr
L_{01}&=\exp\left(i\int_{S}B^2-\frac{i}{2\pi}\int_{\Omega}A^1\wedge dA^2\right)\,,\cr
L_{11}&=\exp\left(i\int_{S}B^1+\frac{i}{2\pi}\int_{\Omega}A^2\wedge dA^2+i\int_{S}B^2-\frac{i}{2\pi}\int_{\Omega}A^1\wedge dA^2\right)\,,
\ea
\ee
which read
\be
\ba
\label{Lmn-Z2}
L_{10}&=\exp\left(\pi i\int_{S}b^1+\pi i\int_{\Omega}a^2\cup a^2\cup a^2\right)\,,\cr
L_{01}&=\exp\left(\pi i\int_{S}b^2+\pi i\int_{\Omega}a^1\cup a^2\cup a^2\right)\,,\cr
L_{11}&=\exp\left(\pi i\int_{S}b^1+\pi i\int_{\Omega}a^2\cup a^2\cup a^2+\pi i\int_{S}b^2+\pi i\int_{\Omega}a^1\cup a^2\cup a^2\right)
\ea
\ee
in the cochain formulation. Here $\Omega$ is a 3-chain with $\ptl\Omega=S$. These operators have the same fusion rules as the 16 topological operators in the untwisted $\mb{Z}_2\times\mb{Z}_2$ BF theory:
\be
S_{\rm BF}=\frac{2}{2\pi}\int_{M_4}\left(B^1\wedge dA^1+B^2\wedge dA^2\right)=\pi\int_{M_4}\left(b^1\cup\delta a^1+b^2\cup\delta a^2\right)\,.
\ee
But these two TQFTs have different linking correlation functions, due to the twist term $\frac{1}{2\pi} A^1\wedge A^2\wedge dA^2$, i.e.\ $\pi\, a^1\cup a^2\cup a^2\cup a^2$ in the cochain formulation.

For the TQFT (\ref{AAdA}), the non-trivial linking correlation functions include the usual surface-line linking
\be
\ba
\label{Linking}
\langle L_{10}(S) P_{10}(C)\rangle &=(-1)^{\langle S,C\rangle}\cr
\langle L_{01}(S) P_{01}(C)\rangle &=(-1)^{\langle S,C\rangle}
\ea
\ee
as well as the triple surface linking~\cite{Putrov:2016qdo}
\be
\label{TLinking}
\langle L_{10}(S_1)L_{01}(S_2)L_{01}(S_3)\rangle=(-1)^{\mathrm{TLink}(S_1,S_2,S_3)}\,.
\ee
$\mathrm{TLink}(S_1,S_2,S_3)$ is the triple linking number.

\subsubsection{Topological boundary conditions}

Let us discuss the different topological boundary conditions for the TQFT (\ref{AAdA}). These are imposed at the topological boundary of the (3+1)D SymTFT, leading to physical theories with different generalized global symmetries. They correspond to the Lagrangian algebra $\mc{L}_1$, $\mc{L}_2^+$, $\mc{L}_2^-$, $\mc{L}_3^+$, $\mc{L}_3^-$, $\mc{L}_4^+$, $\mc{L}_4^-$ in Table~\ref{tab:E1algebra}.

\paragraph{(1) $\mc{L}_1$: Dirichlet b.c. for $P_{mn}$}

Since $P_{mn}$ have no non-trivial linking correlation function among themselves, we can assign Dirichlet b.c. to all the $P_{mn}$ operators, i.e. assign Dirichlet b.c. to $A^1$ and $A^2$. Thus after the restriction to the (2+1)D boundary
\be
A^1|_{M_3}\rightarrow \pi a^1\ ,\ A^2|_{M_3}\rightarrow \pi a^2\,,
\ee
the gauge fields $a^1$, $a^2\in C^1(M_3;\mb{Z}_2)$ correspond to background gauge fields of a $\mb{Z}_2^{(1)}\times\mb{Z}_2^{(2)}$ 0-form symmetry of the boundary (2+1)D system, with mixed 't Hooft anomaly characterized by the anomaly polynomial
\be
\label{anomaly-pol-U1}
\frac{2}{(2\pi)^2}\int_{M_4}A^1\wedge A^2\wedge dA^2
\ee
in the $U(1)$ formulation and
\be
\label{anomaly-pol}
\pi\int_{M_4}a^1\cup a^2\cup a^2\cup a^2=\pi\int_{M_4}a^1\cup a^2\cup \mathrm{Sq}^1a^2
\ee
in the $\mb{Z}_2$-cochain formulation.

As mentioned in \cite{Wang:2017loc}, such an anomaly corresponds to a non-trivial element of the group cohomology $H^4_{\rm grp}(\mb{Z}_2\times\mb{Z}_2;U(1))=\mb{Z}_2\times\mb{Z}_2$, represented by the 4-cocycle
\be
\pi_\beta((g_1,h_1),(g_2,h_2),(g_3,h_3),(g_4,h_4))=(-1)^{g_1 h_2 h_3 h_4}\quad (g_i\in\mb{Z}_2^{(1)}\,,\ h_i\in\mb{Z}_2^{(2)})\,,
\ee
which is $\pi_\beta$ in Section~\ref{sec:cat-no-anomaly} with $\mb{Z}_2^{(1)}\cong\widehat{\Pi}_2$ and $\mb{Z}_2^{(2)}\cong\Pi_1$.
The mixed anomaly obstructs the gauging of the $\mb{Z}_2^{(1)}\times\mb{Z}_2^{(2)}$ 0-form symmetry simultaneously. However, one can still gauge $\mb{Z}_2^{(1)}$, $\mb{Z}_2^{(2)}$ or the diagonal $\mb{Z}_2$ subgroup of $\mb{Z}_2^{(1)}\times\mb{Z}_2^{(2)}$, which leads to the three other topological boundary conditions.

\paragraph{(2) $\mc{L}_2^+$: Dirichlet b.c. for $L_{10}$ and $P_{01}$}

This boundary condition is obtained by gauging the $\mb{Z}_2^{(1)}$ 0-form symmetry from the boundary condition (1).

The operators $L_{10}$ and $P_{01}$ do not have non-trivial linking correlation functions, hence we can assign Dirichlet b.c. to the gauge fields $A^2$ and $B^1$. After restriction to the (2+1)D boundary, 
they become background gauge fields of the boundary (2+1)D system, in terms of $\mb{Z}_2$-cochains:
\be
A^2|_{M_3}\rightarrow \pi a^2\ ,\ B^1|_{M_3}\rightarrow \pi b^1\,,
\ee
$a^2\in C^1(M_3;\mb{Z}_2)$, $b^1\in C^2(M_3;\mb{Z}_2)$. The gauge transformation rules of the $U(1)$-valued gauge fields are
\be
\ba
A^2&\rightarrow A^2+dg^2\cr
B^1&\rightarrow B^1+d\lambda^1-\frac{1}{2\pi}g^2\wedge dA^2\,,
\ea
\ee
and those of the $\mb{Z}_2$-cochains follow from (\ref{SymTFT-gauge-trans-Z2}):
\be
\ba
\label{ab-gauge-trans}
a^2&\rightarrow a^2+\delta g^2\cr
b^1&\rightarrow b^1+\delta\lambda^1+\zeta(a^2,g^2)\,.
\ea
\ee
These gauge transformations exactly give the 2-group symmetry $\mc{G}=(\mb{Z}_2,\mb{Z}_2,\mathrm{triv.},1)$ with non-trivial Postnikov class, after the restriction of the $U(1)$-valued gauge fields to the $\mb{Z}_2$ subgroup: (\ref{ab-gauge-trans}) are the weak 2-gauge transformations in Section~\ref{sec:2-group} with $(a,b)=(a^2,b^1)$. To see this, let us construct the gauge-invariant curvature for $B^1$:
\be
H=dB^1+\frac{1}{2\pi}A^2\wedge dA^2\,.
\ee
In the $\mb{Z}_2$-cochain formulation, the gauge-invariant curvature is $\delta b^1+a^2\cup a^2\cup a^2$. Hence the flatness condition is exactly the same as the flatness condition $\delta b=\beta(a)=a\cup a\cup a$ in the cochain formulation (\ref{weak-flat}). 

In addition, we have $dH=\frac{1}{2\pi}dA^2\wedge dA^2$, which matches the results for the Lie 2-group $(U(1),U(1),\mathrm{triv.},\kappa=1)$ in \cite{Davighi:2023luh}. The restriction to the $\mb{Z}_2$ subgroup exactly reproduces our result.

\paragraph{(3) $\mc{L}_2^-$: Dirichlet b.c. for dressed $L_{10}$ and $P_{01}$}

In this case, the condensed operators include $P_{01}$ and the dressed operator (stacked by SPT phase) $L_{10}^{\psi_1}$:
\be
L_{10}^{\psi_1}=\exp\left(i\int_S B^1+\frac{i}{2\pi}\int_\Omega A^2\wedge dA^2+\frac{i}{2\pi}\int_\Omega A^1\wedge dA^1\right)\,.
\ee
The additional term $\frac{i}{2\pi}\int_\Omega A^1\wedge dA^1$ corresponds to the discrete torsion $\psi_1(a,b,c)=(-1)^{abc}$ in Table~\ref{tab:E1algebra}.

In terms of the $\mb{Z}_2$-valued cochains, it reads
\be
L_{10}^{\psi_1}=\exp\left(\pi i\left(\int_S b^1+\int_\Omega\left(a^2\cup a^2\cup a^2+a^1\cup a^1\cup a^1\right)\right)\right)\,.
\ee

This boundary condition is obtained by gauging the $\mb{Z}_2^{(1)}$ 0-form symmetry in boundary condition $\mc{L}_1$ in the presence of a stacked SPT phase $\pi\int_{M_3}a^1\cup a^1\cup a^1$:
\be
Z_{\mc{L}_2^-}[a^2,b^1]=\frac{1}{\sqrt{|H^1(M_3;\mb{Z}_2)|}}\sum_{a^1\in H^1(M_3;\mb{Z}_2)}(-1)^{\int_{M_3}a^1\cup a^1 \cup a^1}(-1)^{\int_{M_3}a^1\cup b^1}Z_{\mc{L}_1}[a^1,a^2]\,,
\ee
with $\delta b^1=(a^2)^3$. The boundary condition has exactly the same 2-group symmetry $\mc{G}$, as the background gauge fields $a^2$, $b^1$ and their gauge transformation rules are unchanged. We can interpret the topological boundary condition as having a $\mc{G}$ 2-group symmetry with a stacked $\mc{G}$-SPT phase.

\paragraph{(4) $\mc{L}_3^+$: Dirichlet b.c. for $L_{01}$ and $P_{10}$}

Thus the topological operators with Neumann b.c. are $P_{01}$ and $L_{10}$. After pushing to the boundary, they become
\be
\ba
\label{2Rep-gen}
P_{01}(C)|_{M_3}&=W(C)=\exp\left(i\oint_C A^2\right)\cr
L_{10}(S)|_{M_3}&=U(S)=\exp\left(i\int_S B^1+\frac{i}{2\pi}\int_\Omega A^2\wedge dA^2\right)\,,
\ea
\ee
forming a higher-categorical symmetry. In the $\mb{Z}_2$-cochain formulation, they are
\be
\ba
P_{01}(C)|_{M_3}&=W(C)=\exp\left(\pi i\oint_C a^2\right)\cr
L_{10}(S)|_{M_3}&=U(S)=\exp\left(\pi i\left(\int_S b^1+\int_\Omega a^2\cup a^2\cup a^2\right)\right)\,,
\ea
\ee

$L_{10}$ is dressed by a Chern-Simons action $A^2\wedge dA^2$, i.e.\ the Dijkgraaf-Witten action $\pi\int_\Omega a^2\cup a^2\cup a^2$ in the cochain formulation, thus $U(S)$ generates a non-invertible 0-form symmetry in (2+1)D, with the fusion rule
\be
U(S)\otimes \overline{U}(S)=\mc{C}(S)\equiv \int [Da^2][D\bar{a}^2]\exp\left(\pi i\int_\Omega (a^2\cup a^2\cup a^2-\bar{a}^2\cup \bar{a}^2\cup \bar{a}^2)\right)\,.
\ee 
From the perspective of (2+1)D boundary theory, this boundary condition is equivalent to gauging the non-anomalous 2-group symmetry $\mc{G}=(\mb{Z}_2,\mb{Z}_2,\mathrm{triv.},1)$ in the boundary condition (2), hence the 2-categorical symmetry is $\TwoRep(\CG)$, see the definition in (\ref{2Rep-G}). Alternatively, this boundary condition is obtained by gauging the 0-form symmetry $\mb{Z}_2^{(2)}$ in the boundary condition (1).

\paragraph{(5) $\mc{L}_3^-$: Dirichlet b.c. for dressed $L_{01}$ and $P_{10}$}

In this case, the condensed operators include $P_{10}$ and the dressed operator $L_{01}^{\psi_2}$:
\be
L_{01}^{\psi_2}=\exp\left(i\int_S B^2-\frac{i}{2\pi}\int_\Omega A^1\wedge dA^2+\frac{i}{2\pi}\int_\Omega A^2\wedge dA^2\right)\,.
\ee
The additional term $\frac{i}{2\pi}\int_\Omega A^2\wedge dA^2$ corresponds to the twist $\psi_2(a,b,c)=(-1)^{abc}$ in Table~\ref{tab:E1algebra}.

In terms of the $\mb{Z}_2$-valued cochains, it reads
\be
L_{01}^{\psi_2}=\exp\left(\pi i\left(\int_S b^2+\int_\Omega\left(a^1\cup a^2\cup a^2+a^2\cup a^2\cup a^2\right)\right)\right)\,.
\ee

This boundary condition is obtained by gauging the 0-form symmetry $\mb{Z}_2^{(2)}$ in the boundary condition $\mc{L}_1$, with a discrete torsion:
\be
Z_{\mc{L}_3^-}[a^1,b^2]=\frac{1}{\sqrt{|H^1(M_3;\mb{Z}_2)|}}\sum_{a^2\in H^1(M_3;\mb{Z}_2)}(-1)^{\int_{M_3}a^2\cup a^2 \cup a^2}(-1)^{\int_{M_3}a^2\cup b^2}Z_{\mc{L}_1}[a^1,a^2]\,,
\ee
and the 2-categorical symmetry still has the same fusion rules as $\TwoRep(\CG)$, nonetheless with a stacked SPT which we interpret as a $\TwoRep(\CG)$-SPT. 

\paragraph{(6) $\mc{L}_4^+$: Dirichlet b.c. for $P_{11}(C)$ and dressed $L_{11}(S)$ with $\psi_3$}

Starting from the boundary condition 1, this corresponds to the gauging of a non-anomalous, diagonal $\mb{Z}_2$ subgroup of the 0-form symmetry group $\mb{Z}_2^{(1)}\times\mb{Z}_2^{(2)}$. This is possible because after assigning the diagonal value $A^1=A^2$ for the background 1-form gauge fields, the anomaly polynomial (\ref{anomaly-pol-U1}) vanishes. In the cochain formulation, with $a^1=a^2=a$, the anomaly polynomial (\ref{anomaly-pol}) becomes $\pi\int_{M_4}a\cup a\cup a\cup a$, which is non-zero as a cochain but trivial in cohomology:
\be
\frac12\,a\cup a\cup a\cup a=\delta\Big(\frac14\,\tilde a\cup\tilde a\cup\tilde a\Big)\quad\mathrm{mod}\ 1\,,
\ee
where we have used $\delta\tilde a=2\,\tilde a\cup\tilde a$. The two trivializations $\pm\frac14\,\tilde a\cup\tilde a\cup\tilde a$ correspond to $\psi_3$ and $\psi_4$ in Table~\ref{tab:E1algebra}, respectively. This choice is invisible in the $U(1)$ formulation, where $A^1\wedge A^2\wedge dA^2$ vanishes identically for $A^1=A^2$. The condensed operators are $P_{11}(C)$ and the dressed operator
\be
L_{11}^{\psi_3}(S)=\exp\left(i\int_S (B^1+B^2)+\frac{i}{2\pi}\int_\Omega (A^2-A^1)\wedge dA^2+\frac{i}{4\pi}\int_\Omega A^2\wedge dA^2\right)\,,
\ee
which reads
\be
L_{11}^{\psi_3}(S)=L_{11}(S)\exp\left(\frac{\pi i}{2}\int_\Omega \tilde a^2\cup\tilde a^2\cup\tilde a^2\right)
\ee
in the cochain formulation, with $L_{11}(S)$ in (\ref{Lmn-Z2}).

On the (2+1)D boundary, the generalized global symmetry is generated by 
\be
\ba
W(C)&=P_{01}(C)|_{M_3}\cr
U(S)&=L_{10}(S)|_{M_3}\,.
\ea
\ee

The fusion rules of $W(C)$ and $U(S)$ are the same as in the boundary condition (3), hence the 2-categorical symmetry is expected to be $\TwoRep(\CG)$ as well, in this case of diagonal boundary condition, with a stacked $\TwoRep(\CG)$-SPT.

\paragraph{(7) $\mc{L}_4^-$: Dirichlet b.c. for $P_{11}(C)$ and dressed $L_{11}(S)$ with $\psi_4$}

This case is almost the same as $\mc{L}_4^+$, and the only difference is that the condensed dressed operator is
\be
L_{11}^{\psi_4}(S)=\exp\left(i\int_S (B^1+B^2)+\frac{i}{2\pi}\int_\Omega (A^2-A^1)\wedge dA^2-\frac{i}{4\pi}\int_\Omega A^2\wedge dA^2\right)\,,
\ee
which reads $L_{11}^{\psi_4}(S)=L_{11}(S)\exp\big(-\frac{\pi i}{2}\int_\Omega \tilde a^2\cup\tilde a^2\cup\tilde a^2\big)$ in the cochain formulation.
The 2-categorical symmetry is also $\TwoRep(\CG)$, with a stacked $\TwoRep(\CG)$-SPT.

\paragraph{Relation of the topological boundary conditions}

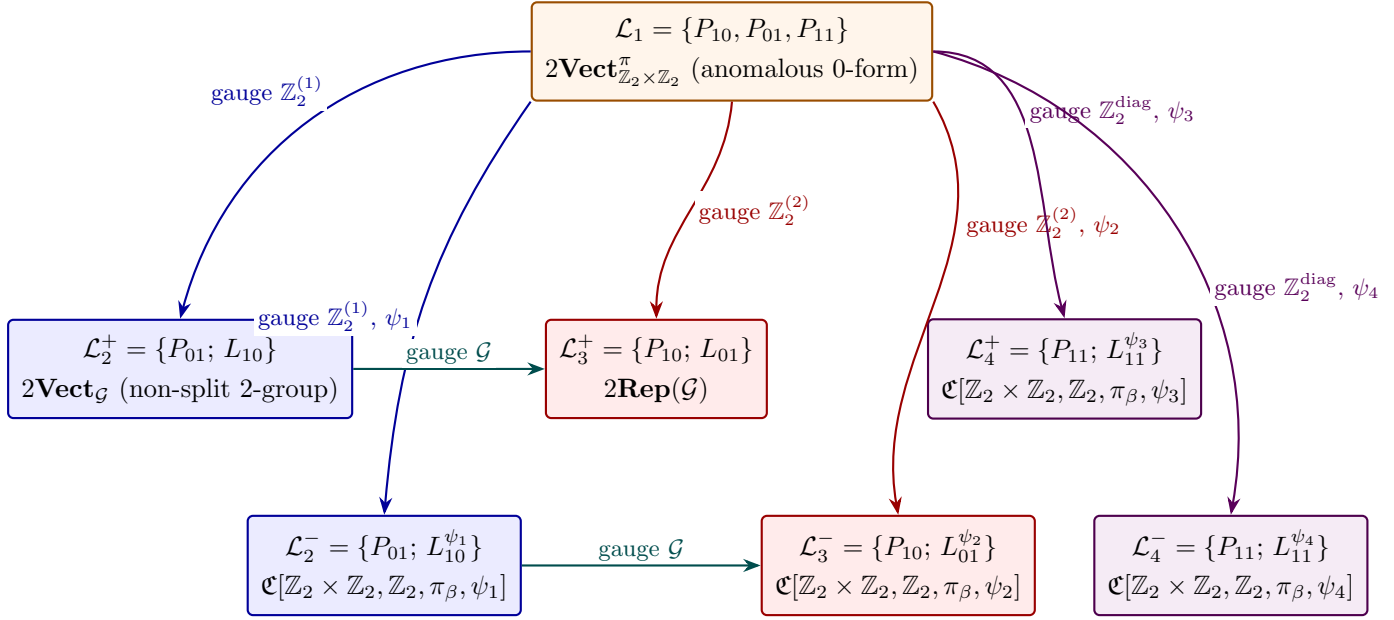
\begin{figure}
\begin{center}
\begin{tikzpicture}[
  every node/.style={font=\small},
  bc/.style={draw, rounded corners=2pt, align=center, inner sep=5pt, fill=#1!8,
             draw=#1!60!black, thick, minimum height=13mm},
  glab/.style={font=\footnotesize, fill=white, inner sep=1.5pt, align=center},
  gA/.style={-{Stealth}, thick, blue!60!black},        
  gB/.style={-{Stealth}, thick, red!60!black},         
  gD/.style={-{Stealth}, thick, violet!70!black},      
  gG/.style={-{Stealth}, thick, teal!60!black},        
  du/.style={{Stealth}-{Stealth}, thick, gray, densely dashed}  
]

\node[bc=orange] (L1) at (0,0)
  {$\CL_1=\{P_{10},P_{01},P_{11}\}$\\[1mm]
   $\TwoVect^{\pi_\beta}_{\Z_2\times\Z_2}$ (anomalous $0$-form)};

\node[bc=blue]  (L2p) at (-7.3,-4.2)
  {$\CL_2^{+}=\{P_{01};\,L_{10}\}$\\[1mm] $\TwoVect_{\CG}$ (non-split 2-group)};
\node[bc=blue]  (L2m) at (-4.6,-6.8)
  {$\CL_2^{-}=\{P_{01};\,L_{10}^{\psi_1}\}$\\[1mm] $\mfr{C}[\bbZ_2\times \bbZ_2, \bbZ_2,\pi_\beta,\psi_1]$};

\node[bc=red]   (L3p) at (-1.0,-4.2)
  {$\CL_3^{+}=\{P_{10};\,L_{01}\}$\\[1mm] $\TwoRep(\CG)$};
\node[bc=red]   (L3m) at (2.2,-6.8)
  {$\CL_3^{-}=\{P_{10};\,L_{01}^{\psi_2}\}$\\[1mm] $\mfr{C}[\bbZ_2\times \bbZ_2, \bbZ_2,\pi_\beta,\psi_2]$};

\node[bc=violet] (L4p) at (4.4,-4.2)
  {$\CL_4^{+}=\{P_{11};\,L_{11}^{\psi_3}\}$\\[1mm] $\mfr{C}[\bbZ_2\times \bbZ_2, \bbZ_2,\pi_\beta,\psi_3]$ };
\node[bc=violet] (L4m) at (6.6,-6.8)
  {$\CL_4^{-}=\{P_{11};\,L_{11}^{\psi_4}\}$\\[1mm] $\mfr{C}[\bbZ_2\times \bbZ_2, \bbZ_2,\pi_\beta,\psi_4]$ };

\draw[gA] (L1.west) to[out=180,in=75]
  node[glab, pos=0.45, above left=-1pt] {gauge $\Z_2^{(1)}$} (L2p.north);
\draw[gA] (L1.south west) to[out=235,in=85]
  node[glab, pos=0.55, left=1pt] {gauge $\Z_2^{(1)}$,\ $\psi_1$} (L2m.north);

\draw[gB] (L1.south) to[out=265,in=90]
  node[glab, pos=0.5, right=1pt] {gauge $\Z_2^{(2)}$} (L3p.north);
\draw[gB] (L1.south east) to[out=300,in=100]
  node[glab, pos=0.3, right=2pt] {gauge $\Z_2^{(2)}$,\ $\psi_2$} (L3m.north);

\draw[gD] (L1.east) to[out=0,in=105]
  node[glab, pos=0.45, above right=-1pt] {gauge $\Z_2^{\rm diag}$,\ $\psi_3$} (L4p.north);
\draw[gD] (L1.east) to[out=-15,in=80]
  node[glab, pos=0.62, right=1pt] {gauge $\Z_2^{\rm diag}$,\ $\psi_4$} (L4m.north);

\draw[gG] (L2p.east) --
  node[glab, above=0pt] {gauge $\CG$} (L3p.west);

\draw[gG] (L2m.east) --
  node[glab, above=0pt] {gauge $\CG$} (L3m.west);

\end{tikzpicture}
\end{center}

\caption{ Gauging routes among the seven minimal topological boundary conditions of
$\mathcal{Z}_1(\TwoVect^{\pi_\beta}_{\Z_2\times\Z_2})=\mathcal{Z}_1(\TwoVect_{\mc{G}})$. Colored solid arrows
denote gauging a subgroup $H\subset\Z_2\times\Z_2$ at the boundary: blue $H=\Z_2^{(1)}$, red $H=\Z_2^{(2)}$, violet
$H=\Z_2^{\rm diag}$, teal the
gauging of the full non-split 2-group $\CG=(\Z_2,\Z_2,\mathrm{triv.},1)$ from the 2-group boundary. Every gauging arrow is invertible --- the reverse arrow gauges the
dual (quantum) symmetry. Gauging the full $\Z_2\times\Z_2$ is obstructed
by the anomaly.}\label{f:AAdA-tbc}
\end{figure}

In conclusion, from the boundary condition $\mc{L}_1$, we see that the SymTFT (\ref{AAdA}) corresponds to the Drinfeld center $\CZ(2\Vect_{\mb{Z}_2\times\mb{Z}_2}^{\pi_\beta})$. The boundary condition $\mc{L}_2^\pm$ describes the 3d theory with the $\mc{G}=(\mb{Z}_2,\mb{Z}_2,\mathrm{triv.},1)$ 2-group symmetry. Due to the presence of the boundary condition $\mc{L}_3^\pm$, there is no obstruction to gauge the $\mc{G}$ 2-group symmetry. Hence the SymTFT also describes the Drinfeld center $\CZ(2\Vect_{\mc{G}})$. The boundary conditions $\mc{L}_3^\pm$ and $\mc{L}_4^\pm$ all give 3d physical theories with $\TwoRep(\CG)$ symmetry. We depict the relation of the topological boundary conditions in Figure~\ref{f:AAdA-tbc}.

\subsubsection{Physical boundary conditions}

Given a fixed boundary condition at the topological boundary, we can classify the different symmetric or SSB phases by assigning different physical boundary conditions, following the philosophy of the categorical Landau paradigm~\cite{Kong:2015flk,Kong:2017etd,Kong:2019byq,Kong:2019cuu,Bhardwaj:2023fca,Bhardwaj:2025piv,Bhardwaj:2024qiv,Bhardwaj:2024qrf}. Note that these physical boundary conditions are the \textit{minimal} ones as in \cite{Bhardwaj:2024qiv}. We discuss the topological boundary conditions $\mc{L}_1$, $\mc{L}_2^+$, $\mc{L}_3^+$ and $\mc{L}_4^+$, while the cases of the other three topological boundary conditions can be carried out similarly by interchanging ``$+$'' with ``$-$''.

\paragraph{Topological boundary condition: $\mc{L}_1$}

The full global symmetry is a $\mb{Z}_2^{(1)}\times\mb{Z}_2^{(2)}$ 0-form symmetry with a mixed anomaly, whose anomaly polynomial is (\ref{anomaly-pol}). We analyze the gapped phases below:

\begin{enumerate}
\item Physical boundary condition: $\mc{L}_1$

The charged operators at the end of the $P_{mn}$ lines all have non-zero vevs, which spontaneously break the full $\mb{Z}_2^{(1)}\times\mb{Z}_2^{(2)}$ 0-form symmetry. There are hence four vacua.

\item Physical boundary condition: $\mc{L}_2^+$

The charged operator at the end of $P_{01}$ line has a non-zero vev, hence $\mb{Z}_2^{(2)}$ is SSB, while $\mb{Z}_2^{(1)}$ is trivially preserved. There are two vacua.

\item Physical boundary condition: $\mc{L}_2^-$

Similar to the $\mc{L}_2^+$ case, the $\mb{Z}_2^{(2)}$ is SSB, while $\mb{Z}_2^{(1)}$ is preserved. Due to the condensation of the dressed operator, the boundary 3d theory is stacked with a $\mb{Z}_2$-SPT, i.e. the topological action $\pi\int_{M_3}a^1\cup a^1\cup a^1$. 

\item Physical boundary condition: $\mc{L}_3^\pm$

The charged operator at the end of $P_{10}$ line has a non-zero vev, hence $\mb{Z}_2^{(1)}$ is SSB, while $\mb{Z}_2^{(2)}$ is preserved. Since the condensed operator $L_{01}$ and $L_{01}^{\psi_2}$ both contain an $\exp(\pi i\int_\Omega a^2\cup a^2\cup a^2)$ factor, the boundary 3d theory is stacked with a $\mb{Z}_2$-SPT, i.e. the topological action $\pi\int_{M_3}a^2\cup a^2\cup a^2$. 

\item Physical boundary condition: $\mc{L}_4^\pm$

A diagonal subgroup of the whole $\mb{Z}_2^{(1)}\times\mb{Z}_2^{(2)}$ 0-form symmetry is spontaneously broken, with another $\mb{Z}_2$ 0-form symmetry preserved with the background gauge field $a_d$. Similar to the cases above, the boundary 3d theory is stacked with a $\mb{Z}_2$-SPT with the same topological action $\pi\int_{M_3}a_d\cup a_d\cup a_d$. 

\end{enumerate}

The fully symmetric, gapped phase with a unique vacuum for the whole $\mb{Z}_2^{(1)}\times\mb{Z}_2^{(2)}$ 0-form symmetry is absent, due to the presence of mixed anomaly (\ref{anomaly-pol}).

\paragraph{Topological boundary condition: $\mc{L}_2^+$}

The full global symmetry is the non-split 2-group symmetry $\mc{G}=(\mb{Z}_2,\mb{Z}_2,\mathrm{triv.},1)$. We analyze the gapped phases below:

\begin{enumerate}
\item Physical boundary condition: $\mc{L}_1$

The charged operator at the end of $P_{01}$ line has a non-zero vev, hence the $\mb{Z}_2$ 0-form symmetry part in the 2-group symmetry $\mc{G}$ is SSB, and only a $\mb{Z}_2$ 1-form symmetry is left. There are two vacua.

\item Physical boundary condition: $\mc{L}_2^+$

The charged operators at the end of $P_{01}$ line and $L_{10}$ surface operators have non-zero vevs, hence the full $\mc{G}$ 2-group symmetry is spontaneously broken. There are two vacua.

\item Physical boundary condition: $\mc{L}_2^-$

This is also a full $\mc{G}$ SSB phase. The difference from the previous case is that the condensed operator at the physical boundary $L_{10}^{\psi_1}$ differs from the condensed operator at the topological boundary $L_{10}$, by an SPT phase $\pi i\int_\Omega a^1\cup a^1\cup a^1$. As a result, the topological spin of the deconfined Wilson loop operator $\exp(\pi i\oint_C a^1)$ is that of a semion, different from that of the physical boundary condition $\mc{L}_2^+$.

\item Physical boundary condition: $\mc{L}_3^+$

The charged operators at the end of $P_{01}$ line and $L_{10}$ surface operators have zero vev. This corresponds to a $\mc{G}$-symmetric phase with a unique, gapped vacuum. 

\item Physical boundary condition: $\mc{L}_3^-$

This is also a $\mc{G}$-symmetric phase with a unique, gapped vacuum.

\item Physical boundary condition: $\mc{L}_4^\pm$

Now since $P_{11}$ and $L_{11}^{\psi_3}$($L_{11}^{\psi_4}$) are both dyons after being pulled back to the boundary, the operators ending at $P_{11}$ and $L_{11}^{\psi_3}$($L_{11}^{\psi_4}$) are non-local, and they cannot gain non-zero vevs. This is analogous to the SPT boundary condition in a (2+1)D $\mb{Z}_2\times\mb{Z}_2$ 0-form SymTFT, where the physical boundary condition with $e_1 m_2$ and $e_2 m_1$ lines condensed describes an SPT phase. Hence in this case, we obtained an SPT phase for the 2-group $\mc{G}$, corresponding to the non-zero element in $H^3(B\mc{G};U(1))=\mb{Z}_2$.

\end{enumerate}

One can see that there is no phase that only preserves the $\mb{Z}_2$ 0-form symmetry, but breaks the $\mb{Z}_2$ 1-form symmetry. This is because in the definition of the Postnikov class for the 2-group symmetry $\mc{G}$, the anomaly for the $\mb{Z}_2$ 0-form symmetry itself is absorbed by a 1-form symmetry transformation, see Figure~\ref{2-group-beta}. Hence a phase that only preserves the $\mb{Z}_2$ 0-form symmetry would suffer from that anomaly.

\paragraph{Topological boundary condition: $\mc{L}_3^+$}

The full global symmetry is the 2-fusion categorical symmetry $2\mathbf{Rep}(\mc{G})$, generated by $P_{01}(C)|_{M_3}:=W(C)$ and $L_{10}(S)|_{M_3}:=U(S)$ in (\ref{2Rep-gen}).

\begin{enumerate}
\item Physical boundary condition: $\mc{L}_1$

The charge for the 0-form symmetry is condensed, hence the non-invertible 0-form symmetry generated by $U(S)$ is SSB, while the $\mb{Z}_2$ 1-form symmetry generated by $W(C)$ is preserved.

\item Physical boundary condition: $\mc{L}_2^+$

None of the charged operators have non-zero vevs, hence this is the symmetric phase for $2\mathbf{Rep}(\mc{G})$.

\item Physical boundary condition: $\mc{L}_2^-$

This is another symmetric phase for $2\mathbf{Rep}(\mc{G})$, nonetheless with a different stacked SPT phase from the previous one.

\item Physical boundary condition: $\mc{L}_3^\pm$

The whole $2\mathbf{Rep}(\mc{G})$ is completely broken, with vacuum degeneracy 2.

\item Physical boundary condition: $\mc{L}_4^+$

The dyons $P_{11}$ and $L_{11}$ are condensed. The categorical symmetry $2\mathbf{Rep}(\mc{G})$ is unbroken, but with a stacked SPT. 

\end{enumerate}

\paragraph{Topological boundary condition: $\mc{L}_4^+$}

In this case the topological operators $P_{11}$ and $L_{11}$ are condensed. The full global symmetry is still the 2-fusion categorical symmetry $2\mathbf{Rep}(\mc{G})$, generated by $W(C)$ and $U(S)$.

\begin{enumerate}
\item Physical boundary condition: $\mc{L}_1$

The non-invertible 0-form symmetry generated by $U(S)$ is SSB, while the $\mb{Z}_2$ 1-form symmetry generated by $W(C)$ is preserved.

\item Physical boundary condition: $\mc{L}_2^+$

The operators attached to $P_{11}$ and $L_{11}$ are not condensed, hence this is a symmetric phase for the full symmetry $2\mathbf{Rep}(\mc{G})$. The exact generators $P_{01}$ and $L_{10}$ for the $2\mathbf{Rep}(\mc{G})$ symmetry are condensed, hence there is no extra stacked SPT.

\item Physical boundary condition: $\mc{L}_2^-$

Similarly to the case before, but there is a $2\mathbf{Rep}(\mc{G})$ symmetry with a stacked SPT.

\item Physical boundary condition: $\mc{L}_3^\pm$

This is another symmetric phase for the full symmetry $2\mathbf{Rep}(\mc{G})$, with stacked SPT. 

\item Physical boundary condition: $\mc{L}_4^\pm$

This is phase where $2\mathbf{Rep}(\mc{G})$ is completely broken, with vacuum degeneracy 2.

\end{enumerate}

\paragraph{Ground state degeneracy:} if we choose the topological boundary as $\mfr{C}[G,H,\pi,\psi]$ and physical boundary as $\mfr{C}[G,K,\pi,\psi]$, the ground state degeneracy is given by~\cite{Wen:2025thg,Bhardwaj:2025piv}
\be  \text{GSD} = \big| H \backslash G/K \big| \ . \ee
Thus we have the ground state degeneracy data presented in Table~\ref{table:GSD}.

\begin{table}[htbp]
\centering
\begin{tabular}{|c|c|c|c|c|}
\hline
\makecell{Topological/Physical \\boundary conditions} & $\CL_1$ & $\CL_2^\pm$ & $\CL_3^\pm$ & $\CL_4^\pm$ \\ \hline
$\CL_1$                                 & 4       & 2           & 2           & 2           \\ \hline
$\CL_2^\pm$                             & 2       & 2           & 1           & 1           \\ \hline
$\CL_3^\pm$                             & 2       & 1           & 2           & 1           \\ \hline
$\CL_4^\pm$                             & 2       & 1           & 1           & 2           \\ \hline
\end{tabular}
\caption{Table of ground state degeneracy when choosing different boundary conditions in $\CZ_1(2\Vect_\CG)$.}
\label{table:GSD}
\end{table}

\paragraph{Hasse diagram}

\begin{figure}
\begin{center}
\begin{tikzpicture}[scale=0.95, every node/.style={transform shape},
  alg/.style   = {font=\small, inner sep=2.5pt},
  lag/.style   = {font=\small, inner sep=2.5pt},
  incl/.style  = {semithick},
  pincl/.style = {semithick},
  base/.style  = {gray!60, thin},
  bridge/.style= {preaction={draw=white, line width=3.2pt}},
  ghost/.style = {draw, dashed, gray, rounded corners=2pt, align=center,
                  font=\scriptsize, inner sep=3pt}]
\node[alg] (one) at (6.6,3.65) {$\mathbf{1}$};
\node[alg] (A10)  at (0.4,1.8)   {$\langle L_{10}\rangle$};
\node[alg] (A10p) at (2.0,1.8)   {$\langle L_{10}^{\psi_1}\rangle$};
\node[alg] (P01)  at (3.5,1.8)   {$\langle P_{01}\rangle$};
\node[alg] (A11a) at (4.9,1.8)   {$\langle L_{11}^{\psi_3}\rangle$};
\node[alg] (P11)  at (6.6,1.8)   {$\langle P_{11}\rangle$};
\node[alg] (A11b) at (8.3,1.8)   {$\langle L_{11}^{\psi_4}\rangle$};
\node[alg] (P10)  at (9.75,1.8)  {$\langle P_{10}\rangle$};
\node[alg] (A01p) at (11.2,1.8)  {$\langle L_{01}^{\psi_2}\rangle$};
\node[alg] (A01)  at (12.8,1.8)  {$\langle L_{01}\rangle$};
\node[lag] (L2p) at (0.4,0.0)  {$\mc{L}_2^{+}$};
\node[lag] (L2m) at (2.0,0.0)  {$\mc{L}_2^{-}$};
\node[lag] (L4p) at (4.9,0.0)  {$\mc{L}_4^{+}$};
\node[lag] (L1)  at (6.6,0.0)  {$\mc{L}_1$};
\node[lag] (L4m) at (8.3,0.0)  {$\mc{L}_4^{-}$};
\node[lag] (L3m) at (11.2,0.0) {$\mc{L}_3^{-}$};
\node[lag] (L3p) at (12.8,0.0) {$\mc{L}_3^{+}$};

\foreach \m in {A10,A10p,P01,A11a,P11,A11b,P10,A01p,A01}
  \draw[base] (one) -- (\m);
\draw[incl] (A10)  -- (L2p);
\draw[incl] (A10p) -- (L2m);
\draw[incl] (A11a) -- (L4p);
\draw[incl] (A11b) -- (L4m);
\draw[incl] (A01p) -- (L3m);
\draw[incl] (A01)  -- (L3p);
\draw[pincl]         (P11) -- (L4p);
\draw[pincl]         (P11) -- (L1);
\draw[pincl]         (P11) -- (L4m);
\draw[pincl]         (P01) -- (L2m);
\draw[pincl,bridge]  (P01) -- (L2p);
\draw[pincl,bridge]  (P01) -- (L1);
\draw[pincl]         (P10) -- (L3m);
\draw[pincl,bridge]  (P10) -- (L3p);
\draw[pincl,bridge]  (P10) -- (L1);
\end{tikzpicture}
\end{center}
\caption{Hasse diagram of the connected condensable algebras of the SymTFT
$\CZ_1(2\Vect_{\bbZ_2\times\bbZ_2}^{\pi_\beta})\cong \CZ_1(2\Vect_\mc{G})$ in the case without 2-group anomaly,
ordered by inclusion relations.  Each algebra is
labelled by the bulk operators it condenses: the Wilson lines $P_{ij}$
and/or the flux strings $L_{ij}$ supported on a subgroup
$H\subseteq\bbZ_2\times\bbZ_2$, the latter decorated by a trivialization
$\psi$ of $\pi_\beta|_H$ on their worldvolume. The maximal elements are the seven Lagrangian algebras
of Table~\ref{tab:E1algebra}, with
$\mc{L}_2^{\pm}=\langle L_{10}^{(\psi_1)},P_{01}\rangle$,
$\mc{L}_3^{\pm}=\langle L_{01}^{(\psi_2)},P_{10}\rangle$,
$\mc{L}_4^{\pm}=\langle L_{11}^{\psi_{3,4}},P_{11}\rangle$ and
$\mc{L}_1=\langle P_{10},P_{01}\rangle$. There is no Lagrangian
algebra with $H=\bbZ_2\times\bbZ_2$ since
$[\pi_\beta|_{\bbZ_2\times\bbZ_2}]=[\pi_\beta]\neq 0$. }
\label{fig:hasse-condensable}
\end{figure}
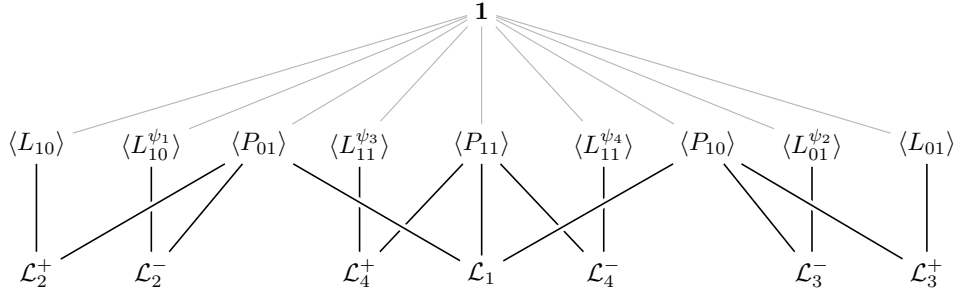

We plot the Hasse diagram connecting different maximal and non-maximal condensable algebras in Figure~\ref{fig:hasse-condensable}. Note that we do not claim that the non-maximal condensable algebras would correspond to a gapless phase.

\subsection{With the 2-group anomaly}
\label{sec:with2-groupanomaly}
\subsubsection{SymTFT action}

In the presence of the 2-group anomaly, the direction of our argument is as follows. The symmetry fusion 2-category is first identified by the monoidal equivalence \eqref{eq:anomalous-2group-fermionic-equivalence}, and the categorical SymTFT is its Drinfeld center $\CZ_1(2\sVect_{(\bbZ_4,\underline{2})})$ in Section~\ref{sec:SymTFT-anomaly}. We then propose the following topological action in the cochain formulation as a Lagrangian realization of this SymTFT:
\be
\label{S-anom}
S =\frac{2\pi}{4} \int_X B\cup \delta A + \pi \int_X (B\text{ mod } 2) \cup w_2(TX) \ , \ A\in C^1(X;\bbZ_4) \ ,\ B\in C^2(X;\bbZ_4) \ . 
\ee
$w_2(TX)\equiv w_2$ is the second Stiefel-Whitney class. The equations of motion are (defined mod 4):
\be
\ba
\delta B&=0\cr
\delta A+2w_2&=0\,.
\ea
\ee
$2w_2=i(w_2)$ is the value after the injection $i:\mb{Z}_2\rightarrow\mb{Z}_4$.

The topological operators include the surface operators
\be
L_n(S)=\exp\left(\frac{2\pi in}{4}\int_S B\right)\quad (n=0,1,2,3)
\ee
and the dressed loop operators
\be
P_m(C)=\exp\left(\frac{2\pi im}{4}\left(\int_C A+\int_S 2w_2\right)\right)\quad (m=0,1,2,3)\quad (C=\ptl S)\,.
\ee
The linking correlation functions are 
\be
\langle L_n(S) P_m(C)\rangle =\exp\left(\frac{2\pi imn}{4}\langle S,C\rangle\right)\,.
\ee
The dressed loop operators have a non-trivial framing when $m$ is odd:
\be
\langle P_m(C)\rangle =(-1)^{m\int_S w_2}\,.
\ee
Hence the four surface operators $L_n$ and the four line operators $P_m$, with fermionic $P_1$ and $P_3$, match the four flux sectors and the line operators $\mathbf{sRep}(\bbZ_4,\underline{2})$ of $\CZ_1(2\sVect_{(\bbZ_4,\underline{2})})$ in Section~\ref{sec:SymTFT-anomaly}. 

\subsubsection{Topological boundary conditions}

The categorical classification above associates a boundary with a
supersubgroup $(H,z)\leq(\bbZ_4,\underline{2})$ and a class
$\mu\in SH^3(B(H,z))$. For the minimal supersubgroup
$H=\langle z\rangle\cong\bbZ_2^f$, the bosonic quotient is a point and
\be
\mu\in SH^3(B\bbZ_2^f)
\equiv SH^3_{w_2}(\mathrm{pt})=0.
\ee
Thus there is no additional choice of $\mu$ for this boundary.

\paragraph{(1) Dirichlet b.c. for $P_2$ and $L_2$: $\mc{L}_1$}

For $H=\langle z\rangle\cong\bbZ_2^f$, the non-trivial character of $\bbZ_4$ that is
trivial on $H$ is labelled by $2$, and the non-trivial flux in $H$ is also
labelled by $2$.  Thus the corresponding boundary condition gives
Dirichlet boundary conditions to $P_2$ and $L_2$.  Its boundary fusion
2-category is
\be
\mfr{C}\big[(\bbZ_4,\underline{2}),\bbZ_2^f,
0,\sVect,0\big]
\simeq 2\Vect_\CG^\omega.
\ee
This is the boundary condition that recovers the original anomalous
non-split 2-group symmetry
$\CG=(\bbZ_2,\bbZ_2,\mathrm{triv.},1)$.

The bulk field $A$ should not be regarded as an ordinary internal
$\bbZ_4$ gauge field.  Since the subgroup
$\langle2\rangle=\bbZ_2^f$ is identified with fermion parity, its on-shell
data form a $\mathrm{Spin}^{\bbZ_4}$ structure, as reviewed in
Appendix~\ref{app:spin-bordism-dual}.  Projecting to
$\bbZ_4^f/\bbZ_2^f\cong\bbZ_2$ gives the ordinary internal background $a$.
After choosing a section of this quotient, decompose the
$\mathrm{Spin}^{\bbZ_4}$ data as
\be
A|_{M_3}=a+2A'\pmod 4,
\qquad
a\in Z^1(M_3;\bbZ_2),
\qquad
A'\in C^1(M_3;\bbZ_2),
\ee
where $a$ on the right-hand side takes the values $0,1\in\bbZ_4$.  Here only
$a$ is an ordinary internal gauge field.  The cochain $A'$ belongs to the
tangential data and records the $\bbZ_2^f$ component.  The bulk equation
$\delta A+2w_2=0$ gives
\be
\delta a=0,
\qquad
\delta A'=a\cup a+w_2(TM_3).
\ee
Thus $A'$ trivializes the combined class $a^2+w_2(TM_3)$.  This is the
cochain description of a $\mathrm{Spin}^{\bbZ_4}$ structure; it does not
require a separate spin structure on $M_3$.

For the 2-cochain $B$, the Dirichlet boundary condition for $L_2$ similarly
decomposes
\be
B|_{M_3}=b_0+2B'\ ,\qquad
b_0=(B\ {\rm mod}\ 2)|_{M_3}\in Z^2(M_3;\bbZ_2)\ ,\qquad
B'\in C^2(M_3;\bbZ_2)\,.
\ee
The bulk equation $\delta B=0$ makes $b_0$ a closed background field, while
the boundary path integral sums over the remaining data $A'$ and $B'$.  No
boundary term is needed for a well-posed variational principle: varying
\eqref{S-anom} with respect to $A$ gives the boundary term
$\pi\int_{M_3}b_0\cup\delta A'$, which vanishes upon integration by parts
because $b_0$ is closed and $M_3$ has no boundary.

To make the 2-group structure manifest, we choose a trivialization
$\eta\in C^1(M_3;\bbZ_2)$ with $\delta\eta=w_2(TM_3)$ and set
\be
\chi=A'+\eta\ ,\qquad \delta\chi=a\cup a\,,
\ee
where the second equality is the on-shell condition
$\delta A'=a\cup a+w_2$.  The combination
\be
\hat a=a+2\chi\in Z^1(M_3;\bbZ_4)
\ee
is then a genuine $\bbZ_4$ cocycle: the boundary theory is a (2+1)D
spin-$\bbZ_4$ gauge theory for $\hat a$ whose mod-2 reduction is locked to
the background $a$.  Shifting the trivialization
$\eta\to\eta+\theta$, $\theta\in Z^1(M_3;\bbZ_2)$, sends
$\hat a\to\hat a+2\theta$: a $\bbZ_2^f$ gauge transformation for exact
$\theta$, and in general a flat shift of the integration variable of the
boundary path integral.  The construction is therefore independent of the
choice of $\eta$.

The flat field $b_0$ is not yet the 2-group background gauge field, since
$\delta b_0=0$ differs from the weak flatness condition
\eqref{Z2Z2-flatness}.  Instead define
\be
b=b_0+a\cup\chi\,,
\ee
which on-shell obeys
\be
\delta b=a\cup\delta\chi=a\cup a\cup a\,,
\ee
precisely the flatness condition of the non-split 2-group
$\CG=(\bbZ_2,\bbZ_2,\mathrm{triv.},1)$.  At fixed $(a,\chi)$ the map
$b_0\mapsto b$ is a bijection from flat 2-cocycles to 2-cochains satisfying
$\delta b=a^3$, realizing the weak 2-group gauge fields.

The bulk gauge symmetries of \eqref{S-anom} induce the weak 2-group gauge
transformations on these backgrounds.  The 1-form gauge symmetry
$B\to B+\delta\Lambda$ acts on the boundary data as
$b_0\to b_0+\delta\lambda$, with
$\lambda=(\Lambda\ {\rm mod}\ 2)|_{M_3}\in C^1(M_3;\bbZ_2)$, hence
\be
a\to a\ ,\qquad b\to b+\delta\lambda\,.
\ee
The 0-form gauge symmetry $A\to A+\delta\tilde g$,
$\tilde g\in C^0(X;\bbZ_4)$, leaves $B$ invariant and decomposes on the
boundary as $\tilde g|_{M_3}=g+2g_1$: the $g_1$ part is the gauge
redundancy $A'\to A'+\delta g_1$ of the boundary theory, while
$g\in C^0(M_3;\bbZ_2)$ acts on the backgrounds as
\be
\label{induced-0form}
a\to a'=a+\delta g\ ,\qquad
\chi\to\chi'=\chi+a\cup_1\delta g+g\cup\delta g\,,
\ee
which preserves $\delta\chi=a\cup a$ by the cup-1 coboundary identity
$\delta(a\cup_1\delta g)=a\cup\delta g+\delta g\cup a$ for closed $a$.
Composing with $b=b_0+a\cup\chi$, the induced transformation of the 2-group
background is $b\to b+\tilde\zeta(a,g)$ with
\be
\label{zeta-induced}
\tilde\zeta(a,g)=\delta g\cup\chi
+(a+\delta g)\cup\bigl(a\cup_1\delta g+g\cup\delta g\bigr)\,.
\ee
Since $\delta(a'\cup\chi')=(a')^3$ and $\delta(a\cup\chi)=a^3$, one has
$\delta\tilde\zeta=(a+\delta g)^3-a^3$, so this is indeed a weak 2-gauge
transformation preserving the flatness condition $\delta b=a^3$.  It differs
from the compensator $\zeta(a,g)$ in
\eqref{zeta-canonical} by $R=\tilde\zeta+\zeta$. Since $\tilde\zeta$ and $\zeta$ satisfy the same coboundary identity, $R$ is a closed 2-cochain,
$\delta R=0$.  This difference is the intrinsic ambiguity of the weak
2-gauge law: $\zeta$ is determined by
$\delta\zeta=(a+\delta g)^3-a^3$ only up to closed 2-cochains, and shifting
$\zeta$ by a closed $R$ amounts to shifting the reference flat background
$b_0\to b_0+R$. The induced law composes correctly under successive
transformations, since it is inherited from the composition of bulk gauge
transformations.

Finally, the 't Hooft anomaly does not appear as a variation of the
boundary partition function on closed $M_3$: the bulk action \eqref{S-anom}
with these boundary conditions is exactly gauge-invariant, and so is the
boundary partition function.  The anomaly is instead visible in the defect
sector: the odd-charge Wilson lines of the boundary spin-$\bbZ_4$ gauge
theory are fermionic, $\langle P_m(C)\rangle=(-1)^{m\int_S w_2}$ for $m$
odd, which is the boundary avatar of the nontrivial extension class
$\upsilon=\mathfrak{a}^2\in H^2(B\bbZ_2;\bbZ_2)$.  Through the equivalence
$2\Vect_\CG^\omega\simeq2\sVect_{(\bbZ_4,\underline{2})}$, the anomaly on
2-group backgrounds is the 4-cocycle
\be
\frac12\bigl(
b\cup b+b\cup_1 a^3+a ^2\cup b
\bigr),
\ee
which is precisely $\frac12v_4$ in~\eqref{eq:v4-weak-gauge}, with the
2-group fields there denoted here by $(a,b)$. It reduces to $\frac12\bigl(\operatorname{Sq}^2b+a^2\cup b\bigr)$ when $\delta b=0$.

\paragraph{(2) Dirichlet b.c. for $L_n$: $\mc{L}_2$}

For $(H,z)=(\bbZ_4,\underline{2})$, all the $L_n$ operators have Dirichlet
boundary conditions and all the $P_m$ operators have Neumann boundary
conditions.  The corresponding boundary fusion 2-category is
\be
\mfr{C}\big[(\bbZ_4,\underline{2}),(\bbZ_4,\underline{2}),
0,\sVect,0\big]
\simeq 2\mathbf{sRep}(\bbZ_4,\underline{2}).
\ee

This represents a non-anomalous fermionic $\bbZ_4^f=(\bbZ_4,\underline{2})$ 1-form symmetry. If the $\bbZ_4^f$ 1-form symmetry is gauged, one obtains a fermionic boundary condition with $\bbZ_4$ 0-form symmetry with all $P_m$ operators condensed, but this is not included in our classification of bosonic boundary conditions.

\subsubsection{Physical boundary conditions}

\paragraph{Topological boundary condition: $\mc{L}_1$}

In this case, the topological operator $L_1$ generates the $\mb{Z}_2$ 0-form symmetry, and $P_1$ generates the $\mb{Z}_2$ 1-form symmetry.

\begin{enumerate}
\item Physical boundary condition: $\mc{L}_1$. 

All the charges attached to $L_2$ and $P_2$ are condensed, hence the whole anomalous $\mc{G}$ 2-group symmetry is spontaneously broken.

\item Physical boundary condition: $\mc{L}_2$.

The $L_n$ operators are condensed, hence this is a phase where the $\mb{Z}_2$ 1-form symmetry is partially SSB, and the $\mb{Z}_2$ 0-form symmetry is preserved.

\end{enumerate}

\paragraph{Topological boundary condition: $\mc{L}_2$}

In this case, the topological operator $P_1$ generates the $\mb{Z}_4^f=(\mb{Z}_4,\underline{2})$ fermionic 1-form symmetry.

\begin{enumerate}
\item Physical boundary condition: $\mc{L}_1$. 

The charges attached to $L_2$ and $P_2$ are condensed, which means that the $\mb{Z}_4^f=(\mb{Z}_4,\underline{2})$ 1-form symmetry is spontaneously broken to a $\mb{Z}_2^f=(\mb{Z}_2,\underline{1})$ subgroup.

\item Physical boundary condition: $\mc{L}_2$.

The $L_n$ operators are condensed, hence this is a complete SSB phase with no global symmetry.

\end{enumerate}

\paragraph{Ground state degeneracy:} For the (3+1)D SymTFT $\CZ_1(2\sVect_{(G,z)}^\varpi)$, if we choose the topological boundary and physical boundary to be
\be
\mathfrak C[(G,z),(H,z),\varpi,\mathbf{sVect},\mu]
\ \text{and} \ \mathfrak C[(G,z),(K,z),\varpi,\mathbf{sVect},\mu']
\ee
respectively, the ground state degeneracy is given by
\be \mathrm{GSD} = |H\backslash G/K| = | H_b\backslash G_b/K_b  | \ , \ee
where $G_b:= G/\expval{z}$. Thus, the data of ground state degeneracy are presented in Table~\ref{table:GSD-anomalous}.
\begin{table}[htbp]
\centering
\begin{tabular}{|c|c|c|}
\hline
\makecell{Topological/Physical \\boundary conditions} & $\CL_1$ & $\CL_2$  \\ \hline
$\CL_1$   & 2       & 1                     \\ \hline
$\CL_2$    & 1     & 1              \\ \hline

\end{tabular}
\caption{Table of ground state degeneracy when choosing different boundary conditions in $\CZ_1(2\Vect_\CG^\omega)$.}
\label{table:GSD-anomalous}
\end{table}

\paragraph{Hasse diagram}

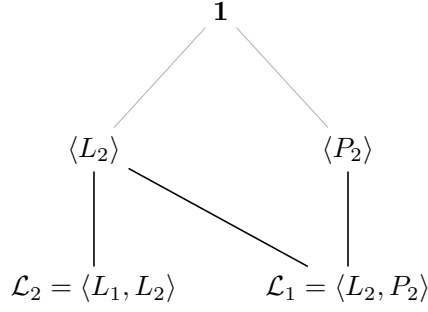
\begin{figure}
\begin{center}
\begin{tikzpicture}[scale=1.05, every node/.style={transform shape},
  alg/.style   = {font=\small, inner sep=2.5pt},
  lag/.style   = {font=\small, inner sep=2.5pt},
  incl/.style  = {semithick},
  base/.style  = {gray!60, thin},
  elab/.style  = {font=\scriptsize, gray!90, inner sep=1.5pt},
  ferm/.style  = {draw, dotted, gray, rounded corners=2pt, align=center,
                  font=\scriptsize, inner sep=3pt}]
\node[alg] (one) at (2.9,3.5) {$\mathbf{1}$};
\node[alg] (L2) at (1.3,1.75) {$\langle L_2\rangle$};
\node[alg] (P2) at (4.5,1.75) {$\langle P_2\rangle$};
\node[lag] (Lag2) at (1.3,0.0) {$\mc{L}_2=\langle L_1,L_2\rangle$};
\node[lag] (Lag1) at (4.5,0.0) {$\mc{L}_1=\langle L_2,P_2\rangle$};
\draw[base] (one) -- (L2) node[elab, midway, above left=-1pt] {};
\draw[base] (one) -- (P2) node[elab, midway, above right=-1pt] {};
\draw[incl] (L2) -- (Lag2) node[elab, midway, left] {};
\draw[incl] (L2) -- (Lag1) node[elab, pos=0.45, below left=-2pt] {};
\draw[incl] (P2) -- (Lag1) node[elab, midway, right] {};
\end{tikzpicture}
\end{center}
\caption{Hasse diagram of the condensable algebras of the anomalous SymTFT
$\CZ_1(2\Vect_\CG^\omega)\simeq\CZ_1(2\mathbf{sVect}_{(\bbZ_4,\underline{2})})$. There are exactly two minimal
Lagrangian algebras: $\mc{L}_1=\langle L_2,P_2\rangle$, the Dirichlet boundary
for $P_2$ and $L_2$ that recovers the anomalous symmetry
$2\Vect_\CG^\omega$, and $\mc{L}_2=\langle L_1\rangle=\{1,L_1,L_2,L_3\}$,
which condenses all surface operators and realizes the non-anomalous
fermionic $\bbZ_4^f=(\bbZ_4,\underline{2})$ $1$-form symmetry. Note that
$\langle L_1\rangle$ is not an intermediate node.}
\label{fig:hasse-anomalous}
\end{figure}

We plot the Hasse diagram connecting the maximal and non-maximal condensable algebras in Figure~\ref{fig:hasse-anomalous}.

\section{Non-split 2-group symmetry in physical systems}
\label{sec:physical}

\paragraph{$\bbZ_2$ Gauge Theory.} In~\cite[Appendix E]{Delmastro:2022pfo}, they analyzed the fractionalization of $\bbZ_2$ gauge theory in (2+1)D with 
\begin{itemize}
    \item $\bbZ_{2,e}^{(1)}\times \bbZ_{2,m}^{(1)}$ 1-form symmetry generated by electric and magnetic line operators (denoted $e$ and $m$ respectively, let $\epsilon :=em$ be the fermionic line), and
    \item $\bbZ_{2,\mathbf{C}}^{(0)}$ 0-form symmetry as an automorphism exchanging $e$ and $m$.
\end{itemize}
There is a choice of fractionalization $\mbf C^2\big|_{\epsilon := em} = \pm 1$. When we take $\mbf C^2\big|_{\epsilon := em} = -1$, the $\bbZ_{2,\mathbf{C}}^{(0)}$ 0-form symmetry with the diagonal subgroup of $\bbZ_{2,e}^{(1)}\times \bbZ_{2,m}^{(1)}$ (which we denote as $\bbZ_{2,\epsilon}^{(1)}$) forms a non-split 2-group $\CG = (\bbZ_2,\bbZ_2,\mathrm{triv.}, 1)$ with non-trivial anomaly. The symmetry category is exactly $2\Vect_\CG^\omega\simeq 2\sVect_{(\bbZ_4,\underline 2)}$ discussed in previous sections.

\paragraph{Non-Interacting Fermion Systems.} Consider a (2+1)D quadratic lattice Hamiltonian with an integer-valued conserved charge $Q$ satisfying $e^{i\pi Q}=(-1)^F$. The transformations $e^{i\theta Q}$ then form a fermionic symmetry $U(1)^f$, whose element $g=e^{i\pi Q/2}$ generates a subgroup $\mathbb{Z}_4^f$ with $g^2=(-1)^F$; for a number-conserving band insulator or integer quantum Hall system one may take $Q=N$, while for a superconductor preserving spin rotations about the $z$ axis one may take $Q=2S_z=N_\uparrow-N_\downarrow$, under which an opposite-spin Cooper pair is neutral \cite[Tables~I--III]{Wen:2011np}\cite[Secs.~2 and 3]{ChenKapustinTurzilloYou2019}. Gauging fermion parity produces the corresponding bosonic shadow with quotient zero-form symmetry $\mathbb{Z}_4^f/\mathbb{Z}_2^f\cong\mathbb{Z}_2$ and a distinguished anomalous fermionic line; this operation can be implemented microscopically using exact lattice bosonization, which maps two-dimensional free-fermion models to modified $\mathbb{Z}_2$ lattice gauge theories while preserving locality \cite[Secs.~3.1--3.2]{ChenKapustinRadicevic2018}.

\section*{Acknowledgments}
We thank Hank Chen, Yu-An Chen, Qing-Rui Wang, Hao Xu, Peng Ye, Hao Y. Zhang and Yunqin Zheng for useful discussions. RL and YNW are supported by the National Natural Science Foundation of China under Grant No.~12422503. ZBG is supported in part by the National Natural Science Foundation of China under Grants No.~12275004 and No.~12588101. YZ is supported by WPI Initiative, MEXT, Japan at Kavli IPMU, the University of Tokyo.

\appendix

\section{Fusion 2-Categories}\label{app:fusion2cat}
In this appendix, we introduce the classification of fusion 2-categories up to monoidal equivalence, and provide examples crucial to this work.

\subsection{Classification of fusion 2-categories}\label{app:classificationfusion2cat}
A handy tool for categorical analysis is the $E_k$-centers. This tool depicts the categorical commutativity of different levels. Commonly used examples include the $E_1$-center, also known as Drinfeld center and the $E_2$-center, also known as the M\"uger center. We will denote $E_k$-center of a (multi)-fusion $n$-category $\mfr{L}$ as $\CZ_k(\mfr{L})$. A convenient definition of $E_k$-center is given as follows~\cite{Kong:2020iek}:
\begin{definition}
    Let $\CX$ be an $E_k$ multifusion $n$-category. The $E_k$-center of $\CX$ is the $E_{k+1}$ multifusion $n$-category \be\label{eq:Ekcenter}\CZ_{k}(\CX)\equiv \Fun(\Sigma^k\CX,\Sigma^k\CX)\cong \Omega^k\CZ_0(\Sigma^k\CX) \ . \ee
\end{definition}
The definition of $\CZ_0(\CX)$ is
\be \CZ_0(\CX) = \CX\boxtimes \CX^{\rm op} \ , \ee
and $\Sigma$ is the composition of delooping $B$ and Karoubi completion ${\rm Kar}$. Generically, for a fusion $n$-category $\CX$, we have
\be\ba
  \Omega\Sigma \CX &= \CX \\
  \Sigma\Omega \CX &\ne \CX \ .
\ea\ee

Let $\mathfrak{C}$ be a fusion 2-category. The first classifying datum is the M\"uger center of the braided fusion 1-category of endomorphisms of the tensor unit,
\begin{equation}
\Omega\mathfrak{C} := \operatorname{End}_{\mathfrak{C}}(\mathbf 1),
\qquad
\CZ_2(\Omega\mathfrak{C}) := \text{the M\"uger center of }\Omega\mathfrak{C}.
\end{equation}
The classification splits according to the symmetric fusion category
$\CZ_2(\Omega\mathfrak{C})$:
\begin{equation}
\CZ_2(\Omega\mathfrak{C})\simeq
\begin{cases}
\Rep(H)\text{ for some finite group } H, & \text{bosonic case},\\
\Rep(H,z)\text{ for some finite supergroup } (H,z), & \text{fermionic case}.
\end{cases}
\end{equation}

The Drinfeld center supplies the canonical symmetric monoidal functor
\begin{equation}
\Omega \CZ_1(\mathfrak{C})\longrightarrow \CZ_2(\Omega\mathfrak{C}) \ .
\end{equation}
Notice that through~\eqref{eq:Ekcenter}, we have
\be  \Omega\CZ_1(\mfr{C})\cong  \Omega^2\CZ_0(\Sigma\mfr{C}) \hookrightarrow \CZ_2(\Omega\mfr{C}) \cong \Omega^2\CZ_0(\Sigma^2\Omega\mfr{C}) \ .  \ee
By Deligne reconstruction, this corresponds to an inclusion of finite groups
or supergroups:
\begin{equation}
\iota : H\hookrightarrow G
\qquad\text{or}\qquad
\iota : (H,z)\hookrightarrow (G,z).
\end{equation}

Next, one de-equivariantizes $\Omega\mathfrak{C}$:
\begin{equation}
\CA_{\mathrm{bos}}=\Omega\mathfrak{C}\boxtimes_{\Rep(H)}\Vect,
\qquad
\CA_{\mathrm{ferm}}=\Omega\mathfrak{C}\boxtimes_{\Rep(H,z)}\sVect.
\end{equation}
The remaining higher coherence is encoded by a Delphic square. In the bosonic
case it has the form
\[
\begin{tikzcd}
BH
    \arrow[r, "{\rho}"]
    \arrow[d, "{\iota}"']
&
B\Aut_{\br}(A)
    \arrow[d, "{[-]}"]
\\
BG
    \arrow[r, "{\pi}"']
&
B^4\mathbb{C}^{\times}
\end{tikzcd}
\ \mathpunct{.}
\]
Thus the obstruction, or anomaly, of the action $\rho$ is matched with
$\pi|_H$; after choosing a cochain $\mu$, this can be written as
\begin{equation}
    [\rho]=\pi|_H\,d\mu.
\end{equation}

With the above information, we are able to classify fusion 2-categories as~\cite{Decoppet:2024htz}:
\begin{theorem}[Bosonic fusion 2-category]\label{def:bosonicfusion2cat}
Bosonic fusion 2-categories are parameterized by the following data:
\begin{itemize}
\item a finite group $G$;
\item a subgroup $H\le G$ up to conjugation;
\item a class $\pi\in H^4(BG;\mathbb C^\times)$;
\item a non-degenerate braided fusion 1-category $\CA$ with $H$-action $\rho$ (induced by equivariantization);
\item a class $\psi\in C^3(H;U(1))$ such that $d\psi = \CO^4(\rho)\pi\big|_H$, where $\CO^4(\rho)$ is the anomaly of $H$-action $\rho$.
\end{itemize}
\end{theorem}
We will denote such fusion 2-category as $\mfr{C} [G,H,\pi,\CA,\psi]$. When $\CA = \Vect$, the action $\rho$ is automatically trivial, and $\psi$ will just have to satisfy $d\psi = \pi\big|_H$. We refer to the $\CA = \Vect$ cases as group-theoretical fusion 2-categories, and we abbreviate the notion to be $\mfr{C}[G,H,\pi,\psi]$.

\begin{theorem}[Fermionic fusion 2-category]\label{def:fermionicfusion2cat}
Fermionic fusion 2-categories are parameterized by the following data:
\begin{itemize}
\item a finite supergroup $(G,z)$;
\item a sub-supergroup $(H,z)\le (G,z)$;
\item an $\sVect$-non-degenerate braided fusion 1-category $\CA$ with action by $(H,z)$ given by $\rho$;
\item a class $\varpi$ in a torsor for $\SH^4(B(G,z))$;
\item a homotopy between the superanomaly $[\rho]$ and
$\varpi|_{(H,z)}$.
\end{itemize}
\end{theorem}

Conversely, for a given set of fusion 2-category classification data, we can reconstruct $\mathfrak{C}$ by forming a bimodule category,
\be\ba
\mathfrak{D}&=\TwoVect_G^\pi\boxtimes \Mod(\CA),
\qquad
R(h)=X_h\boxtimes \rho(h)^{-1} \ , \\ 
\mathscr A&=\bigoplus_{h\in H}R(h),
\qquad
\mathfrak{C}\simeq \Bimod_{\mfr{D}}(\mathscr A).
\ea\ee
From this construction we can calculate the species of codimension-1 defects labeled by connected components of $\mfr{C}$.
\begin{definition}[\cite{Johnson-Freyd:2020ivj}]\label{def:connectedcomponents}
    The set of components of $\mfr C$, denoted $\pi_0(\mfr C)$, is the set of equivalence classes of indecomposable objects for the equivalence relation related by a nonzero morphism.
\end{definition}
\begin{proposition}[\cite{Decoppet:2024htz}]\label{prop:pi0group}
    Let $\mfr C$ be a fusion 2-category such that $\Omega\mfr C$ is either a non-degenerate or a slightly degenerate braided fusion 1-category. Then the set of connected components $\pi_0(\mfr C)$ inherits a group structure from the monoidal product. In particular, $\mfr{C}$ is faithfully graded by its connected components.
\end{proposition}
\begin{theorem}[\cite{Decoppet:2024htz}]\label{thm:doublecoset}
Let $\mathfrak{C}$ be a fusion $2$-category such that
$G \coloneqq \pi_0(\mathfrak{C})$ forms a group. (For instance, this holds if every connected component of $\mathfrak{C}$ contains an invertible object.) Let $A$ be a strongly connected rigid algebra (which implies separable by~\cite{Decoppet2022Drinfeld}). Let $H$ be the support of $A$ in $\pi_0(\mathfrak{C})$, which is necessarily a group. Then
\be
    \pi_0\!\left(\mathbf{Bimod}_{\mathfrak{C}}(A)\right)
    \cong H \backslash G/H.
\ee
\end{theorem}

\subsection{Examples of fusion 2-categories}
Here we present several examples of fusion 2-categories.
\begin{example}[$2\Vect_G^\pi$]\label{exm:0-formGsym}
    In (2+1)D system, a 0-form $G$ symmetry with 't Hooft anomaly $\pi\in H^4(BG;U(1))$ is described by the fusion 2-category $2\Vect_G^\pi$, which corresponds to the classifying data $\mfr{C}[G, \{e\},\pi,\Vect,0]$. The explicit categorical construction is given by
    \begin{enumerate}
        \item each object is $G$-graded, $X = \oplus_{g\in G} X_g$, where each $X_g$ is a $\bbC$-linear 1-category,
        \be  X_g \cong \Vect^{\oplus n_g}  \ee
        for some $n_g\in\bbN$.
        \item Each hom space is $\Vect$.
        \item There is a $\pi\in H^4_{\rm grp}(G;U(1))$ data giving the 2-associativity of simple object fusion rules.
    \end{enumerate}
    The illustration is given in Figure~\ref{fig:illustration2vecg}.
    
    \begin{figure}[htbp]
        \centering
        \begin{tikzpicture}[
    vec/.style={font=\large},
    loop/.style={-{Stealth[length=1.5mm]}, line width=0.7pt}
]
    \node[vec] (v1) at (0,0)   {$\Vect_{e}$};
    \node[vec] (g1) at (2.4,0) {$\Vect_{g_1}$};
    \node[vec] (g2) at (4.8,0) {$\Vect_{g_2}$};
    \node[vec]      at (7.0,0) {$\cdots$};

    \foreach \x in {0,2.4,4.8}{
        \node at (\x,1.15) {$\Vect$};
        \draw[loop]
            (\x-0.25,0.38)
            .. controls (\x-0.75,1.25) and (\x+0.75,1.25) ..
            (\x+0.25,0.38);
    }
\end{tikzpicture}
        \caption{Illustration of the fusion 2-category $2\Vect_G^\pi$.}
        \label{fig:illustration2vecg}
    \end{figure}
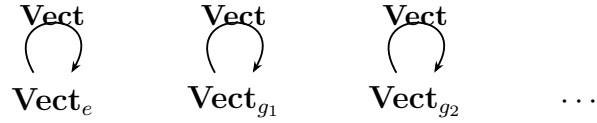
\end{example}
\begin{example}[$2\Vect_\CG$]\label{exm:2-groupsym}
     In (2+1)D system, a 2-group symmetry 
    \be\CG = \big(\Pi_1,\Pi_2,\alpha:\Pi_1\to \Aut(\Pi_2),\beta\in H^3_{\rm grp}(\Pi_1,\Pi_2)\big)\ee
    is given by a fusion 2-category $2\Vect_\CG$. The classifying data of $2\Vect_\CG$ is~\cite{Decoppet:2023bay}
    \be\ba\label{eq:2grpclassifying data}
    &2\Vect_\CG = \mfr{C}[\Pi_1\ltimes \widehat{\Pi}_2, \widehat{\Pi}_2, \pi, \Vect, 0 ]\\
    &\widehat{\Pi}_2 := \Hom(\Pi_2,U(1)) \\
    &\pi\big( (g_1,h_1),(g_2,h_2),(g_3,h_3),(g_4,h_4) \big) = h_1\big(\beta(g_2,g_3,g_4)\big) \ .
    \ea\ee
    Explicitly, its objects are similar to that of $2\Vect_{\Pi_1}$, but the morphism wrapping around the objects are different, the illustrations are given in Figure~\ref{fig:illustration2Vect2-group}.
    \begin{figure}[htbp]
        \centering
        \begin{tikzpicture}[
    vec/.style={font=\large},
    loop/.style={-{Stealth[length=1.5mm]}, line width=0.7pt}
]
    \node[vec] (v1) at (0,0)   {$\Vect_{e}$};
    \node[vec] (g1) at (2.4,0) {$\Vect_{g_1}$};
    \node[vec] (g2) at (4.8,0) {$\Vect_{g_2}$};
    \node[vec]      at (7.0,0) {$\cdots$};

    \foreach \x in {0,2.4,4.8}{
        \node at (\x,1.15) {$\Vect_{\Pi_2}$};
        \draw[loop]
            (\x-0.25,0.38)
            .. controls (\x-0.75,1.25) and (\x+0.75,1.25) ..
            (\x+0.25,0.38);
    }
\end{tikzpicture}
        \caption{Illustration of the fusion 2-category $2\Vect_\CG$. Here $g_i\in \Pi_1$ and $e$ is the identity in $\Pi_1$.}
        \label{fig:illustration2Vect2-group}
    \end{figure}
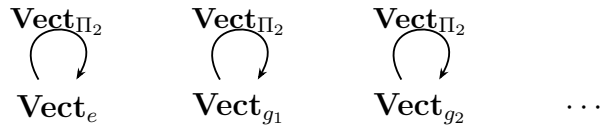
\end{example}
\begin{example}[$2\mbf{Rep}(\CG)$]\label{exm:2-repsym}
    For a 2-group $\CG= (\Pi_1,\Pi_2,\alpha,\beta)$, the 2-representation category $\TwoRep(\CG)$ is given by the classifying data
    \be \label{2Rep-G}\TwoRep(\CG) = \mfr{C} [ \Pi_1\ltimes \widehat{\Pi}_2, \Pi_1,\pi,\Vect,0 ] \ ,  \ee
    where $\pi$ is the same as in~\eqref{eq:2grpclassifying data}.
\end{example}

\subsection{Morita equivalence and SymTFT}

In the categorical framework, gauging takes a fusion $n$-category to another fusion $n$-category. The equivalence relation established through gauging is called Morita equivalence.
\begin{definition}[Morita equivalence]
    Suppose $\CC$ and $\CD$ are fusion $n$-categories, they are Morita equivalent if
    \be \mathbf{Mod}(\CC) \simeq \mathbf{Mod}(\CD) \ . \ee
\end{definition}
Direct consequences of Morita equivalence include the existence of invertible domain wall and a shared Drinfeld center.
\begin{proposition}
    Suppose $\CC$ and $\CD$ are fusion $n$-categories, then they are Morita equivalent if and only if there is an invertible bimodule category $\CM\in \mathbf{Bimod}(\CC,\CD)$.
\end{proposition}
\begin{proposition}
    If $\CC$ and $\CD$ are Morita equivalent fusion $n$-categories, then $\CZ_1(\CC)\simeq \CZ_1(\CD)$.
\end{proposition}

The data of the Drinfeld center for fusion 2-categories is rather coarse. The Drinfeld center of fusion 2-categories are determined by their group data $G$ and the anomaly class $\pi \in H^4(BG;U(1))$ (or for the fermionic case, supergroup $(G,z)$ and $\varpi\in SH^4(B(G,z))$)~\cite{Decoppet:2024htz,Decoppet:2023uoy,Xu:2026EtaleAlgebras}. In other words, we have
\be\ba
&{\rm Bosonic:} \ \CZ_1(\mfr{C}[G,H,\pi,\CA , \psi]) \simeq \CZ_1(2\Vect_G^\pi) \\
&{\rm Fermionic:} \ \CZ_1(\mfr{C}[(G,z),(H,z),\varpi,\CA , \mu]) \simeq \CZ_1(2\sVect_{(G,z)}^\varpi) \ .
\ea\ee

The boundary conditions of the SymTFT are given by the condensable $E_2$ algebras, the thorough definitions are given in ~\cite[Appendix B]{Wen:2025thg}. Here we introduce the classification:
\begin{theorem}[\cite{Xu:2024pwd,Wen:2025thg}]
    A condensable $E_2$ algebra of $\CZ_1 (2\Vect_G^\pi)$ is given by a $\pi$-twisted $G$-crossed braided multi-fusion category such that the $G$-action permutes the simple summands of the monoidal unit transitively.
\end{theorem}

\subsection{Gauging and full center}\label{app:gaugingfullcenter}
Generically we can consider taking the fusion 2-category of $\mathtt{A}-\mathtt{A}$-bimodules, where  $\mathtt{A}$ is an $E_1$ algebra in $\mfr C$~\cite{Decoppet:2023rlx},
\be  \mfr D_\mathtt{A} = \Bimod_{\mathtt{A}-\mathtt{A}} (\mfr C) \ .  \ee
This process is called gauging in physics literature. Here we intend to show the identification between the $E_1$ algebras $\mathtt{A}\in {\rm Alg}_{E_1}(\mfr C)$ and the $E_2$ algebras $\mathscr{A} \in  {\rm Alg}_{E_2}\left(\CZ_1\left(\mfr C\right)\right)$. Before that we introduce the definition of a full center,
\begin{definition}
    The full center of an algebra $\mathtt{A}$ is the terminal object in the 2-category $\mfr C$ of commuting half-braidings of $\mathtt A$, if it exists. We denote it by $\mbf{Z}^{\mfr C}_1(\mathtt{A})$.
\end{definition}
We also have the braided equivalence of their SymTFT~\cite[Theorem 2.3.2]{Decoppet2022Drinfeld},
\be  F_\mathtt{A} :  \CZ_1(\mfr C)\stackrel{\simeq}{\longrightarrow}  \CZ_1(\mfr D_\mathtt{A})\ .  \ee
The action of $\mathcal Z_1(\mathfrak C)$ on $\mfr D_\mathtt{A}$ is induced through the 2-functor
\be  V_\mathtt{A} : \CZ_1(\mfr C) \to \mfr D_\mathtt{A} \  , \ X\mapsto X\boxtimes \mathtt{A} \ , \ee
with the enriched endomorphism object $[\mathbf1_{\mathfrak D_\mathtt{A}},\mathbf1_{\mathfrak D_\mathtt{A}}]$ such that
\be
\operatorname{Hom}_{\mathcal Z_1(\mathfrak C)} \left( X, [\mathbf1_{\mathfrak D_\mathtt{A}},\mathbf1_{\mathfrak D_\mathtt{A}}] \right) \simeq \operatorname{Hom}_{\mathfrak D_\mathtt{A}} \left(V_\mathtt{A}(X)\boxtimes\mathbf1_{\mathfrak D_\mathtt{A}} ,\mathbf1_{\mathfrak D_\mathtt{A}}\right) \simeq \operatorname{Hom}_{\mathfrak D_\mathtt{A}} \left(V_\mathtt{A}(X),\mathtt{A}\right).
\ee
Thus through~\cite[Theorem 6.12]{Xu:2026centers} we have
\be
F_\mathtt{A}\left(\mathbf Z_1^{\mathfrak C}\left(\mathtt{A}\right)\right) \simeq [\mathbf1_{\mathfrak D_\mathtt{A}},\mathbf1_{\mathfrak D_\mathtt{A}}].
\ee
The enriched object satisfies the property~\cite[Proposition 6.3]{Xu:2024pwd}
\be \ba 
\mfr D_\mathtt{A} &\simeq \mathbf{Mod}_{\CZ_1(\mfr{D}_\mathtt{A})}\left([\mathbf1_{\mathfrak D_\mathtt{A}},\mathbf1_{\mathfrak D_\mathtt{A}}]\right)\\  &\simeq  \mathbf{Mod}_{\CZ_1(\mfr{D}_\mathtt{A})}\left(F_\mathtt{A}\left(\mathbf Z_1^{\mathfrak C}\left(\mathtt{A}\right)\right) \right) \\ &\simeq  \mathbf{Mod}_{\CZ_1(\mfr C)}\left(\mathbf Z_1^{\mathfrak C}\left(\mathtt{A}\right) \right)\ea\ee
Thus $\mathbf Z_1^{\mathfrak C}\left(\mathtt{A}\right)$ is the Lagrangian algebra corresponding to $\mfr D_\mathtt{A}$ the gauged symmetry category we desire.

\section{The three-layer approximation to spin bordism and supercohomology}
\label{app:spin-bordism-dual}

In this appendix we review the three-stage generalized cohomology theory
$SH^\bullet$ obtained by retaining the first three layers of the Pontryagin
dual of spin bordism.  For an ordinary spin structure, it agrees with the
full reduced spin bordism dual in degrees three and four.  We then introduce
the $v$-twisted theory associated with a central extension by $\bbZ_2^f$.
For $v\neq0$, we use $SH_v$ as a three-stage theory and do not assume an
identification with the full twisted spin bordism dual.

\subsection{Untwisted supercohomology}

For a space $Y$, the Pontryagin dual of spin bordism is the generalized
cohomology theory whose degree-$n$ group is
\begin{equation}
    \operatorname{Hom}\bigl(
    \Omega_n^{\mathrm{Spin}}(Y),U(1)
    \bigr).
\end{equation}
For pointed $Y$, replacing $\Omega_n^{\mathrm{Spin}}(Y)$ by
$\widetilde\Omega_n^{\mathrm{Spin}}(Y)$ gives the reduced theory.  Its
coefficient groups are the Pontryagin duals of the spin bordism groups of a
point.  The first three spin bordism groups are
\begin{equation}
    \Omega^{\mathrm{Spin}}_0(\mathrm{pt})=\bbZ,\qquad
    \Omega^{\mathrm{Spin}}_1(\mathrm{pt})=\bbZ_2,\qquad
    \Omega^{\mathrm{Spin}}_2(\mathrm{pt})=\bbZ_2.
    \label{eq:first-three-spin-layers}
\end{equation}
Keeping the Pontryagin duals of these three groups, together with their
Postnikov invariants, defines the three-stage generalized cohomology theory
$SH^\bullet$, called extended supercohomology.  The word ``extended'' refers
to the third layer, dual to $\Omega_2^{\mathrm{Spin}}(\mathrm{pt})$, which is
added to the two layers of Gu--Wen supercohomology
\cite{Gu:2012ib,WangGu:2017gsc,Wang:2018pdc}. We nevertheless use
``supercohomology'' below for the three-layer theory unless the original
two-layer construction is explicitly meant. The AHSS takes the form
\begin{equation}
    E_2^{p,q}
    =
    H^p\bigl(Y;SH^q(\mathrm{pt})\bigr)
    \Longrightarrow SH^{p+q}(Y),
    \label{eq:SH-AHSS}
\end{equation}
where, for $q=0,1,2$,
\begin{equation}
    SH^q(\mathrm{pt})
    =
    \operatorname{Hom}\bigl(
    \Omega_q^{\mathrm{Spin}}(\mathrm{pt}),U(1)
    \bigr)
    =
    \begin{cases}
        U(1), & q=0,\\
        \bbZ_2, & q=1,2.
    \end{cases}
\end{equation}
The Postnikov invariants determine the differentials in this spectral
sequence. They also determine how the three layers are coupled in the
cochain description.

On the total-degree-$n$ diagonal, the three coefficients occur in degrees
$n$, $n-1$, and $n-2$. Correspondingly, the cochain data take the form
\begin{equation}
    (\nu_n,n_{n-1},n_{n-2})
    \in
    C^n(Y;\bbR/\bbZ)
    \times C^{n-1}(Y;\bbZ_2)
    \times C^{n-2}(Y;\bbZ_2).
    \label{eq:three-layer-data}
\end{equation}
The three cochains are not independent.  The explicit untwisted
degree-three and degree-four models used below are reviewed in
Subsection~\ref{subsec:explicit-three-layer-models}
\cite{Brumfiel:2016vpy,BrumfielMorgan4}.  Their equivalence classes give
$SH^n(Y)$.

Let us now compare $SH^n(Y)$ with the full Pontryagin dual of spin bordism.
The next two spin bordism groups of a point are
$\Omega_3^{\mathrm{Spin}}(\mathrm{pt})=0$ and
$\Omega_4^{\mathrm{Spin}}(\mathrm{pt})=\bbZ$. Therefore, no omitted row
contributes in reduced total degree three. In degree four, the omitted
$q=4$ term is the point contribution and is removed by taking reduced
bordism. Brumfiel and Morgan constructed a pairing and proved
\begin{equation}
    SH^n(Y)
    \xrightarrow{\ \simeq\ }
    \operatorname{Hom}\bigl(
    \widetilde\Omega_n^{\mathrm{Spin}}(Y),U(1)
    \bigr),
    \qquad n=3,4.
    \label{eq:BM-spin-dual-isomorphism}
\end{equation}
Thus, in degrees three and four, the three-layer construction gives the full
reduced spin bordism dual.

\subsection{Supergroups and twisted spin structures}
\label{app:twistedspin}
Let $G_{\mathrm b}$ be a finite group.  The supergroup in this subsection
specifies tangential data.  Identifying it with supergroup data in a
fusion-2-categorical description requires additional categorical input and
is not assumed here.  A finite supergroup over $G_{\mathrm b}$ is a pair
$(G_{\mathrm f},z)$ given by a central extension
\begin{equation}
    1\longrightarrow \bbZ_2^f=\{1,z\}
    \longrightarrow G_{\mathrm f}
    \longrightarrow G_{\mathrm b}
    \longrightarrow 1
    \label{eq:fermionic-central-extension}
\end{equation}
where $z$ is a specified central element of order two.  Such extensions are
classified by
\begin{equation}
    v\in H^2(BG_{\mathrm b};\bbZ_2).
\end{equation}
Let $\widehat v\in Z^2(G_{\mathrm b};\bbZ_2)$ be a normalized cocycle
representing $v$.  A normalized section of
\eqref{eq:fermionic-central-extension} noncanonically identifies the
underlying set of $G_{\mathrm f}$ with $G_{\mathrm b}\times\bbZ^f_2$, with
multiplication
\begin{equation}
    (g,\epsilon)(h,\eta)
    =
    \bigl(gh,\epsilon+\eta+\widehat v(g,h)\bigr).
\end{equation}
Changing the section changes $\widehat v$ by a coboundary.  Thus only its
cohomology class $v$ is invariant.  For $v=0$, the extension splits:
$G_{\mathrm f}\cong\bbZ_2^f\times G_{\mathrm b}$. This is the case considered
in \cite{Gu:2012ib}.

Let us next explain how this extension changes the spin structure. Recall
that the double covering $\mathrm{Spin}(d)\to\mathrm{SO}(d)$ has kernel
$\{1,-1\}$. Thus $-1$ denotes the nontrivial central element of
$\mathrm{Spin}(d)$ that projects to the identity in $\mathrm{SO}(d)$. We
identify this element with $z\in G_{\mathrm f}$. This gives the tangential structure
\cite{Freed:2016rqq}
\begin{equation}
    \mathrm{Spin}^{G_{\mathrm f}}(d)
    :=
    \frac{\mathrm{Spin}(d)\times G_{\mathrm f}}
    {\langle(-1,z)\rangle}.
    \label{eq:spin-Gf-structure}
\end{equation}
Here the denominator is the order-two central subgroup
$\langle(-1,z)\rangle=\{(1,1),(-1,z)\}$, so the quotient identifies
$(s,g)$ with $((-1)s,zg)$.
Let $M$ be an oriented manifold and let
$A:M\to BG_{\mathrm b}$ classify a principal $G_{\mathrm b}$-bundle.  A
$\mathrm{Spin}^{G_{\mathrm f}}$-structure exists
if and only if
\begin{equation}
    w_2(TM)+A^*v=0,
    \label{eq:twisted-spin-condition}
\end{equation}
where $w_2(TM)$ and $A^*v$ are the obstructions to lifting the frame
bundle and the $G_{\mathrm b}$-bundle, respectively. They enter the same
equation because $-1$ and $z$ are identified in
\eqref{eq:spin-Gf-structure}. For $v=0$, this is an ordinary spin
structure independent of $A$. For $v\neq0$, $A^*v$ can cancel
$w_2(TM)$.

We denote the bordism group for this tangential structure by
\begin{equation}
    \Omega_n^{\mathrm{Spin}}(BG_{\mathrm b};v)
    :=
    \Omega_n^{\mathrm{Spin}^{G_{\mathrm f}}}
    =
    \pi_n\bigl(M\mathrm{Spin}^{G_{\mathrm f}}\bigr).
\end{equation}
Here $M\mathrm{Spin}^{G_{\mathrm f}}$ is the Thom spectrum for this
tangential structure, defining the $v$-twisted spin bordism theory over
$BG_{\mathrm b}$.  Its AHSS is
\begin{equation*}
    E^2_{p,q}
    =
    H_p\bigl(BG_{\mathrm b};
    \Omega_q^{\mathrm{Spin}}(\mathrm{pt})\bigr)
    \Longrightarrow
    \Omega_{p+q}^{\mathrm{Spin}}(BG_{\mathrm b};v).
\end{equation*}
The first three rows therefore have the usual coefficients
$\bbZ,\bbZ_2,\bbZ_2$ in degrees $q=0,1,2$. The twist $v$ appears in
the differentials, while the coefficient groups are unchanged.

Let $i:\mathrm{pt}\to BG_{\mathrm b}$ be the inclusion of the basepoint,
corresponding to the trivial $G_{\mathrm b}$-bundle. Since $i^*v=0$, it
induces the map
\begin{equation*}
    i_*:\Omega_n^{\mathrm{Spin}}(\mathrm{pt})
    \longrightarrow\Omega_n^{\mathrm{Spin}}(BG_{\mathrm b};v),
    \qquad
    [M]\longmapsto[M,A=0].
\end{equation*}
Since we are interested only in the part depending on the internal symmetry,
we define
\begin{equation}
    \widetilde\Omega_n^{\mathrm{Spin}}(BG_{\mathrm b};v)
    :=
    \frac{\Omega_n^{\mathrm{Spin}}(BG_{\mathrm b};v)}
    {\operatorname{im}(i_*)}.
\end{equation}
For $v=0$, these recover the usual groups
$\Omega_n^{\mathrm{Spin}}(BG_{\mathrm b})$ and
$\widetilde\Omega_n^{\mathrm{Spin}}(BG_{\mathrm b})$.

\subsection{Twisted supercohomology}

Let us now include the same twist in supercohomology.  We denote the
three-stage theory by
\begin{equation}
    SH_v^n(BG_{\mathrm b}).
\end{equation}
The notation $SH^{n+v}(BG_{\mathrm b})$ is also used. Here
$n+v$ means degree $n$ with twist $v$; it is not a sum of
degrees. The coefficient groups and the $E_2$-page are the same as in
\eqref{eq:SH-AHSS}, but the differentials depend on $v$.

For a unitary $G_{\mathrm b}$, choose a representative
$\widehat v\in Z^2(G_{\mathrm b};\bbZ_2)$ of $v$ and define
\begin{equation}
    D_{\widehat v}(x):=\operatorname{Sq}^2(x)+\widehat v\cup x,
    \label{eq:twisted-Sq2}
\end{equation}
The twisted three-layer cochain model in the two degrees used in this paper
is as follows.

\paragraph{Degree three.}
The cochain data are
\begin{equation}
    (\nu_3,n_2,n_1)
    \in
    C^3(BG_{\mathrm b};\bbR/\bbZ)
    \times C^2(BG_{\mathrm b};\bbZ_2)
    \times C^1(BG_{\mathrm b};\bbZ_2),
\end{equation}
with cocycle equations
\begin{equation}
    \delta n_1=0,\qquad
    \delta n_2=D_{\widehat v}(n_1)
    =\widehat v\cup n_1,\qquad
    \exp\bigl(2\pi i\,\delta\nu_3\bigr)
    =
    (-1)^{D_{\widehat v}(n_2)}f_{\widehat v}(n_1).
    \label{eq:twisted-degree-three-equations}
\end{equation}
The second equality uses $\operatorname{Sq}^2(n_1)=0$.

\paragraph{Degree four.}
The cochain data are
\begin{equation}
    (\nu_4,n_3,n_2)
    \in
    C^4(BG_{\mathrm b};\bbR/\bbZ)
    \times C^3(BG_{\mathrm b};\bbZ_2)
    \times C^2(BG_{\mathrm b};\bbZ_2)
\end{equation}
obeying
\begin{equation}
    \delta n_2=0,\qquad
    \delta n_3=D_{\widehat v}(n_2),\qquad
    \exp\bigl(2\pi i\,\delta\nu_4\bigr)
    =
    (-1)^{D_{\widehat v}(n_3)}f_{\widehat v}(n_2).
    \label{eq:twisted-degree-four-equations}
\end{equation}
In each degree, $f_{\widehat v}$ is the secondary correction that makes the
final right-hand side closed.  The cochain model and the associated twisted
AHSS differentials are given in \cite[Appendix~B]{Debray:2025kfg}; explicit
formulas for $f_{\widehat v}$ through degree four are given in
\cite[Eq.~(136)]{Wang:2018pdc}.  The degree-four application in
Section~\ref{sec:cat-anomaly} has $n_2=0$ and uses only the first two
equations in \eqref{eq:twisted-degree-four-equations}.

Since $SH_v$ keeps only three layers, there is a comparison with the
full Pontryagin dual,
\begin{equation}
    SH_v^n(BG_{\mathrm b})
    \longrightarrow
    \operatorname{Hom}\bigl(
    \widetilde\Omega_n^{\mathrm{Spin}}
    (BG_{\mathrm b};v),U(1)\bigr).
    \label{eq:twisted-SH-comparison}
\end{equation}
For $v=0$ and $n=3,4$, this is an isomorphism by
\eqref{eq:BM-spin-dual-isomorphism}. Brumfiel and Morgan did not treat
$v\neq0$. We therefore use twisted supercohomology as the three-layer
approximation in the non-split case. Its comparison with the full twisted
spin bordism dual has to be checked separately.

\subsection{Cochain models and examples}
\label{subsec:explicit-three-layer-models}

Let us now write the untwisted Brumfiel--Morgan models used in
\eqref{eq:BM-spin-dual-isomorphism}. We fix an ordered simplicial model for
$Y$, write $U(1)$ additively as $\bbR/\bbZ$, and denote the coboundary by
$\delta$. Except for the final twisted examples, this subsection has $v=0$.
The twisted formulas needed in this paper are
\eqref{eq:twisted-degree-three-equations} and
\eqref{eq:twisted-degree-four-equations}.
The degree-three examples illustrate the non-componentwise stacking law,
while the degree-four examples illustrate the successive obstruction
equations.

\paragraph{The three-layer model in degree three.}
A representative is a triple
\begin{equation}
    (\nu_3,n_2,n_1)
    \in
    \frac{C^3(Y;\bbR/\bbZ)}
    {\delta C^2(Y;\bbR/\bbZ)}
    \times
    Z^2(Y;\bbZ_2)\times
    Z^1(Y;\bbZ_2).
    \label{eq:spin-three-cochains}
\end{equation}
Thus $\delta n_1=\delta n_2=0$, while the remaining cochain satisfies
\begin{equation}
    \delta\nu_3+\frac12\,n_2\cup n_2=0.
    \label{eq:spin-three-equation}
\end{equation}
Here $\frac12:\bbZ_2\rightarrow\bbR/\bbZ$ is the coefficient homomorphism
sending $1$ to $\frac12$. In this convention, $n_1$ enters through the
equivalence relation and the stacking law.\footnote{Equivalently, let
$\widetilde n_1$ be the $\{0,1\}$-valued integral lift of $n_1$ and set
$\nu_3^{\mathrm K}=\nu_3-\frac18\widetilde n_1^{\,3}$. Then
$\delta\nu_3^{\mathrm K}+\frac12n_2\cup n_2
+\frac14\mathcal P(n_1\cup n_1)=0$, where $\mathcal P$ is the Pontryagin
square.}

The triples are taken modulo changes of cochain representatives. In the
convention~\eqref{eq:spin-three-equation}, for
$t\in C^1(Y;\bbZ_2)$ and $x\in C^0(Y;\bbZ_2)$ one imposes
\begin{equation}
    \left(\frac12\,t\cup\delta t,\delta t,\delta x\right)
    \sim (0,0,0).
\end{equation}
Addition is not componentwise: its lower entries are
\begin{equation}
    n_1''=n_1+n_1',
    \qquad
    n_2''=n_2+n_2'+n_1\cup n_1',
    \label{eq:spin-three-stacking}
\end{equation}
Writing $\widetilde n_1$ and $\widetilde n_1'$ for the $\{0,1\}$-valued
integral lifts, the top entry is
\begin{equation}
    \begin{aligned}
    \nu_3''={}&\nu_3+\nu_3'
    +\frac12\Bigl[
        n_2\cup_1n_2'
        +(n_2+n_2')\cup_1(n_1\cup n_1')\\
    &\hspace{34mm}
        +n_1\cup(n_1\cup_1n_1')\cup n_1'
    \Bigr]
    +\frac14\,\widetilde n_1\cup\widetilde n_1'
        \cup\widetilde n_1'.
    \end{aligned}
    \label{eq:spin-three-stacking-top}
\end{equation}
This is the product of \cite[Theorem~1.1]{Brumfiel:2016vpy} written in our
notation.
The equivalence classes, with this addition law, define $SH^3(Y)$. Setting
$n_1=0$ recovers the two-layer model $(\nu_3,n_2)$.

\paragraph{Examples in degree three.}
For $Y=B\bbZ_2$, choose a cocycle
$\mathfrak{a}\in Z^1(B\bbZ_2;\bbZ_2)$ representing the standard generator, and let
$x=[(0,0,\mathfrak{a})]\in SH^3(B\bbZ_2)$. Although $\mathfrak{a}+\mathfrak{a}=0$, the class $x$ does not
have order two because addition of the triples is not componentwise.
Substituting $n_1=n_1'=\mathfrak{a}$ and $n_2=n_2'=0$ into
\eqref{eq:spin-three-stacking}, we find that $2x$ has lower entry zero and
middle entry $\mathfrak{a}^2$. The full stacking law
\eqref{eq:spin-three-stacking-top} further gives
\begin{equation}
    4x=\left[\left(\frac12\mathfrak{a}^3,0,0\right)\right]\neq0,
    \qquad 8x=0,
\end{equation}
where $\frac12\mathfrak{a}^3$ represents the nonzero element of
$H^3(B\bbZ_2;\bbR/\bbZ)$. Hence $x$ has order eight and generates
\begin{equation}
    SH^3(B\bbZ_2)\cong\bbZ_8.
\end{equation}
This agrees with
$\widetilde\Omega_3^{\mathrm{Spin}}(B\bbZ_2)\cong\bbZ_8$, since
$SH^3(B\bbZ_2)$ is its Pontryagin dual.

For $Y=B\bbZ_4$, write
\begin{equation}
    H^\bullet(B\bbZ_4;\bbZ_2)
    \cong\frac{\bbZ_2[x,y]}{(x^2)},
    \qquad |x|=1,\quad |y|=2.
\end{equation}
The lowest-layer class $(0,0,x)$ generates a $\bbZ_2$ summand. The class $y$
can occur in the middle layer because its obstruction $\frac12y^2$ vanishes in
$H^4(B\bbZ_4;\bbR/\bbZ)=0$. One may therefore choose a cochain $\nu_y$
satisfying
\begin{equation}
    \delta\nu_y=-\frac12\,y^2.
\end{equation}
The lift $(\nu_y,y,0)$ generates a $\bbZ_8$ summand, while its square
generates the top-layer subgroup
$H^3(B\bbZ_4;\bbR/\bbZ)\cong\bbZ_4$. Consequently,
\begin{equation}
    SH^3(B\bbZ_4)\cong\bbZ_8\oplus\bbZ_2.
\end{equation}
These results agree with the Adams-spectral-sequence computation of
\cite{GuoEtAlSpinBordism}.

\paragraph{The three-layer model in degree four.}
A representative is a triple
\begin{equation}
    (\nu_4,n_3,n_2)
    \in
    \frac{C^4(Y;\bbR/\bbZ)}
    {\delta C^3(Y;\bbR/\bbZ)}
    \times
    C^3(Y;\bbZ_2)\times
    Z^2(Y;\bbZ_2).
    \label{eq:spin-four-cochains}
\end{equation}
The degree-three entry need not be a cocycle. Its coboundary is fixed by the
degree-two entry:
\begin{equation}
    \delta n_3=n_2\cup n_2.
    \label{eq:spin-four-first-equation}
\end{equation}
Let $\widetilde n_2$ be the integral lift of $n_2$ taking the values $0$ and
$1$, and define
\begin{equation}
    \operatorname{Sq}^2(n_3)
    :=
    n_3\cup_1n_3+n_3\cup_2\delta n_3.
\end{equation}
Brumfiel and Morgan give a natural choice of cochain
$x(n_2)\in C^5(Y;\bbZ_2)$ \cite[Sec.~6]{BrumfielMorgan4} satisfying
\begin{equation}
    \delta x(n_2)
    =
    (n_2\cup n_2)\cup_2(n_2\cup n_2)
    +(n_2\cup_1n_2)\cup(n_2\cup_1n_2).
    \label{eq:x-n2-equation}
\end{equation}
The equation for $\nu_4$ is
\begin{equation}
    \delta\nu_4
    =
    \frac12\,\operatorname{Sq}^2(n_3)
    +\frac14\,\widetilde n_2\cup
    \bigl(\widetilde n_2\cup_1\widetilde n_2\bigr)
    +\frac12\,x(n_2).
    \label{eq:spin-four-second-equation}
\end{equation}
Here $\frac14$ sends an integral cochain to an $\bbR/\bbZ$-valued cochain by
$m\mapsto m/4$. The first equation determines whether a given $n_2$ admits an $n_3$. After
choosing $n_3$, the second equation determines whether $\nu_4$ exists. A
change of a lower cochain also changes the higher cochains. The lower entries
of the addition law are
\begin{equation}
    n_2''=n_2+n_2',
    \qquad
    n_3''=n_3+n_3'+n_2\cup_1n_2',
    \label{eq:spin-four-stacking}
\end{equation}
with the corresponding top-cochain correction and the full representative
changes given in \cite[Secs.~6--7]{BrumfielMorgan4}. The resulting equivalence
classes, with this addition law, define $SH^4(Y)$. Setting $n_2=0$ recovers
the two-layer model, for which the equations reduce to
\begin{equation}
    \delta n_3=0,\qquad
    \delta\nu_4=\frac12\,\operatorname{Sq}^2(n_3).
\end{equation}

\paragraph{Examples in degree four.}
We solve the two obstruction equations from the lowest layer upward. For
$Y=B\bbZ_2$, the only nonzero degree-two candidate is $n_2=\mathfrak{a}^2$. The first
equation would require $\delta n_3=\mathfrak{a}^4$, which is impossible because $\mathfrak{a}^4$
is not a coboundary. Hence $n_2=0$. The remaining candidate
$n_3=\mathfrak{a}^3$ is excluded by the second equation, since
\begin{equation}
    \frac12\,\operatorname{Sq}^2(\mathfrak{a}^3)
    =\frac12\,\mathfrak{a}^5
    \neq0
    \quad\text{in}\quad
    H^5(B\bbZ_2;\bbR/\bbZ).
\end{equation}
Finally, $H^4(B\bbZ_2;\bbR/\bbZ)=0$, and hence
\begin{equation}
    SH^4(B\bbZ_2)=0.
\end{equation}

For $Y=B\bbZ_4$, the same two steps apply. The relation $y^2\neq0$ excludes
the degree-two choice $n_2=y$. After setting $n_2=0$, the degree-three
candidate $n_3=xy$ is excluded because
\begin{equation}
    \frac12\,\operatorname{Sq}^2(xy)
    =\frac12\,xy^2
    \neq0
    \quad\text{in}\quad
    H^5(B\bbZ_4;\bbR/\bbZ).
\end{equation}
Since $H^4(B\bbZ_4;\bbR/\bbZ)=0$, this gives
\begin{equation}
    SH^4(B\bbZ_4)=0.
\end{equation}

For an example with a nonzero answer, take
$Y=B(\bbZ_2\times\bbZ_2)$ and let
$\mathfrak{a},\mathfrak{b}\in H^1(Y;\bbZ_2)$ be the two standard generators. Since
$H^\bullet(Y;\bbZ_2)=\bbZ_2[\mathfrak{a},\mathfrak{b}]$ has no nilpotents, the first obstruction
$n_2^2=0$ forces $n_2=0$. The middle layer is also killed by the second
obstruction. Indeed,
\begin{equation}
    \begin{aligned}
    \operatorname{Sq}^2(\mathfrak{a}^3)&=\mathfrak{a}^5,&
    \operatorname{Sq}^2(\mathfrak{a}^2\mathfrak{b})&=\mathfrak{a}^4\mathfrak{b},\\
    \operatorname{Sq}^2(\mathfrak{a}\mathfrak{b}^2)&=\mathfrak{a}\mathfrak{b}^4,&
    \operatorname{Sq}^2(\mathfrak{b}^3)&=\mathfrak{b}^5.
    \end{aligned}
\end{equation}
The kernel of the coefficient map
\begin{equation}
    H^5(Y;\bbZ_2)\xlongrightarrow{\,1/2\,}H^5(Y;\bbR/\bbZ)
\end{equation}
is the image of
$\operatorname{Sq}^1:H^4(Y;\bbZ_2)\to H^5(Y;\bbZ_2)$, namely
\begin{equation}
    \left\langle
    \mathfrak{a}^4\mathfrak{b}+\mathfrak{a}^3\mathfrak{b}^2,\,
    \mathfrak{a}^2\mathfrak{b}^3+\mathfrak{a}\mathfrak{b}^4
    \right\rangle.
\end{equation}
The four classes above are linearly independent modulo this image. Hence
\begin{equation}
    \frac12\operatorname{Sq}^2:
    H^3(Y;\bbZ_2)\longrightarrow H^5(Y;\bbR/\bbZ)
\end{equation}
is injective, so $n_3=0$. The remaining top layer is
\begin{equation}
    H^4(Y;\bbR/\bbZ)
    \cong\bbZ_2\langle\eta_1\rangle
    \oplus\bbZ_2\langle\eta_2\rangle.
\end{equation}
Here $\eta_1,\eta_2$ are fixed generators.
Thus the four classes admit representatives
\begin{equation}
    \left(
    r\eta_1+s\eta_2,\,
    0,\,
    0
    \right),
    \qquad r,s\in\bbZ_2,
\end{equation}
and hence
\begin{equation}
    SH^4\bigl(B(\bbZ_2\times\bbZ_2)\bigr)
    \cong\bbZ_2\oplus\bbZ_2.
\end{equation}
This agrees with the independent spin bordism computation in
\cite[Theorem~3]{GuoEtAlSpinBordism} through the identification
\eqref{eq:BM-spin-dual-isomorphism}. Note that
\cite[Example~4.14]{Decoppet:2024moc} claims a larger group, retaining
$\mathfrak{a}^2\mathfrak{b}$ and $\mathfrak{a}\mathfrak{b}^2$; those classes are killed by
$d_2=(-1)^{\operatorname{Sq}^2}$, as computed above.

\paragraph{Twisted examples in degrees three and four.}
Let $G_{\mathrm b}=\bbZ_2$, let
$\mathfrak{a}\in H^1(B\bbZ_2;\bbZ_2)$ be the standard generator, and take
\begin{equation}
    v=\mathfrak{a}^2.
\end{equation}
This class determines the non-split extension
\begin{equation}
    1\longrightarrow\bbZ_2^f\longrightarrow\bbZ_4
    \longrightarrow\bbZ_2\longrightarrow1.
\end{equation}
The twisted degree-three group used for the class $\mu$ of the
$\mc{L}_2$ boundary condition is
\begin{equation}
    SH_v^3(B\bbZ_2)=0
    \label{eq:twisted-degree-three-example}
\end{equation}
as computed in \cite[Table~III]{Wang:2018pdc}.  In degree four,
\cite[Example~4.13]{Decoppet:2024moc} finds
\begin{equation}
    SH_v^4(B\bbZ_2)\cong\bbZ_2.
    \label{eq:twisted-degree-four-example}
\end{equation}
The nonzero group in \eqref{eq:twisted-degree-four-example} does not
contradict $\varpi=0$ in Section~\ref{sec:cat-anomaly}: $\varpi$ is the
particular degree-four class identified there, and it is trivial.

\section{Cohomology and Bordism of the 2-group Classifying Space}
\label{app:spectral}

This appendix computes the integral and mod-$2$ (co)homology of the
classifying space $X=B\mathcal G$, followed by its oriented and spin bordism
groups.  Throughout this appendix,
\begin{equation}
\mathcal G=(\Pi_1,\Pi_2,\alpha,\beta)
=(\ztwo,\ztwo,\mathrm{triv},1),
\end{equation}
and we write
$k:=\beta\in H^3(B\Pi_1;\Pi_2)$ for the nontrivial Postnikov class.  Thus
$X$ is the total space of the Postnikov
fibration classified by $k$,
\begin{equation}\label{eq:fibration}
 F=K(\Pi_2,2)\longrightarrow X=B\mathcal G
 \xrightarrow{\,p\,}B=K(\Pi_1,1).
\end{equation}
Because the action
$\alpha$ of $\Pi_1$ on $\Pi_2$ is trivial, all local coefficient systems in
the Serre spectral sequences below are untwisted.
 
\subsection{Cohomology and homology data}
\paragraph{The base.}
Write $B=K(\ztwo,1)$ and let $x\in H^1(B;\ztwo)$ be the standard generator.
Then
\begin{equation}
 H^\bullet(B;\ztwo)=\ztwo[x],
 \qquad |x|=1\,.
\end{equation}
Let $\Bock$ denote the Bockstein map associated to
\begin{equation}\label{eq:integral-bockstein-sequence}
 0\longrightarrow \ringz\xrightarrow{\,2\,}\ringz
  \longrightarrow\ztwo\longrightarrow 0\,,
\end{equation}
and $\rho_2$ denote the mod-2 reduction.
We also define
\begin{equation}
 c_B:=\Bock(x)\in H^2(B;\ringz).
\end{equation}
Then
\begin{equation}\label{eq:base-cohomology}
 H^\bullet(B;\ringz)=\ringz[c_B]/(2c_B),
 \qquad \rho_2(c_B)=x^2.
\end{equation}
We will need
\begin{equation}\label{eq:base-z4}
H^p(B;\ringz_4)=
\begin{cases}
\ringz_4,&p=0,\\
\ringz_2,&p>0.
\end{cases}
\end{equation}
For later use, the coefficient reduction
$H^p(B;\ringz_4)\to H^p(B;\ringz_2)$ is nonzero for even $p$ and is
zero for odd $p$; these statements follow from the standard periodic
resolution for cyclic groups~\cite{brown2012cohomology}.  Let
\begin{equation}
 \zeta_1\in H^1(B;\ringz_4)\cong\ringz_2
\end{equation}
denote its generator, so that $\rho_2(\zeta_1)=0$.

\paragraph{The fiber.}
Let $F=K(\ztwo,2)$, let $u_2\in H^2(F;\ztwo)$ be its fundamental class,
and set
\begin{equation}\label{eq:fiber-mod2-generators}
 u_3:=\StSq^1u_2,
 \qquad
 u_5:=\StSq^2\StSq^1u_2.
\end{equation}
The mod-$2$ cohomology ring in the required range is~\cite{Cartan1954EM}
\begin{equation}\label{eq:fiber-mod2-ring}
 H^{\bullet\leq 7}(F;\ztwo)
 =\ztwo[u_2,u_3,u_5]^{\bullet\leq 7}.
\end{equation}

Together with the Bockstein calculation, this determines the required
integral groups~\cite{HatcherSpectralSequences,clementintegral}.  The two
coefficient systems are summarized in Table~\ref{tab:fiber-cohomology}.
\begin{table}[h]
\centering
\small
\setlength{\tabcolsep}{8pt}
\begin{tabular}{c|l|l}
$q$ & $H^q(F;\ztwo)$ & $H^q(F;\ringz)$ \\ \midrule
$0$ & $\ztwo$ & $\ringz$ \\
$1$ & $0$ & $0$ \\
$2$ & $\ztwo\,u_2$ & $0$ \\
$3$ & $\ztwo\,u_3$ & $\ztwo\,\eta_3$ \\
$4$ & $\ztwo\,u_2^2$ & $0$ \\
$5$ & $\ztwo\langle u_5,u_2u_3\rangle$ & $\ringz_4\,\eta_5$ \\
$6$ & $\ztwo\langle u_3^2,u_2^3\rangle$ & $\ztwo\,\eta_6$ \\
$7$ & $\ztwo\langle u_2u_5,u_2^2u_3\rangle$ & $\ztwo\,\eta_7$.
\end{tabular}
\caption{Low-degree cohomology of $F=K(\ztwo,2)$.}
\label{tab:fiber-cohomology}
\end{table}

We normalize the integral generators by
\begin{align}
 \eta_3&=\Bock(u_2),
 &\rho_2(\eta_3)&=u_3,\label{eq:eta3}\\
 \rho_2(\eta_5)&=u_5+u_2u_3,
 &2\eta_5&=\Bock(u_2^2),\label{eq:eta5}\\
 \eta_6&=\eta_3^2,
 &\rho_2(\eta_6)&=u_3^2,\label{eq:eta6}\\
 \eta_7&=\Bock(u_2^3),
 &\rho_2(\eta_7)&=u_2^2u_3.\label{eq:eta7}
\end{align}
The notation $x,c_B$ is reserved for classes on the base, while
$u_j,\eta_j$ denotes classes on the fiber.  We introduce separate symbols
for classes on $X$ below.  Higher-degree (co)homology groups of the fiber can
be found in~\cite{clementintegral}.

\paragraph{The mod-$2$ Serre spectral sequence.}
The $E_2$-page is the bigraded algebra
\begin{equation}\label{eq:mod2-e2}
 \overline E_2^{p,q}
 =H^p(B;\ztwo)\otimes H^q(F;\ztwo)
 =\ztwo[x]\otimes\ztwo[u_2,u_3,u_5]
\end{equation}
in the range under consideration~\cite{McCleary_2000}, shown in Table~\ref{tab:E2modtwo}.

\begin{sseqdata}[
 name = E2modtwo,
 classes = { draw = none },
 axes type = frame,
 scale = 0.68
]
\class["\ztwo"](0,0)
\class["\ztwo"](1,0)
\class["\ztwo"](2,0)
\class["\ztwo"](3,0)
\class["\ztwo"](4,0)
\class["\ztwo"](5,0)
\class["\ztwo"](6,0)
\class["\ztwo"](7,0)

\class["\ztwo"](0,2)
\class["\ztwo"](1,2)
\class["\ztwo"](2,2)
\class["\ztwo"](3,2)
\class["\ztwo"](4,2)
\class["\ztwo"](5,2)

\class["\ztwo"](0,3)
\class["\ztwo"](1,3)
\class["\ztwo"](2,3)
\class["\ztwo"](3,3)
\class["\ztwo"](4,3)

\class["\ztwo"](0,4)
\class["\ztwo"](1,4)
\class["\ztwo"](2,4)
\class["\ztwo"](3,4)

\class["\ztwo^2"](0,5)
\class["\ztwo^2"](1,5)
\class["\ztwo^2"](2,5)

\class["\ztwo^2"](0,6)
\class["\ztwo^2"](1,6)

\class["\ztwo^2"](0,7)
\end{sseqdata}

\begin{table}[htbp]
 \centering
 \printpage[
 name = E2modtwo,
 grid = chess,
 ]
 \caption{The relevant $\overline E_2$-page of
 $\overline E_2^{p,q}=H^p(K(\ztwo,1);H^q(K(\ztwo,2);\ztwo))$.
 Only entries with $p+q\leq 7$ are displayed.}
 \label{tab:E2modtwo}
\end{table}

The group in bidegree $(p,q)$ is generated by $x^p$ times the generators in
the $q$th row of Table~\ref{tab:fiber-cohomology}. The Postnikov class is
precisely the transgression of the fundamental class of the
fiber~\cite{McCleary_2000,singer2006steenrod}:
\begin{equation}\label{eq:basic-mod2-differential}
 \overline d_3(u_2)=k=\beta=x^3.
\end{equation}
This is the first place where the non-split $2$-group differs from the product
$K(\ztwo,1)\times K(\ztwo,2)$. The class $k$ does not change the
$E_2$-page; it changes its differential.

The differential is a derivation. In the degrees needed here,
\begin{equation}
 \overline d_3(u_2u_3)=x^3u_3,
 \qquad
 \overline d_3(u_2^3)=x^3u_2^2,
 \qquad
 \overline d_3(u_3)=\overline d_3(u_5)=0.
\end{equation}
Consequently,
\begin{equation}\label{eq:mod2-e4}
 \overline E_4^{\bullet\leq 6}
 \cong
 \left(\ztwo[x,u_3,u_2^2,u_5]/(x^3)\right)^{\bullet\leq 6}.
\end{equation}

The Kudo transgression theorem also determines what happens to the
Steenrod operations on $u_2$~\cite{singer2006steenrod}.  Starting from
$\overline d_3(u_2)=x^3$, it says that the first differential on which
$\StSq^i u_2$ can transgress has target $\StSq^i(x^3)$, with the page shifted
by the degree of the Steenrod operation.  In the present calculation,
\begin{equation*}
 u_3=\StSq^1u_2,
 \qquad
 u_2^2=\StSq^2u_2,
 \qquad
 u_5=\StSq^2u_3=\StSq^2\StSq^1u_2.
\end{equation*}
Applying the theorem first to $u_2$, and then to $u_3$, gives
\begin{align}
 \overline d_4(u_3)
  &=\StSq^1(x^3)=x^4=0
  &&\text{in }\overline E_4,\label{eq:d4-u3}\\
 \overline d_5(u_2^2)
  &=\StSq^2(x^3)=x^5=0
  &&\text{in }\overline E_5,\label{eq:d5-u22}\\
 \overline d_6(u_5)
  &=\StSq^2\StSq^1(x^3)
  =\StSq^2(x^4)=0.\label{eq:d6-u5}
\end{align}
Here the first two displayed targets vanish because
$x^4=x\,x^3$ and $x^5=x^2x^3$, while $x^3$ has already been killed by
$\overline d_3(u_2)=x^3$.  The last target vanishes directly from
$\StSq^2(x^4)=\binom{4}{2}x^6=0$ over $\ztwo$.  Thus no further differential
removes $u_3$, $u_2^2$, or $u_5$ in the stated degree range, and
Eq.~\eqref{eq:mod2-e4} is the associated graded algebra of
$H^{\bullet\leq 6}(X;\ztwo)$.

We record the degree-five calculation separately because it will resolve the
integral extension in degree six. On the $\overline E_2$-page, the diagonal is
\begin{equation}\label{eq:mod2-degree-five-e2}
\begin{aligned}
\bigoplus_{p+q=5}\overline E_2^{p,q}
={}&\ztwo\,x^5
 \oplus\ztwo\,(x^3u_2)
 \oplus\ztwo\,(x^2u_3)
 \oplus\ztwo\,(xu_2^2)\\
&\oplus\ztwo\langle u_5,u_2u_3\rangle.
\end{aligned}
\end{equation}
The relevant $\overline d_3$-differentials are
\begin{equation}\label{eq:mod2-degree-five-d3}
\overline d_3(x^2u_2)=x^5,
\qquad
\overline d_3(x^3u_2)=x^6,
\qquad
\overline d_3(u_2u_3)=x^3u_3.
\end{equation}
Thus $x^5$ is a boundary, while $x^3u_2$ and $u_2u_3$ are not cycles. The
remaining three classes are $x^2u_3$, $xu_2^2$, and $u_5$. Their only possible
outgoing higher differentials are
\begin{equation}\label{eq:mod2-degree-five-higher}
\begin{aligned}
\overline d_4(x^2u_3)&=x^6=0 &&\text{in }\overline E_4,\\
\overline d_5(xu_2^2)&=x^6=0 &&\text{in }\overline E_5,\\
\overline d_6(u_5)&=\StSq^2\StSq^1(x^3)=\StSq^2(x^4)=0.
\end{aligned}
\end{equation}
In the first two lines, $x^6$ is already a $\overline d_3$-boundary. No higher
differential can enter any of these three bidegrees. Hence
\begin{equation}\label{eq:mod2-h5-associated}
\bigoplus_{p+q=5}\overline E_\infty^{p,q}
=\ztwo\langle x^2u_3,xu_2^2,u_5\rangle.
\end{equation}
The induced filtration of $H^5(X;\ztwo)$ is a filtration by $\ztwo$-vector
spaces. Its additive extensions therefore split, although not canonically and
not necessarily multiplicatively. Consequently,
\begin{equation}\label{eq:mod2-h5-result}
H^5(X;\ztwo)\cong\ztwo^{\oplus3}.
\end{equation}

We now distinguish actual classes on $X$ from their associated-graded
representatives.  Define
\begin{equation}\label{eq:mod2-x-generators}
 \mathfrak{a}:=p^*x,
 \qquad
 \left.v_3\right|_F=u_3,
 \qquad
 \left.v_4\right|_F=u_2^2,
 \qquad
 \left.v_5\right|_F=u_5.
\end{equation}
These choices are possible because the indicated base and fiber classes
survive to $\overline E_\infty$.  Table~\ref{tab:x-mod2-cohomology} gives an
additive basis through degree $6$.
\begin{table}[h]
\centering
\begin{tabular}{c|l}
$n$ & $H^n(X;\ztwo)$ and a chosen basis \\ \midrule
$0$ & $\ztwo$ \\
$1$ & $\ztwo\langle \mathfrak{a} \rangle$ \\
$2$ & $\ztwo\langle \mathfrak{a}^2 \rangle$ \\
$3$ & $\ztwo\langle v_3 \rangle$ \\
$4$ & $\ztwo\langle \mathfrak{a}v_3,v_4\rangle$ \\
$5$ & $\ztwo\langle \mathfrak{a}^2v_3,\mathfrak{a}v_4,v_5\rangle$ \\
$6$ & $\ztwo\langle v_3^2,\mathfrak{a}^2v_4,\mathfrak{a}v_5\rangle$.
\end{tabular}
\caption{Mod-$2$ cohomology of $X$ through degree $6$.}
\label{tab:x-mod2-cohomology}
\end{table}
Here $\mathfrak{a}$ is the universal class introduced in
Section~\ref{sec:anomaly-pol}.  In particular, $\mathfrak{a}^3=0$ in cohomology.  This is compatible
with the cochain equation $\delta\mathfrak{b}=\mathfrak{a}^3$: the cochain $\mathfrak{a}^3$ need not vanish,
but it is exact on $X$. The choices of the $v_j$ are not canonical, and
Table~\ref{tab:x-mod2-cohomology} is an additive statement rather than a
complete presentation of the cohomology ring. The displayed products are
nonzero and independent because their representatives lie in different
nonzero bidegrees on the $\overline E_\infty$-page.

\paragraph{The integral Serre spectral sequence.}
The integral spectral sequence is
\begin{equation}\label{eq:integral-sss}
 E_2^{p,q}=H^p\bigl(B;H^q(F;\ringz)\bigr)
 \Longrightarrow H^{p+q}(X;\ringz).
\end{equation}
The nonzero row patterns relevant through total degree $7$ are
\begin{equation}\label{eq:integral-e2-rows}
\begin{aligned}
E_2^{p,0}&=
\begin{cases}
\ringz,&p=0,\\
\ztwo,&p>0\text{ even},\\
0,&p\text{ odd},
\end{cases}
\\[3pt]
E_2^{p,3}&=E_2^{p,6}=E_2^{p,7}=\ztwo
\qquad (p\geq 0),\\[3pt]
E_2^{p,5}&=
\begin{cases}
\ringz_4,&p=0,\\
\ztwo,&p>0.
\end{cases}
\end{aligned}
\end{equation}
The rows $q=1,2,4$ vanish.

\begin{sseqdata}[
 name = E2integral,
 classes = { draw = none },
 axes type = frame,
 scale = 0.68
]
\class["\ringz"](0,0)
\class["\ztwo"](2,0)
\class["\ztwo"](4,0)
\class["\ztwo"](6,0)
\class["\ztwo"](8,0)

\class["\ztwo"](0,3)
\class["\ztwo"](1,3)
\class["\ztwo"](2,3)
\class["\ztwo"](3,3)
\class["\ztwo"](4,3)
\class["\ztwo"](5,3)

\class["\ringz_4"](0,5)
\class["\ztwo"](1,5)
\class["\ztwo"](2,5)
\class["\ztwo"](3,5)

\class["\ztwo"](0,6)
\class["\ztwo"](1,6)
\class["\ztwo"](2,6)

\class["\ztwo"](0,7)
\class["\ztwo"](1,7)
\end{sseqdata}

\begin{table}[htbp]
 \centering
 \printpage[
 name = E2integral,
 grid = chess,
 ]
 \caption{The relevant $E_2$-page of
 $E_2^{p,q}=H^p(K(\ztwo,1);H^q(K(\ztwo,2);\ringz))$.
 The nonzero rows used below are displayed through total degree $8$.}
 \label{tab:E2integral}
\end{table}

For $q=3,6,7$, the entry at $(p,q)$ is denoted by $x^p\eta_q$ in the
corresponding $\ztwo$ coefficient system. For $q=5$, the entry at $(0,5)$
is generated by $\eta_5$, while the positive-filtration entries are generated
over the groups in Eq.~\eqref{eq:base-z4}.  This notation refers to classes
on the $E_2$-page, not to products of integral classes on $B$.

The relevant $E_2$-page for space $X$ is displayed in Table~\ref{tab:E2integral}. 

All relevant $d_2$-differentials vanish. Indeed, the differentials leaving
the $q=3$ and $q=5$ rows land in the zero rows $q=2$ and $q=4$, while
$d_2(\eta_6)=d_2(\eta_3^2)=0$. Moreover,
$\rho_2(\eta_7)=u_2^2u_3$ has zero mod-$2$ differential, and coefficient
reduction is injective on its target. Hence $d_2(\eta_7)=0$ and $E_2=E_3$
in the displayed range.

By Eq.~\eqref{eq:eta5} and naturality of coefficient reduction,
Eq.~\eqref{eq:basic-mod2-differential} gives
\begin{equation}\label{eq:integral-d3-eta5}
\begin{aligned}
\rho_2\bigl(d_3(\eta_5)\bigr)
&=\overline d_3(u_5+u_2u_3)\\
&=x^3u_3.
\end{aligned}
\end{equation}
The target reduction is an isomorphism, hence
\begin{equation}\label{eq:integral-d3-list}
\begin{aligned}
 d_3(\eta_5)&=x^3\eta_3,
  &&E_3^{0,5}=\ringz_4\longrightarrow E_3^{3,3}=\ztwo,
  \\[2pt]
 d_3(\zeta_1\eta_5)&=0,
  &&E_3^{1,5}=\ztwo\longrightarrow E_3^{4,3}=\ztwo,
  \\[2pt]
 d_3(c_B\eta_5)&=x^5\eta_3,
  &&E_3^{2,5}=\ztwo\longrightarrow E_3^{5,3}=\ztwo.
\end{aligned}
\end{equation}
The first map is surjective with kernel
$2\ringz_4\langle \eta_5 \rangle \cong\ztwo\langle 2\eta_5 \rangle$, the second map vanishes because
$\rho_2(\zeta_1)=0$, and the third map is an isomorphism.

The degree-$7$ fiber class is also killed. Using
$\eta_7=\Bock(u_2^3)$ and compatibility of the Bockstein with the spectral
sequence differential~\cite{McCleary_2000,singer2006steenrod}
,
\begin{equation}\label{eq:integral-d3-eta7}
\begin{aligned}
d_3(\eta_7)
 &=\Bock\bigl(\overline d_3(u_2^3)\bigr)\\
 &=\Bock(x^3u_2^2)
  =x^3(2\eta_5).
\end{aligned}
\end{equation}
Thus
\begin{equation}
  d_3:E_3^{0,7}\overset{\sim}{\to}E_3^{3,5}.
\end{equation}
We collect the entries survived to $E_4$-page in Table~\ref{tab:E4integral}.
\begin{sseqdata}[
 name = E4integral,
 classes = { draw = none },
 axes type = frame,
 scale = 0.68
]
\class["\ringz"](0,0)
\class["\ztwo"](2,0)
\class["\ztwo"](4,0)
\class["\ztwo"](6,0)
\class["\ztwo"](8,0)

\class["\ztwo"](0,3)
\class["\ztwo"](1,3)
\class["\ztwo"](2,3)
\class["\ztwo"](4,3)

\class["\ztwo"](0,5)
\class["\ztwo"](1,5)

\class["\ztwo"](0,6)
\class["\ztwo"](1,6)

\class[""](0,7)
\end{sseqdata}

\begin{table}[htbp]
 \centering
 \printpage[
 name = E4integral,
 grid = chess,
 ]
 \caption{The integral $E_4$-page through total degree $7$.}
 \label{tab:E4integral}
\end{table}

The integral class $\eta_3=\Bock(u_2)$ detects the same Postnikov class one
page later:
\begin{equation}\label{eq:integral-d4-basic}
\begin{aligned}
    d_4(\eta_3)
      &=\Bock\bigl(\overline d_3(u_2)\bigr)\\
      &=\Bock(x^3)=c_B^2.
\end{aligned}
\end{equation}
Therefore $d_4:E_4^{0,3}\to E_4^{4,0}$ is an isomorphism.  Multiplicativity
also gives
\begin{equation}\label{eq:integral-d4-family}
    d_4(c_B\eta_3)=c_B^3,
    \qquad
    d_4(c_B^2\eta_3)=c_B^4,
\end{equation}
so the corresponding maps from bidegrees $(2,3)$ and $(4,3)$ are
isomorphisms as well.  By contrast,
\begin{equation}\label{eq:d4-eta6}
\begin{aligned}
    d_4(\eta_6)
      &=d_4(\eta_3^2)\\
      &=d_4(\eta_3)\eta_3-\eta_3d_4(\eta_3)=0.
\end{aligned}
\end{equation}
Hence the class in $E_4^{0,6}$ survives.

No later differential can affect the surviving terms in total degree at
most $7$. The nonzero $E_\infty$-terms are listed in
Table~\ref{tab:integral-einfty}.
\begin{table}[h]
\centering
\begin{tabular}{c|c|c}
total degree & bidegree & surviving group and representative \\ \midrule
$0$ & $(0,0)$ & $\ringz$ \\
$2$ & $(2,0)$ & $\ztwo\langle c_B\rangle$ \\
$4$ & $(1,3)$ & $\ztwo\langle x\eta_3\rangle$ \\
$5$ & $(0,5)$ & $\ztwo\langle 2\eta_5\rangle$ \\
$6$ & $(0,6)$ & $\ztwo\langle\eta_6\rangle$ \\
    & $(1,5)$ & $\ztwo\langle\zeta_1\eta_5\rangle$ \\
$7$ & $(1,6)$ & $\ztwo\langle x\eta_6\rangle$.
\end{tabular}
\caption{Nonzero integral $E_\infty$-terms through total degree $7$.}
\label{tab:integral-einfty}
\end{table}

Here $x\eta_3$ and $x\eta_6$ denote generators in the cohomology of the base
with the indicated $\ztwo$ coefficient modules.  They are not products with
an integral degree-$1$ class, since no such class exists on $B$.

First consider total degree five in the integral spectral sequence.  The class
$\eta_5\in E_3^{0,5}\cong\ringz_4$ has kernel
$\ztwo\,(2\eta_5)$ under $d_3$, while
$d_4:E_4^{2,3}\to E_4^{6,0}$ is an isomorphism.  Therefore the only nonzero
$E_\infty$-term in total degree five is
\begin{equation}\label{eq:integral-h5-filtration}
E_\infty^{0,5}=\ztwo\,(2\eta_5).
\end{equation}
There is no additive extension to solve in this degree, and hence
\begin{equation}\label{eq:integral-h5-result}
H^5(X;\ringz)\cong\ztwo.
\end{equation}

In total degree six, the integral spectral sequence gives only
\begin{equation}\label{eq:h6-extension}
0\longrightarrow E_\infty^{1,5}\cong\ztwo
\longrightarrow H^6(X;\ringz)
\longrightarrow E_\infty^{0,6}\cong\ztwo
\longrightarrow 0.
\end{equation}
Thus it determines the order of $H^6(X;\ringz)$ but does not distinguish
$\ringz_4$ from $\ztwo\oplus\ztwo$.  We now use the independently computed
mod-$2$ result~\eqref{eq:mod2-h5-result}. The coefficient universal
coefficient theorem gives
\begin{equation}\label{eq:coefficient-uct}
0\longrightarrow H^5(X;\ringz)\otimes\ztwo
\longrightarrow H^5(X;\ztwo)
\longrightarrow \operatorname{Tor}\bigl(H^6(X;\ringz),\ztwo\bigr)
\longrightarrow 0.
\end{equation}
Substituting Eqs.~\eqref{eq:integral-h5-result} and
\eqref{eq:mod2-h5-result} gives a short exact sequence of $\ztwo$-vector spaces
\begin{equation}\label{eq:coefficient-uct-groups}
0\longrightarrow\ztwo
\longrightarrow\ztwo^{\oplus3}
\longrightarrow\operatorname{Tor}\bigl(H^6(X;\ringz),\ztwo\bigr)
\longrightarrow0.
\end{equation}
Therefore
\begin{equation}
\operatorname{Tor}\bigl(H^6(X;\ringz),\ztwo\bigr)\cong\ztwo^{\oplus2}.
\end{equation}
On the other hand,
\begin{equation}
\operatorname{Tor}(\ringz_4,\ztwo)\cong\ztwo,
\qquad
\operatorname{Tor}(\ztwo\oplus\ztwo,\ztwo)\cong\ztwo^{\oplus2}.
\end{equation}
Comparison with Eq.~\eqref{eq:h6-extension} therefore resolves the extension:
\begin{equation}\label{eq:h6-split}
    H^6(X;\ringz)\cong\ztwo\oplus\ztwo.
\end{equation}

We name the integral classes on $X$ by their associated-graded
representatives in
Table~\ref{tab:integral-einfty}:
\begin{equation}\label{eq:integral-x-generators}
\begin{array}{c|ccccc}
\text{class on }X & c & A_4 & A_5 & A_6 & B_6 \\ \midrule
E_\infty\text{-representative}
& c_B & x\eta_3 & 2\eta_5 & \eta_6 & \zeta_1\eta_5.
\end{array}
\end{equation}
Thus $c=p^*c_B$, while
\begin{equation}
    \left.A_4\right|_F=\left.B_6\right|_F=0,
    \qquad
    \left.A_5\right|_F=2\eta_5,
    \qquad
    \left.A_6\right|_F=\eta_6.
\end{equation}
The coefficient reductions also satisfy
\begin{equation}
    \rho_2(c)=\mathfrak{a}^2,
    \qquad
    \rho_2(A_4)=\mathfrak{a}v_3.
\end{equation}
Then
\begin{equation}\label{eq:integral-cohomology-result}
H^n(X;\ringz)=
\begin{cases}
\ringz,&n=0,\\
\ztwo\,c,&n=2,\\
\ztwo\,A_4,&n=4,\\
\ztwo\,A_5,&n=5,\\
\ztwo\langle A_6,B_6\rangle,&n=6,\\
0,&n=1,3.
\end{cases}
\end{equation}
For example, $c^2=0$ because $c_B^2$ is killed by
Eq.~\eqref{eq:integral-d4-basic}.  Equation
\eqref{eq:integral-cohomology-result} is an additive result; a full integral
ring presentation is not required for the bordism calculation.

The spaces in the Postnikov fibration are of finite type. Since the
positive-degree integral cohomology groups above are finite, the universal
coefficient theorem shows that the corresponding homology groups have no
free part and gives, noncanonically,
\begin{equation}\label{eq:homology-from-cohomology}
    H_n(X;\ringz)\cong H^{n+1}(X;\ringz),
    \qquad n\geq 1,
\end{equation}
in this range.  To obtain $H_6(X;\ringz)$ we use the additional result from
Table~\ref{tab:integral-einfty},
\begin{equation}
    H^7(X;\ringz)\cong\ztwo.
\end{equation}
Over the field $\ztwo$, cohomology is dual to homology, so the mod-$2$
homology groups have the same dimensions as those in
Table~\ref{tab:x-mod2-cohomology}.

The resulting integral and mod-$2$ (co)homology groups are collected in
Table~\ref{tab:final-cohomology}.
\begin{table}[h]
\centering
\small
\renewcommand{\arraystretch}{1.15}
\setlength{\tabcolsep}{5pt}
\begin{tabular}{c|ccccccc}
$n$ & $0$ & $1$ & $2$ & $3$ & $4$ & $5$ & $6$ \\ \midrule
$H^n(X;\ringz)$
& $\ringz$ & $0$ & $\ztwo$ & $0$ & $\ztwo$ & $\ztwo$ & $\ztwo^2$ \\
$H_n(X;\ringz)$
& $\ringz$ & $\ztwo$ & $0$ & $\ztwo$ & $\ztwo$ & $\ztwo^2$ & $\ztwo$ \\
$H^n(X;\ztwo)$
& $\ztwo$ & $\ztwo$ & $\ztwo$ & $\ztwo$
& $\ztwo^2$ & $\ztwo^3$ & $\ztwo^3$ \\
$H_n(X;\ztwo)$
& $\ztwo$ & $\ztwo$ & $\ztwo$ & $\ztwo$
& $\ztwo^2$ & $\ztwo^3$ & $\ztwo^3$.
\end{tabular}
\caption{Integral and mod-$2$ (co)homology of $X=B\mathcal{G}$ through
degree $6$.}
\label{tab:final-cohomology}
\end{table}

\subsection{Oriented bordism}
The oriented-bordism AHSS collapses in the required range. We begin with the
coefficient groups through degree $9$~\cite{MilnorStasheff1974}:
\begin{equation}\label{eq:oriented-bordism-point}
\begin{tabular}{c|cccccccccc}
$q$ & $0$ & $1$ & $2$ & $3$ & $4$ & $5$ & $6$ & $7$ & $8$ & $9$ \\ \midrule
$\Omega_q^{\mathrm{SO}}(\mathrm{pt})$
& $\ringz$ & $0$ & $0$ & $0$ & $\ringz$ & $\ztwo$ & $0$ & $0$
& $\ringz^2$ & $\ztwo^2$.
\end{tabular}
\end{equation}
The AHSS for reduced oriented bordism is~\cite{adams1974stable}
:
\begin{equation}\label{eq:oriented-ahss}
E^2_{p,q}
=\widetilde H_p\bigl(X;\Omega_q^{\mathrm{SO}}(\mathrm{pt})\bigr)
\Longrightarrow \widetilde\Omega_{p+q}^{\mathrm{SO}}(X),
\end{equation}
with homological differential
\begin{equation}
d^r_{p,q}:E^r_{p,q}\longrightarrow E^r_{p-r,q+r-1}.
\end{equation}
Using Table~\ref{tab:final-cohomology}, the part of the $E^2$-page needed
to compute total degree at most $6$ is shown below.  We also display the
total-degree-$7$ sources of possible incoming differentials; the value of
$H_7(X;\ringz)$ will not be needed.

\begin{sseqdata}[
  name = E2oriented,
  classes = { draw = none },
  axes type = frame,
  scale = 0.75
]
\class(0,7)

\class["\ztwo"](1,0)
\class["\ztwo"](3,0)
\class["\ztwo"](4,0)
\class["\ztwo^2"](5,0)
\class["\ztwo"](6,0)
\class["H_7"](7,0)

\class["\ztwo"](1,4)
\class["\ztwo"](3,4)

\class["\ztwo"](1,5)
\class["\ztwo"](2,5)
\end{sseqdata}

\begin{table}[h]
\centering
\printpage[
  name = E2oriented,
  grid = chess,
]
\caption{The $E^2$-page of
$E^2_{p,q}=\widetilde H_p(X;\Omega_q^{\mathrm{SO}}(\mathrm{pt}))$.
The total-degree-$7$ entries are included to show possible incoming
differentials to total degree $6$.}
\label{tab:E2oriented}
\end{table}

The rows $q=0$ and $q=4$ use integral homology, while the row $q=5$
uses mod-$2$ homology.  The first possible differential between the rows
$q=0$ and $q=4$ is $d^5$.  In the present range,
\begin{equation}
d^5_{5,0}:E^5_{5,0}\longrightarrow E^5_{0,4}
\end{equation}
vanishes because the zeroth column is absent in the reduced AHSS.  The next
such differential is
\begin{equation}
d^5_{6,0}:H_6(X;\ringz)\longrightarrow H_1(X;\ringz),
\end{equation}
which also vanishes.  Indeed, the oriented $d^5$ is detected by a mod-$3$
cohomology operation, whereas the source and target here consist only of
elements of order $2$.

The low-degree Postnikov invariants of $MSO$ controlling the other possible
incoming differentials are described in~\cite{Troue}. On the groups appearing
here they give zero maps, so
\begin{equation}
d^2_{3,4}:E^2_{3,4}\longrightarrow E^2_{1,5},
\qquad
d^6_{7,0}:E^6_{7,0}\longrightarrow E^6_{1,5},
\end{equation}
vanish, and $\Omega_5^{\mathrm{SO}}(\mathrm{pt})=\ztwo$ contributes a
decoupled copy of $\widetilde H_p(X;\ztwo)$. The same low-degree Postnikov
decomposition separates the three displayed rows, so there is no additive
extension between them.
Therefore
\begin{equation}\label{eq:oriented-direct-sum}
\widetilde\Omega_n^{\mathrm{SO}}(X)
\cong \widetilde H_n(X;\ringz)
\oplus \widetilde H_{n-4}(X;\ringz)
\oplus \widetilde H_{n-5}(X;\ztwo),
\qquad n\leq6.
\end{equation}
Groups in negative degrees are understood to vanish. Hence
\begin{equation}\label{eq:reduced-oriented-result}
\begin{tabular}{c|cccccc}
$n$ & $1$ & $2$ & $3$ & $4$ & $5$ & $6$ \\ \midrule
$\widetilde\Omega_n^{\mathrm{SO}}(X)$
& $\ztwo$ & $0$ & $\ztwo$ & $\ztwo$ & $\ztwo^3$ & $\ztwo^2$.
\end{tabular}
\end{equation}
Finally,
\begin{equation}
\Omega_n^{\mathrm{SO}}(X)
\cong\widetilde\Omega_n^{\mathrm{SO}}(X)
\oplus\Omega_n^{\mathrm{SO}}(\mathrm{pt}),
\end{equation}
so the unreduced groups are
\begin{equation}\label{eq:unreduced-oriented-result}
\begin{tabular}{c|ccccccc}
$n$ & $0$ & $1$ & $2$ & $3$ & $4$ & $5$ & $6$ \\ \midrule
$\Omega_n^{\mathrm{SO}}(X)$
& $\ringz$ & $\ztwo$ & $0$ & $\ztwo$ & $\ringz\oplus\ztwo$
& $\ztwo^4$ & $\ztwo^2$.
\end{tabular}
\end{equation}

\paragraph{The Pontryagin dual.}
We next identify the generator of
$\operatorname{Hom}(\widetilde{\Omega}_4^{\mathrm{SO}}(X),U(1))$.  Write
$U(1)=\mathbb{R}/\ringz$ additively.  The cohomological AHSS for the
Pontryagin dual of reduced oriented bordism is~\cite{SaitoTachikawaZhang2026}
:
\begin{equation}\label{eq:oriented-dual-ahss}
E_2^{p,q}
=\widetilde H^p\!\left(
X;\operatorname{Hom}\bigl(\Omega_q^{\mathrm{SO}}(\mathrm{pt}),U(1)\bigr)
\right)
\Longrightarrow
\operatorname{Hom}\bigl(\widetilde\Omega_{p+q}^{\mathrm{SO}}(X),U(1)\bigr),
\end{equation}
with differential
\begin{equation}
d_r:E_r^{p,q}\longrightarrow E_r^{p+r,q-r+1}.
\end{equation}
The rows $q=0$ and $q=4$ have $U(1)$ coefficients, while the row $q=5$
has $\ztwo$ coefficients.  Since $U(1)$ is divisible,
\begin{equation}
H^p(X;U(1))
\cong\operatorname{Hom}\bigl(H_p(X;\ringz),U(1)\bigr).
\end{equation}
The reduced $E_2$-page through total degree $6$ is therefore

\begin{sseqdata}[
  name = E2orienteddual,
  classes = { draw = none },
  axes type = frame,
  scale = 0.75
]
\class(0,6)

\class["\ztwo"](1,0)
\class["\ztwo"](3,0)
\class["\ztwo"](4,0)
\class["\ztwo^2"](5,0)
\class["\ztwo"](6,0)

\class["\ztwo"](1,4)

\class["\ztwo"](1,5)
\end{sseqdata}

\begin{table}[h]
\centering
\printpage[
  name = E2orienteddual,
  grid = chess,
]
\caption{The $E_2$-page for the Pontryagin dual of reduced oriented
bordism.  Only entries with $p+q\leq6$ are displayed.}
\label{tab:E2orienteddual}
\end{table}

The first differential from the row $q=4$ to the row $q=0$ is
given explicitly in~\cite{SaitoTachikawaZhang2026}
:
\begin{equation}\label{eq:oriented-dual-d5}
\begin{aligned}
d_5:H^p(X;U(1))
&\xrightarrow{\ \operatorname{Bock}_{U(1)\to\ringz_3}\ }
H^{p+1}(X;\ringz_3)
\xrightarrow{\ \mathcal P^1\ }H^{p+5}(X;\ringz_3)\\
&\longrightarrow H^{p+5}(X;U(1)).
\end{aligned}
\end{equation}
Here $\operatorname{Bock}_{U(1)\to\ringz_3}$ is the Bockstein for
\begin{equation}
0\longrightarrow\ringz_3\longrightarrow U(1)
\xrightarrow{\,\times3\,}U(1)\longrightarrow0,
\end{equation}
and the last arrow in~\eqref{eq:oriented-dual-d5} is induced by the
coefficient homomorphism $\ringz_3\to U(1)$ sending $1$ to $1/3$.
The only such differential on the displayed page is
$d_5:E_5^{1,4}\to E_5^{6,0}$.  Multiplication by $3$ is an automorphism on
$H^1(X;U(1))\cong\ztwo$, so
$\operatorname{Bock}_{U(1)\to\ringz_3}$ and hence this
differential vanish.
The row $q=5$ is decoupled for the same reason as in the homological AHSS.

On the diagonal $p+q=4$, the reduced page has only
\begin{equation}
E_2^{4,0}=H^4(X;U(1)),
\end{equation}
because the point contribution $E_2^{0,4}$ is absent.  No differential can
enter or leave this term, and there is no additive extension.  The coefficient
sequence
\begin{equation}
0\longrightarrow\ringz\longrightarrow\mathbb{R}\longrightarrow U(1)
\longrightarrow0
\end{equation}
gives the isomorphism
\begin{equation}\label{eq:u1-integral-connecting}
\operatorname{Bock}_{U(1)\to\ringz}:
H^4(X;U(1))\xrightarrow{\ \cong\ }H^5(X;\ringz)
=\ztwo\langle A_5\rangle,
\end{equation}
since $H^4(X;\mathbb{R})=H^5(X;\mathbb{R})=0$.

The generator can be identified explicitly.  Under the coefficient
homomorphism $\ztwo\to U(1)$ sending $1$ to $1/2$, denote the image of
$v_4\in H^4(X;\ztwo)$ by $\tfrac12v_4$.  Naturality gives
\begin{equation}
\operatorname{Bock}_{U(1)\to\ringz}\!\left(\frac12v_4\right)
=\Bock(v_4).
\end{equation}
The restriction of this integral class to the fiber is
\begin{equation}
\left.\Bock(v_4)\right|_F
=\Bock(u_2^2)=2\eta_5
=\left.A_5\right|_F\neq0.
\end{equation}
Thus $\Bock(v_4)=A_5$, and
\begin{equation}\label{eq:oriented-dual-result}
\operatorname{Hom}\bigl(\widetilde\Omega_4^{\mathrm{SO}}(X),U(1)\bigr)
\cong\ztwo\langle\tfrac12v_4\rangle.
\end{equation}
For a closed oriented $4$-manifold $M$ and a map $f:M\to X$, the nonzero
character is
\begin{equation}
\chi_4([M,f])
=\frac12\left\langle f^*v_4,[M]\right\rangle
\in\mathbb{R}/\ringz,
\end{equation}
or, multiplicatively,
\begin{equation}
\exp\bigl(2\pi i\,\chi_4([M,f])\bigr)
=(-1)^{\langle f^*v_4,[M]\rangle}.
\end{equation}

\subsection{Spin bordism}
We use the AHSS to determine reduced spin bordism through degree $4$ and then
use the Adams spectral sequence (ASS) to extend the computation through
degree $6$.  The first method makes the geometric origin of the low-degree
classes transparent, while the second resolves the remaining extensions and
higher-degree groups.

\paragraph{AHSS through degree four.}
The spin bordism groups of a point needed for the AHSS are~\cite{ABP1967}:
\begin{equation}\label{eq:spin-bordism-point}
\begin{tabular}{c|ccccc}
$q$ & $0$ & $1$ & $2$ & $3$ & $4$ \\ \midrule
$\Omega_q^{\mathrm{Spin}}(\mathrm{pt})$
& $\ringz$ & $\ztwo$ & $\ztwo$ & $0$ & $\ringz$.
\end{tabular}
\end{equation}
The AHSS for reduced spin bordism is
\begin{equation}\label{eq:spin-ahss}
E^2_{p,q}
=\widetilde H_p\bigl(X;\Omega_q^{\mathrm{Spin}}(\mathrm{pt})\bigr)
\Longrightarrow \widetilde\Omega_{p+q}^{\mathrm{Spin}}(X),
\end{equation}
with differential
\begin{equation}
d^r_{p,q}:E^r_{p,q}\longrightarrow E^r_{p-r,q+r-1}.
\end{equation}
The part of the $E^2$-page needed through total degree $4$, together with
the possible incoming sources, is
\begin{sseqdata}[
  name = E2spinlow,
  classes = { draw = none },
  axes type = frame,
  scale = 0.78
]
\class["\ztwo"](1,2)
\class["\ztwo"](2,2)

\class["\ztwo"](1,1)
\class["\ztwo"](2,1)
\class["\ztwo"](3,1)
\class["\ztwo^2"](4,1)

\class["\ztwo"](1,0)
\class["\ztwo"](3,0)
\class["\ztwo"](4,0)
\class["\ztwo^2"](5,0)
\end{sseqdata}

\begin{table}[h]
\centering
\printpage[
  name = E2spinlow,
  grid = chess,
]
\caption{The part of
$E^2_{p,q}=\widetilde H_p(X;\Omega_q^{\mathrm{Spin}}(\mathrm{pt}))$
used to compute reduced spin bordism through degree $4$.}
\label{tab:E2spinlow}
\end{table}

There are two families of $d^2$-differentials between the displayed rows.
For the untwisted spin AHSS, their description follows from
Proposition~1 and the lemma immediately following it in~\cite{Teichner1993}.
The differential
\begin{equation}\label{eq:spin-d2-operations}
d^2_{p,1}:H_p(X;\ztwo)\longrightarrow H_{p-2}(X;\ztwo)
\end{equation}
is dual to
\begin{equation}
\StSq^2:H^{p-2}(X;\ztwo)\longrightarrow H^p(X;\ztwo).
\end{equation}
For the integral row,
\begin{equation}
d^2_{p,0}:H_p(X;\ringz)\longrightarrow H_{p-2}(X;\ztwo)
\end{equation}
is dual, under the $U(1)$-valued pairing, to
\begin{equation}
\frac12\StSq^2:
H^{p-2}(X;\ztwo)\longrightarrow H^p(X;U(1)).
\end{equation}
Here $\tfrac12$ denotes the coefficient homomorphism
$\ztwo\to U(1)$ sending $1$ to $1/2$.  Equivalently, on homology one first
reduces the integral class modulo $2$ and then applies the homological
operation dual to $\StSq^2$.

In the present calculation, the required Steenrod squares are
\begin{equation}\label{eq:spin-required-squares}
\StSq^2(\mathfrak{a})=0,
\qquad
\StSq^2(\mathfrak{a}^2)=\mathfrak{a}^4=0,
\qquad
\StSq^2(v_3)=v_5.
\end{equation}
The first equality follows from instability and the second from $\mathfrak{a}^3=0$.
Since $\left.\StSq^2(v_3)\right|_F=\StSq^2(u_3)=u_5$, the choice of $v_5$
in Eq.~\eqref{eq:mod2-x-generators} can be made so that the last equality
holds.

\paragraph{Degrees one and two.}
In total degree $1$, the only nonzero term is
\begin{equation}
E^2_{1,0}=H_1(X;\ringz)\cong\ztwo.
\end{equation}
No differential can affect it, so
\begin{equation}\label{eq:spin-omega1-result}
\widetilde\Omega_1^{\mathrm{Spin}}(X)\cong\ztwo.
\end{equation}

In total degree $2$, the only nonzero term is
\begin{equation}
E^2_{1,1}=H_1(X;\ztwo)\cong\ztwo.
\end{equation}
The only possible incoming differential is
\begin{equation}
d^2_{3,0}:H_3(X;\ringz)\longrightarrow H_1(X;\ztwo).
\end{equation}
Its dual is $\tfrac12\StSq^2(\mathfrak{a})=0$, and hence it vanishes.  Therefore
\begin{equation}\label{eq:spin-omega2-result}
\widetilde\Omega_2^{\mathrm{Spin}}(X)\cong\ztwo.
\end{equation}

\paragraph{Degree three.}
The total-degree-$3$ diagonal is
\begin{equation}
E^2_{3,0}\cong\ztwo,
\qquad
E^2_{2,1}\cong\ztwo,
\qquad
E^2_{1,2}\cong\ztwo.
\end{equation}
By Eq.~\eqref{eq:spin-required-squares}, the relevant $d^2$-differentials
\begin{equation}
\begin{split}
d^2_{3,0}&:E^2_{3,0}\longrightarrow E^2_{1,1},\\
d^2_{4,0}&:E^2_{4,0}\longrightarrow E^2_{2,1},\\
d^2_{3,1}&:E^2_{3,1}\longrightarrow E^2_{1,2}
\end{split}
\end{equation}
all vanish.

There is one possible higher differential entering this diagonal,
\begin{equation}\label{eq:spin-d3-degree-one}
d^3_{4,0}:E^3_{4,0}\longrightarrow E^3_{1,2}.
\end{equation}
However, Brumfiel and Morgan show that on the total-degree-$3$ diagonal the
$E^3$-page already equals the $E^\infty$-page: the only differential involving
this diagonal is the $d^2$-differential described above~\cite[Sec.~2.1]{Brumfiel:2016vpy}.
Hence Eq.~\eqref{eq:spin-d3-degree-one} vanishes.

Consequently, the associated graded group on this diagonal is
\begin{equation}\label{eq:spin-omega3-associated}
\operatorname{gr}\widetilde\Omega_3^{\mathrm{Spin}}(X)
\cong\ztwo\oplus\ztwo\oplus\ztwo.
\end{equation}
It remains to resolve the extensions. Let
$p:X\to B\ztwo$ be the projection in Eq.~\eqref{eq:fibration}.
The known result
\begin{equation}\label{eq:spin-bz2-degree-three}
\widetilde\Omega_3^{\mathrm{Spin}}(B\ztwo)\cong\ringz_8
\end{equation}
has a filtration with three successive $\ztwo$ quotients
in bidegrees $(1,2)$, $(2,1)$, and $(3,0)$
respectively~\cite{Brumfiel:2016vpy}
.  The map induced by $p$ is an
isomorphism on the first two associated quotients because
\begin{equation}
p^*(x)=\mathfrak{a},
\qquad
p^*(x^2)=\mathfrak{a}^2.
\end{equation}
It vanishes on the quotient in bidegree $(3,0)$.  Indeed, if
$c_B=\Bock(x)$ and $c=p^*(c_B)$, then
\begin{equation}
p^*(c_B^2)=c^2=0,
\end{equation}
and therefore
\begin{equation}
p_*:H_3(X;\ringz)\longrightarrow H_3(B\ztwo;\ringz)
\end{equation}
vanishes.  Here we used the natural universal-coefficient identification
\begin{equation}
H^4(-;\ringz)\cong\operatorname{Ext}\bigl(H_3(-;\ringz),\ringz\bigr)
\end{equation}
in this torsion range.

It follows that the image of
$p_*:\widetilde\Omega_3^{\mathrm{Spin}}(X)\to\ringz_8$
is the order-$4$ filtration subgroup of $\ringz_8$.  The group in
Eq.~\eqref{eq:spin-omega3-associated} cannot be $\ztwo^3$, since the image of
any homomorphism from $\ztwo^3$ has exponent at most $2$.
It also cannot be $\ringz_8$: a homomorphism $\ringz_8\to\ringz_8$ with image of
order $4$ vanishes on the unique order-$2$ subgroup, whereas $p_*$ is
an isomorphism on the filtration quotient in bidegree $(1,2)$.
Therefore
\begin{equation}\label{eq:spin-omega3-result}
\widetilde\Omega_3^{\mathrm{Spin}}(X)
\cong\ringz_4\oplus\ztwo.
\end{equation}

\paragraph{Degree four.}

The total-degree-$4$ diagonal is
\begin{equation}\label{eq:spin-degree-four-diagonal}
E^2_{4,0}\cong\ztwo,
\qquad
E^2_{3,1}\cong\ztwo,
\qquad
E^2_{2,2}\cong\ztwo.
\end{equation}
The differential
\begin{equation}\label{eq:spin-d2-surjective}
d^2_{5,0}:E^2_{5,0}\cong\ztwo^2
\longrightarrow E^2_{3,1}\cong\ztwo
\end{equation}
is surjective.  Its dual is the nonzero operation
\begin{equation}
\frac12\StSq^2(v_3)=\frac12v_5
\in H^5(X;U(1)).
\end{equation}
This class is nonzero because its coefficient Bockstein restricts to
$\Bock(u_5)=\eta_6\neq0$ on the fiber.
By contrast,
\begin{equation}
d^2_{4,0}:E^2_{4,0}\longrightarrow E^2_{2,1},
\qquad
d^2_{4,1}:E^2_{4,1}\longrightarrow E^2_{2,2}
\end{equation}
both vanish because they are dual to $\StSq^2(\mathfrak{a}^2)=0$.  Therefore
\begin{equation}\label{eq:spin-degree-four-e3}
E^3_{4,0}\cong\ztwo,
\qquad
E^3_{3,1}=0,
\qquad
E^3_{2,2}\cong\ztwo,
\qquad
E^3_{5,0}\cong\ker d^2_{5,0}\cong\ztwo.
\end{equation}

The remaining differential into the total-degree-$4$ diagonal is
\begin{equation}\label{eq:spin-secondary-d3}
d^3_{5,0}:E^3_{5,0}\longrightarrow E^3_{2,2}.
\end{equation}
Its Pontryagin dual is the secondary operation
\begin{equation}\label{eq:spin-theta}
\Theta=
\left\langle\frac12\StSq^2,\StSq^2\right\rangle:
\ker\!\left(
\StSq^2:H^2(X;\ztwo)\longrightarrow H^4(X;\ztwo)
\right)
\longrightarrow
\frac{H^5(X;U(1))}
{\frac12\StSq^2H^3(X;\ztwo)}.
\end{equation}
The explicit form of this operation and its identification with the
$d^3$-differential are given in~\cite[Sec.~4.2]{BrumfielMorgan4}.
For a degree-$2$ cocycle $z$ with $[z^2]=0$, choose a degree-$3$ cochain $P$
satisfying
\begin{equation}\label{eq:spin-theta-cochain-choice}
\delta P=z^2.
\end{equation}
Let $\widehat z$ be the integral lift of $z$ taking values $0$ and $1$, and
let $\chi(z)\in C^5(X;\ztwo)$ be the natural correction cochain satisfying
\begin{equation}
\delta\chi(z)=z^2\cup_2z^2+(z\cup_1z)^2.
\end{equation}
Then $\Theta(z)$ is represented by the cocycle
\begin{equation}\label{eq:spin-theta-cochain-formula}
k(P,z)=
\frac12\StSq^2(P)
+\frac14\widehat z\bigl(\widehat z\cup_1\widehat z\bigr)
+\frac12\chi(z).
\end{equation}
Changing $P$ by a degree-$3$ cocycle changes this class by an element of
$\frac12\StSq^2H^3(X;\ztwo)$, which gives the quotient in
Eq.~\eqref{eq:spin-theta}.

In the present case, the domain of Eq.~\eqref{eq:spin-theta} is generated
by $\mathfrak{a}^2$.  The coefficient Bockstein and the integral Serre filtration give
\begin{equation}
H^5(X;U(1))
\cong\ztwo\langle\tfrac12\mathfrak{a}v_4\rangle
\oplus\ztwo\langle\tfrac12v_5\rangle.
\end{equation}
Under the connecting isomorphism
$H^5(X;U(1))\cong H^6(X;\ringz)$, these two classes have respective
$E_\infty$-representatives $\zeta_1\eta_5$ in bidegree $(1,5)$ and
$\eta_6$ in bidegree $(0,6)$, so they are nonzero and independent.
Since $\StSq^2(v_3)=v_5$, the target of $\Theta$ is therefore
\begin{equation}\label{eq:spin-theta-target}
\frac{H^5(X;U(1))}
{\langle\tfrac12v_5\rangle}
\cong\ztwo,
\qquad
\tfrac12\mathfrak{a}v_4\longmapsto 1.
\end{equation}

We now apply Eq.~\eqref{eq:spin-theta-cochain-choice} to $z=\mathfrak{a}^2$, using the
same letters for cohomology classes and chosen cocycle representatives.
The transgression $\overline d_3(u_2)=x^3$ means that on $X$ one can choose
a degree-$2$ cochain $\mathfrak{b}$ satisfying
\begin{equation}
\delta\mathfrak{b}=\mathfrak{a}^3,
\end{equation}
whose restriction to the fiber represents $u_2$.  Set
\begin{equation}
P=\mathfrak{a}\mathfrak{b}.
\end{equation}
Then
\begin{equation}
\delta P=\mathfrak{a}^4=(\mathfrak{a}^2)^2,
\end{equation}
so $P$ is precisely the degree-$3$ cochain required by
Eq.~\eqref{eq:spin-theta-cochain-choice}.  Since $\mathfrak{a}$ comes from the base
class $x$ and $\mathfrak{b}$ restricts to $u_2$, the associated graded term of $P$
is $xu_2$.

The part of $\frac12\StSq^2(P)$ with base degree $1$ and fiber degree $4$ is
\begin{equation}
\frac12\left[\StSq^2(xu_2)\right]_{(1,4)}
=\frac12xu_2^2,
\end{equation}
because
\begin{equation}
\StSq^2(xu_2)=xu_2^2+x^2u_3.
\end{equation}
The term $xu_2^2$ has base degree $1$ and fiber degree $4$.  Since
$z=\mathfrak{a}^2$ is pulled back from the base, the last two terms in
Eq.~\eqref{eq:spin-theta-cochain-formula} are also pulled back from the base.
They have no positive-degree fiber component and therefore cannot cancel
$xu_2^2$.  Under the naming convention
\eqref{eq:mod2-x-generators}, $\frac12xu_2^2$ represents
$\frac12\mathfrak{a}v_4$.  Hence
\begin{equation}\label{eq:spin-theta-evaluation}
\Theta(\mathfrak{a}^2)
=\frac12\mathfrak{a}v_4
\pmod{\langle\tfrac12v_5\rangle}
\neq0
\end{equation}
in Eq.~\eqref{eq:spin-theta-target}.  Thus
Eq.~\eqref{eq:spin-secondary-d3} is an isomorphism and
\begin{equation}
E^\infty_{2,2}=0.
\end{equation}

The other possible $d^3$ from $E^3_{4,0}$ is
Eq.~\eqref{eq:spin-d3-degree-one}, which was shown above to vanish.
The only possible later outgoing differential from $E^4_{4,0}$ is
\begin{equation}
d^4_{4,0}:E^4_{4,0}\longrightarrow E^4_{0,3}=0.
\end{equation}
Therefore the total-degree-$4$ diagonal at the limiting page is
\begin{equation}
E^\infty_{4,0}\cong\ztwo,
\qquad
E^\infty_{3,1}=E^\infty_{2,2}=0.
\end{equation}
There is no additive extension, and hence
\begin{equation}\label{eq:spin-omega4-result}
\widetilde\Omega_4^{\mathrm{Spin}}(X)
\cong\ztwo.
\end{equation}

Combining the preceding calculations gives
\begin{equation}\label{eq:spin-bordism-summary}
\begin{array}{c|cccc}
n & 1 & 2 & 3 & 4 \\ \midrule
\widetilde\Omega_n^{\mathrm{Spin}}(X)
& \ztwo & \ztwo & \ringz_4\oplus\ztwo & \ztwo.
\end{array}
\end{equation}

\paragraph{Four-dimensional characters.}
The surviving filtration-$4$ character group is
\begin{equation}
\frac{H^4(X;U(1))}
{\frac12\StSq^2H^2(X;\ztwo)}.
\end{equation}
Since $H^2(X;\ztwo)=\ztwo\langle \mathfrak{a}^2\rangle$ and
$\StSq^2(\mathfrak{a}^2)=0$, this quotient is simply $H^4(X;U(1))$.
The calculation in Eq.~\eqref{eq:oriented-dual-result} therefore gives
\begin{equation}\label{eq:spin-dual-result}
\operatorname{Hom}\bigl(
\widetilde\Omega_4^{\mathrm{Spin}}(X),U(1)
\bigr)
\cong\ztwo\langle\tfrac12v_4\rangle.
\end{equation}
For a closed spin $4$-manifold $M$ and a map $f:M\to X$, its nonzero
character is
\begin{equation}
\chi_4^{\mathrm{Spin}}([M,f])
=\frac12\left\langle f^*v_4,[M]\right\rangle
\in\mathbb{R}/\ringz.
\end{equation}
Equivalently, the nontrivial multiplicative bordism invariant is
\begin{equation}\label{eq:spin-four-dimensional-phase}
\exp\bigl(2\pi i\,\chi_4^{\mathrm{Spin}}([M,f])\bigr)
=(-1)^{\langle f^*v_4,[M]\rangle}.
\end{equation}
This is the symmetry-dependent $4$-dimensional invertible phase relevant
to anomaly inflow for a $3$-dimensional theory.  Passing to reduced bordism
removes the symmetry-independent contribution
$\Omega_4^{\mathrm{Spin}}(\mathrm{pt})\cong\ringz$.

\paragraph{ASS through degree six.}
The ASS extends the result to degree $6$ and resolves the degree-$3$ extension
directly on its $E_\infty$-page.  Background on the ASS may be found
in~\cite{HatcherSpectralSequences,McCleary_2000,beaudry2018guide}.  For $X$,
the spectral sequence takes the form
\begin{equation}\label{eq:ass-e2}
E_2^{s,t}
\cong
\operatorname{Ext}_{\mathcal A}^{s,t}
\bigl(\widetilde H^*(M\mathrm{Spin}\wedge X;\ztwo),\ztwo\bigr)
\Longrightarrow
\bigl(\widetilde\Omega_{t-s}^{\mathrm{Spin}}(X)\bigr)^{\wedge}_2,
\end{equation}
where the superscript $\wedge_2$ denotes completion at the prime $2$. In the
degrees considered here, the reduced groups contain only torsion of order a
power of $2$, so this determines the full groups. The algebra $\mathcal A$ is
the Steenrod algebra. The Anderson--Brown--Peterson
splitting~\cite{ABP1967}, together with the K\"unneth theorem and change of
rings, reduces its $E_2$-page for $t-s<8$ to
\begin{equation}\label{eq:ass-e2-simplified}
\operatorname{Ext}_{\mathcal A_1}^{s,t}
\bigl(\widetilde H^*(X;\ztwo),\ztwo\bigr),
\end{equation}
where $\mathcal A_1\subset\mathcal A$ is generated by $\StSq^1$ and
$\StSq^2$.
Using the results in Table~\ref{tab:x-mod2-cohomology} and the fact that Steenrod squares commute with differentials in Serre spectral sequences and $E_\infty$-page extension,we decompose the $\mathcal{A}_1$-module $\widetilde H^*(X;\ztwo)$ as~\cite{singer2006steenrod}:
\begin{equation}
\widetilde{H}^{\ast}(X;\ztwo)\simeq\widetilde{H}^{\ast}(\mathbb{R}\textrm{P}^2)\oplus \Sigma^3\mathcal{P}\oplus \Sigma^4 J'\oplus\cdots,
\end{equation}\label{eq:Z2-cohomology-decomposition}
where the suspension $\Sigma^r$ for a graded module $M$ stands for
\begin{displaymath}
(\Sigma^r M)^t=M^{t-r},
\end{displaymath}
and the $\mathcal{A}_1$-submodules are:
\begin{eqnarray}
\widetilde{H}^{\ast}(\mathbb{R}\textrm{P}^2)&\simeq&\ztwo \mathfrak{a}\oplus\ztwo \mathfrak{a}^2,\\
\Sigma^3\mathcal{P}&\simeq&\ztwo v_3\oplus\ztwo v_4\oplus\ztwo v_5\oplus\ztwo v_6,\qquad \textrm{and}\\
\Sigma^4 J'&\simeq&\ztwo \mathfrak{a}v_3\oplus\ztwo \mathfrak{a}^2v_3\oplus\ztwo \mathfrak{a}v_5\oplus\ztwo \mathfrak{a}^2v_5\oplus\ztwo(\mathfrak{a}v_3^2+\mathfrak{a}^2v_5)\oplus\ztwo \mathfrak{a}^2v_3^2.
\end{eqnarray}\label{eq:Z2-cohomology-components}
For later computation, we also list the non-trivial actions of $\mathcal{A}_1$ generators:
\begin{equation}
\StSq^1\mathfrak a=\mathfrak a^2;
\end{equation}
\begin{equation}
\StSq^2v_3=v_5,\qquad\StSq^1v_5=v_6,\qquad\StSq^2v_4=v_6;
\end{equation}
\begin{equation}
\begin{aligned}
\StSq^1(\mathfrak av_3)=\mathfrak a^2v_3,\quad&\StSq^2(\mathfrak av_3)=\mathfrak av_5,\quad\StSq^1(\mathfrak av_5)=\mathfrak a^2v_5,\\\StSq^2(\mathfrak a^2v_3)=\mathfrak av_3^2+\mathfrak a^2v_5,\quad&\StSq^2(\mathfrak av_5)=\StSq^1(\mathfrak av_3^2+\mathfrak a^2v_5)=\mathfrak a^2v_3^2.
\end{aligned}
\end{equation}
We also mention that there is the joker module $J$ used
in the ensuing long exact sequence~\cite{beaudry2018guide}, which can be regarded as the cokernel of the embedding $\Sigma^3\ztwo\hookrightarrow J'$ whose image is killed by $\StSq^1$.
\begin{figure}[!h]
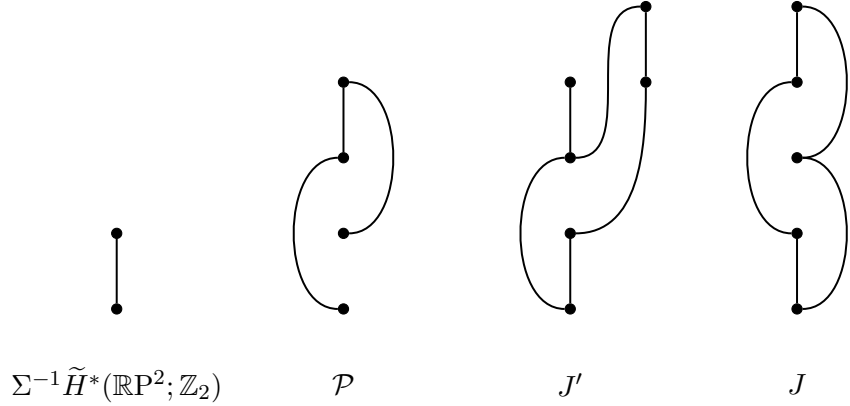
\begin{center}
\begin{sseqpage}[classes={circle,fill},no axes,x range={-1}{11}]
\class(0,0)\class(0,1)
\structline(0,0)(0,1)
\class(3,0)\class(3,1)\class(3,2)\class(3,3)
\structline[bend left = 90](3,0)(3,2)
\structline[bend right = 90](3,1)(3,3)
\structline(3,2)(3,3)
\class(6,0)\class(6,1)\class(6,2)\class(6,3)\class(7,3)\class(7,4)
\structline(6,0)(6,1)\structline(6,2)(6,3)
\structline[bend left = 90](6,0)(6,2)
\structline(7,3)(7,4)
\structline[out=0,in=south](6,1)(7,3)
\structline[out=0,in=180](6,2)(7,4)
\class(9,0)\class(9,1)\class(9,2)\class(9,3)\class(9,4)
\structline(9,0)(9,1)\structline(9,3)(9,4)
\structline[bend left = 90](9,1)(9,3)
\structline[bend right = 90](9,0)(9,2)
\structline[bend right = 90](9,2)(9,4)
\node[background] at (0,-1) {\Sigma^{-1}\widetilde{H}^{\ast}(\mathbb{R}\textrm{P}^2;\ztwo)};
\node[background] at (3,-1) {\mathcal{P}};
\node[background] at (6,-1) {J'};
\node[background] at (9,-1) {J};
\end{sseqpage}\end{center}
\caption{$\mathcal A_1$-module cell diagrams used in the ASS.  Straight and
curved lines represent the actions of $\StSq^1$ and $\StSq^2$, respectively.}
\label{fig:cell-diagrams}
\end{figure}
The cell diagrams are illustrated in Figure~\ref{fig:cell-diagrams}, where straight lines denote the action of $\StSq^1$ and curved lines denote the action $\StSq^2$.

For each summand
$M=\widetilde H^*(\mathbb{R}\mathrm P^2;\ztwo),\Sigma^3\mathcal P$, or
$\Sigma^4J'$, we compute $\operatorname{Ext}_{\mathcal A_1}^{s,t}(M,\ztwo)$
from the long exact sequences associated with
\begin{equation}\label{eq:3ses}
\begin{aligned}
&0\longrightarrow\Sigma^2\ztwo\longrightarrow\widetilde{H}^{\ast}(\mathbb{R}\textrm{P}^2;\ztwo)\longrightarrow\Sigma\ztwo\longrightarrow 0,\\
&0\longrightarrow\Sigma\mathcal{P}\longrightarrow J\longrightarrow\ztwo\longrightarrow 0,\qquad\textrm{and}\\
&0\longrightarrow\Sigma^4\widetilde{H}^{\ast}(\mathbb{R}\textrm{P}^2;\ztwo)\longrightarrow\mathcal{A}_1\longrightarrow J'\longrightarrow 0.
\end{aligned}
\end{equation}
The AHSS for $\widetilde\Omega_*^{\mathrm{Spin}}(\mathbb{R}\mathrm P^2)$
fixes the connecting morphism in the first long exact sequence: it pairs the
$h_0$-towers in stems $t-s=1$ and $2$, and likewise those in stems $5$ and
$6$.

To see the resulting $h_0$-extension explicitly, consider the multiplication
connecting
\begin{equation*}
\operatorname{Ext}_{\mathcal A_1}^{1,4}(\Sigma^2\ztwo,\ztwo)
\quad\text{and}\quad
\operatorname{Ext}_{\mathcal A_1}^{2,5}(\Sigma\ztwo,\ztwo).
\end{equation*}
This is the
extension through which the $\mathbb{R}\mathrm P^2$ summand contributes a
$\ringz_4$ in stem $3$; the additional $\ztwo$ summand in
Eq.~\eqref{eq:spin-omega3-result} comes from the remaining modules.  Write
$\mathbb{R}\mathrm P^2$ as the stunted projective space
$\mathbb{R}\mathrm P^2_1$ and work in the Spanier--Whitehead category
~\cite{barnes2020foundations}. Spanier--Whitehead duality gives
\begin{equation}\label{eq:ko-duality}
\mathrm{ko}_3(\mathbb{R}\mathrm P^2)
=\mathrm{ko}^{-3}\bigl(\mathbb D(\mathbb{R}\mathrm P^2_1)\bigr)
=\widetilde{\mathrm{ko}}^0
\bigl(\Sigma^3\mathbb D(\mathbb{R}\mathrm P^2_1)\bigr).
\end{equation}
Here $\mathbb D$ denotes Spanier--Whitehead duality
~\cite{hopkins2012spectra,adams1974stable}. Using the identities
\begin{equation*}
\mathbb D(\mathbb{R}\mathrm P^2_1)
=\Sigma\mathbb{R}\mathrm P^{-2}_{-3},
\qquad
\Sigma^2\mathbb{R}\mathrm P^l_{l-1}
=\mathbb{R}\mathrm P^{l+2}_{l+1},
\end{equation*}
the right-hand side becomes
\begin{equation*}
\widetilde{\mathrm{ko}}^0(\mathbb{R}\mathrm P^2_1)
=\widetilde{\mathrm{KO}}^0(\mathbb{R}\mathrm P^2_1)
\cong\ringz_4
\end{equation*}
by~\cite{hopkins2012spectra}. This gives the required nontrivial
$h_0$-extension.

It remains to determine whether the map
$\operatorname{Ext}_{\mathcal A_1}^{2,4}(\ztwo,\ztwo)
\to\operatorname{Ext}_{\mathcal A_1}^{2,4}(J,\ztwo)$
is an isomorphism.  If it were, one class in stem $4$ would disappear from
the total Adams chart, leaving one fewer cancellation in stem $3$.  The
resulting group in degree $3$ would then have order greater than $8$, in
contradiction with the AHSS $E_2$-page above. Therefore this map is not an
isomorphism. This selects the required connecting map in the long exact
$\operatorname{Ext}$ sequence induced by
$0\to\Sigma\mathcal P\to J\to\ztwo\to0$ in
Eq.~\eqref{eq:3ses}.
\DeclareSseqGroup\towerL {} {
\class(0,0)
\foreach \y in {1,...,4} {
\class(0,\y)
\structline
}
}
\DeclareSseqGroup\towerS {} {
\class(0,0)
\foreach \y in {1,2} {
\class(0,\y)
\structline
}
}

\begin{figure}\begin{center}
\begin{sseqpage}[Adams grading, grid = crossword, classes = fill, x range={0}{8}, y range={0}{4}]
\class(1,0)\class(2,1)\class(3,2)
\class(3,1)\class(4,2)
\class(5,3)
\structline(1,0)(2,1)\structline(2,1)(3,2)
\structline(3,1)(4,2)\structline(3,1)(3,2)
\end{sseqpage}\end{center}
\caption{Contribution of $\widetilde H^*(\mathbb{R}\mathrm P^2;\ztwo)$ to the Adams $E_2$-page.}
\label{fig:RP2-Adams}
\end{figure}
\begin{figure}[!h]
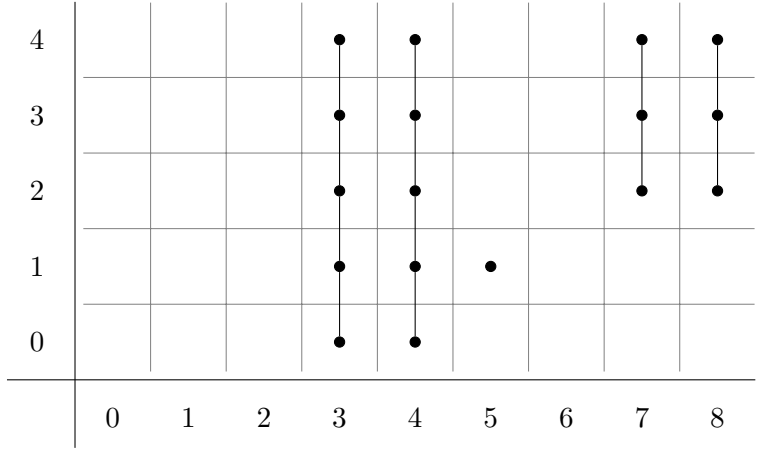
\begin{center}
\begin{sseqpage}[Adams grading, grid = crossword, classes = fill, x range={0}{8}, y range={0}{4}]
\towerL(3,0)\towerL(4,0)
\class(5,1)
\towerS(7,2)\towerS(8,2)
\end{sseqpage}\end{center}
\caption{Contribution of $\Sigma^3\mathcal P$ to the Adams $E_2$-page.}
\label{fig:P-Adams}
\end{figure}
\begin{figure}
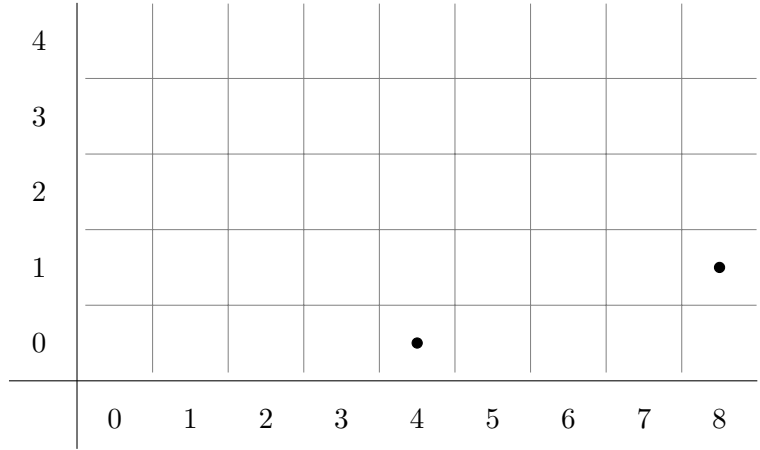
\begin{center}
\begin{sseqpage}[Adams grading, grid = crossword, classes = fill, x range={0}{8}, y range={0}{4}]
\class(4,0)\class(8,1)
\end{sseqpage}\end{center}
\caption{Contribution of $\Sigma^4J'$ to the Adams $E_2$-page.}
\label{fig:J'-Adams}
\end{figure}
The three summands contribute the following Adams charts in Figure~\ref{fig:RP2-Adams}, \ref{fig:P-Adams} and \ref{fig:J'-Adams}.

\begin{figure}[!htp]
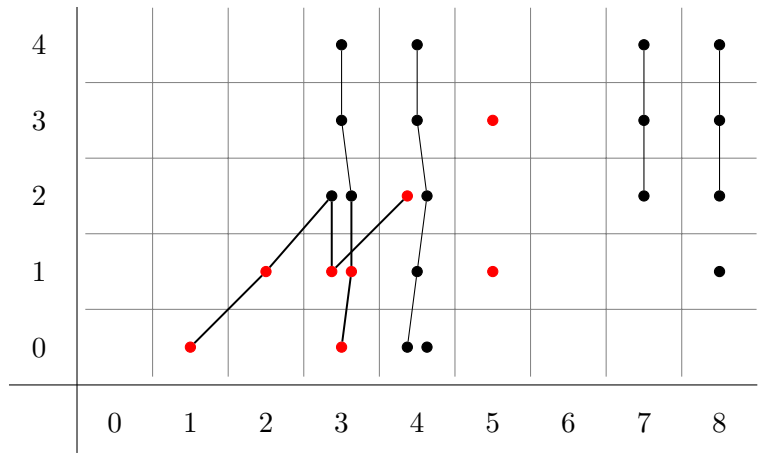
\begin{center}
\begin{sseqpage}[Adams grading, grid = crossword, classes = fill, x range={0}{8}, y range={0}{4}]
\class[red](1,0)\class[red](2,1)\class(3,2)
\class[red](3,1)\class[red](4,2)
\class[red](5,3)
\structline(1,0)(2,1)\structline(2,1)(3,2)
\structline(3,1)(4,2)\structline(3,1)(3,2)
\towerS[tag=b](3,2)\class[tag=a,red](3,1)\class[red](3,0)\structline(3,1,a)(3,2,b)\structline(3,0)(3,1,a)
\towerL(4,0)\class(4,0)
\class[red](5,1)
\towerS(7,2)\towerS(8,2)
\class(8,1)
\end{sseqpage}\end{center}
\caption{The combined Adams $E_2$-page.  Black classes are removed by the
$d_2$-differentials fixed by comparison with the low-degree AHSS.}
\label{fig:Adams-E2}
\end{figure}

\begin{figure}[!htp]
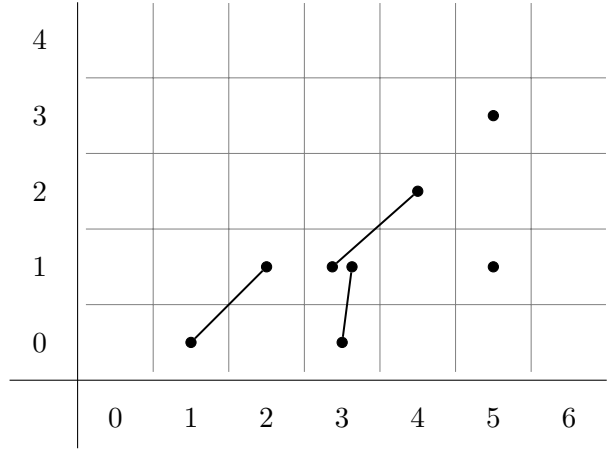
\begin{center}
\begin{sseqpage}[Adams grading, grid = crossword, classes = fill, x range={0}{6}, y range={0}{4}]
\class(1,0)\class(2,1)
\class(3,1)\class(4,2)\structline(3,1)(4,2)
\class(5,3)
\class[tag=a](3,1)\class(3,0)
\class(5,1)
\structline(1,0)(2,1)\structline(3,0)(3,1,a)
\end{sseqpage}\end{center}
\caption{The Adams $E_3=E_\infty$-page in stems $t-s\leq6$.}
\label{fig:Adams-Einf}
\end{figure}

Combining these contributions gives the Adams $E_2$-page through stem $8$ as shown in Figure~\ref{fig:Adams-E2}.
The degree-$3$ order bound obtained from the AHSS fixes the remaining
$d_2$-differentials from stem $4$ to stem $3$; the affected classes are
shown in red below.  After these differentials, no further differential can
occur in stems $t-s\leq6$, so $E_3=E_\infty$ in the range used here, which is displayed in Figure~\ref{fig:Adams-Einf}.

The vertical line from $(3,0)$ to $(3,1)$ represents multiplication by $2$;
it is the nontrivial extension producing the $\ringz_4$ summand in degree
$3$.  Reading off the limiting page gives
\begin{equation}\label{eq:spin-bordism-ass-summary}
\begin{array}{c|cccccc}
n & 1 & 2 & 3 & 4 & 5 & 6 \\ \midrule
\widetilde\Omega_n^{\mathrm{Spin}}(X)
& \ztwo & \ztwo & \ringz_4\oplus\ztwo
& \ztwo & \ztwo\oplus\ztwo & 0.
\end{array}
\end{equation}
In degrees $n\leq4$, this agrees with the AHSS result
~\eqref{eq:spin-bordism-summary}; the ASS supplies the additional groups in
degrees $5$ and $6$.

\bibliographystyle{references/JHEP}
\bibliography{references/FM}

@misc{may2009spectral,
    author = "MAY, JP and SHULMAN, MEGAN GUICHARD",
    title = "{SPECTRAL SEQUENCES FOR LOCAL COEFFICIENTS}",
    howpublished = "\url{https://math.uchicago.edu/~may/PEOPLE/MEG/localcoeffJan26.pdf}",
    year = "2009"
}

@article{Cartan1954EM,
  author  = {Cartan, Henri},
  title   = {D\'etermination des alg\`ebres $H^*(\pi,n;\mathbf Z_p)$ et $H^*(\pi,n;\mathbf Z)$, $p$ premier impair},
  journal = {S\'eminaire Henri Cartan},
  volume  = {7},
  number  = {9},
  year    = {1954--1955},
  pages   = {1--10}
}

@article{DATUASHVILI2001352,
title = {On (Co)Homology of 2-Types and Crossed Modules},
journal = {Journal of Algebra},
volume = {244},
number = {1},
pages = {352-365},
year = {2001},
issn = {0021-8693},
doi = {https://doi.org/10.1006/jabr.2001.8919},
url = {https://www.sciencedirect.com/science/article/pii/S0021869301989197},
author = {T Datuashvili and T Pirashvili}
}

@article{BROWN1994235,
title = {The Serre spectral sequence theorem for continuous and ordinary cohomology},
journal = {Topology and its Applications},
volume = {56},
number = {3},
pages = {235-248},
year = {1994},
issn = {0166-8641},
doi = {https://doi.org/10.1016/0166-8641(94)90077-9},
url = {https://www.sciencedirect.com/science/article/pii/0166864194900779},
author = {Edgar H. Brown}
}

@article{BrumfielMorgan4,
    title="{The Pontrjagin Dual of 4-Dimensional Spin Bordism}",
    author={Brumfiel, Greg and  Morgan, John},
    year={2018},
    eprint={1803.08147},
    archivePrefix={arXiv},
    primaryClass={math.GT}
}

@article{Brumfiel:2016vpy,
    author = "Brumfiel, Greg and Morgan, John",
    title = "{The Pontrjagin Dual of 3-Dimensional Spin Bordism}",
    eprint = "1612.02860",
    archivePrefix = "arXiv",
    primaryClass = "math.AT",
    month = "12",
    year = "2016"
}

@article{GuoEtAlSpinBordism,
    author = "Guo, Meng and Ohmori, Kantaro and Putrov, Pavel and Wan, Zheyan and Wang, Juven",
    title = "{Fermionic Finite-Group Gauge Theories and Interacting Symmetric/Crystalline Orders via Cobordisms}",
    eprint = "1812.11959",
    archivePrefix = "arXiv",
    primaryClass = "hep-th",
    doi = "10.1007/s00220-019-03671-6",
    journal = "Commun. Math. Phys.",
    volume = "376",
    pages = "1073--1154",
    year = "2020"
}

@article{WangGu:2017gsc,
    author = "Wang, Qing-Rui and Gu, Zheng-Cheng",
    title = "{Towards a Complete Classification of Symmetry-Protected Topological Phases for Interacting Fermions in Three Dimensions and a General Group Supercohomology Theory}",
    eprint = "1703.10937",
    archivePrefix = "arXiv",
    primaryClass = "cond-mat.str-el",
    doi = "10.1103/PhysRevX.8.011055",
    journal = "Phys. Rev. X",
    volume = "8",
    number = "1",
    pages = "011055",
    year = "2018"
}

@article{Yonekura:2018ufj,
    author = "Yonekura, Kazuya",
    title = "{On the cobordism classification of symmetry protected topological phases}",
    eprint = "1803.10796",
    archivePrefix = "arXiv",
    primaryClass = "hep-th",
    reportNumber = "IPMU-18-0040",
    doi = "10.1007/s00220-019-03439-y",
    journal = "Commun. Math. Phys.",
    volume = "368",
    number = "3",
    pages = "1121--1173",
    year = "2019"
}

@article{Kapustin:2014tfa,
    author = "Kapustin, Anton",
    title = "{Symmetry Protected Topological Phases, Anomalies, and Cobordisms: Beyond Group Cohomology}",
    eprint = "1403.1467",
    archivePrefix = "arXiv",
    primaryClass = "cond-mat.str-el",
    month = "3",
    year = "2014"
}

@article{Wang:2018pdc,
    author = "Wang, Qing-Rui and Gu, Zheng-Cheng",
    title = "{Construction and classification of symmetry protected topological phases in interacting fermion systems}",
    eprint = "1811.00536",
    archivePrefix = "arXiv",
    primaryClass = "cond-mat.str-el",
    doi = "10.1103/PhysRevX.10.031055",
    journal = "Phys. Rev. X",
    volume = "10",
    number = "3",
    pages = "031055",
    year = "2020"
}

@article{Kapustin:2014dxa,
    author = "Kapustin, Anton and Thorngren, Ryan and Turzillo, Alex and Wang, Zitao",
    title = "{Fermionic Symmetry Protected Topological Phases and Cobordisms}",
    eprint = "1406.7329",
    archivePrefix = "arXiv",
    primaryClass = "cond-mat.str-el",
    doi = "10.1007/JHEP12(2015)052",
    journal = "JHEP",
    volume = "12",
    pages = "052",
    year = "2015"
}

@article{Kong:2019brm,
    author = "Kong, Liang and Tian, Yin and Zhou, Shan",
    title = "{The center of monoidal 2-categories in 3+1D Dijkgraaf-Witten theory}",
    eprint = "1905.04644",
    archivePrefix = "arXiv",
    primaryClass = "math.QA",
    doi = "10.1016/j.aim.2019.106928",
    journal = "Adv. Math.",
    volume = "360",
    pages = "106928",
    year = "2020"
}

@article{Douglas:2018qfz,
    author = "Douglas, Christopher L. and Reutter, David J.",
    title = "{Fusion 2-categories and a state-sum invariant for 4-manifolds}",
    eprint = "1812.11933",
    archivePrefix = "arXiv",
    primaryClass = "math.QA",
    note = "accepted for publication in Mem. Amer. Math. Soc.",
    month = "12",
    year = "2018"
}

@article{Wen:2025thg,
    author = "Wen, Rui",
    title = "{Topological Holography for 2+1-D Gapped and Gapless Phases with Generalized Symmetries}",
    eprint = "2503.13685",
    archivePrefix = "arXiv",
    primaryClass = "hep-th",
    month = "3",
    year = "2025"
}

@article{Decoppet:2023uoy,
    author = "D{\'e}coppet, Thibault D.",
    title = "{The Morita theory of fusion 2-categories.}",
    eprint = "2311.16827",
    archivePrefix = "arXiv",
    primaryClass = "math.CT",
    doi = "10.4171/QT/224",
    journal = "High.  Struct.",
    volume = "7",
    number = "1",
    pages = "234--292",
    year = "2023"
}

@article{Decoppet:2023rlx,
    author = "D{\'e}coppet, Thibault D. and Xu, Hao",
    title = "{Local modules in braided monoidal 2-categories}",
    eprint = "2307.02843",
    archivePrefix = "arXiv",
    primaryClass = "math.CT",
    doi = "10.1063/5.0172042",
    journal = "J. Math. Phys.",
    volume = "65",
    number = "6",
    pages = "061702",
    year = "2024"
}

@article{Johnson-Freyd:2020ivj,
    author = "Johnson-Freyd, Theo and Yu, Matthew",
    title = "{Fusion 2-categories With no Line Operators are Grouplike}",
    eprint = "2010.07950",
    archivePrefix = "arXiv",
    primaryClass = "math.QA",
    doi = "10.1017/S0004972721000095",
    journal = "Bull. Austral. Math. Soc.",
    volume = "104",
    number = "3",
    pages = "434--442",
    year = "2021"
}

@phdthesis{Xu:2026EtaleAlgebras,

    author = "Xu, Hao",

    title = "{On {\'E}tale Algebras and Fusion 2-Categories}",

    school = "University of G{\"o}ttingen",

    year = "2026",

    doi = "10.53846/goediss-12041",

    url = "https://ediss.uni-goettingen.de/handle/11858/16728"

}

@article{Xu:2024pwd,
    author = "Xu, Hao",
    title = "{On {\'E}tale Algebras and Bosonic Fusion 2-Categories}",
    eprint = "2411.13367",
    archivePrefix = "arXiv",
    primaryClass = "math.CT",
    month = "11",
    year = "2024"
}

@article{Decoppet:2024moc,
    author = "D{\'e}coppet, Thibault Didier",
    title = "{Extension Theory and Fermionic Strongly Fusion 2-Categories (with an Appendix by Thibault Didier D{\'e}coppet and Theo Johnson-Freyd)}",
    eprint = "2403.03211",
    archivePrefix = "arXiv",
    primaryClass = "math.CT",
    doi = "10.3842/SIGMA.2024.092",
    journal = "SIGMA",
    volume = "20",
    pages = "092",
    year = "2024"
}

@article{Barkeshli:2022edm,
    author = "Barkeshli, Maissam and Chen, Yu-An and Hsin, Po-Shen and Kobayashi, Ryohei",
    title = "{Higher-group symmetry in finite gauge theory and stabilizer codes}",
    eprint = "2211.11764",
    archivePrefix = "arXiv",
    primaryClass = "cond-mat.str-el",
    doi = "10.21468/SciPostPhys.16.4.089",
    journal = "SciPost Phys.",
    volume = "16",
    number = "4",
    pages = "089",
    year = "2024"
}

@article{Xi:2023djc,
    author = "Xi, Wenjie and Lan, Tian and Wang, Longye and Wang, Chenjie and Chen, Wei-Qiang",
    title = "{On a class of fusion 2-category symmetry: condensation completion of braided fusion category}",
    eprint = "2312.15947",
    archivePrefix = "arXiv",
    primaryClass = "hep-th",
    doi = "10.4310/atmp.250524031646",
    journal = "Adv. Theor. Math. Phys.",
    volume = "29",
    number = "1",
    pages = "151--204",
    year = "2025"
}

@article{Decoppet:2023bay,
    author = "D{\'e}coppet, Thibault D. and Yu, Matthew",
    title = "{Fiber 2-Functors and Tambara{\textendash}Yamagami Fusion 2-Categories}",
    eprint = "2306.08117",
    archivePrefix = "arXiv",
    primaryClass = "math.CT",
    doi = "10.1007/s00220-025-05249-x",
    journal = "Commun. Math. Phys.",
    volume = "406",
    number = "3",
    pages = "64",
    year = "2025"
}

@article{Tachikawa:2017gyf,
    author = "Tachikawa, Yuji",
    title = "{On gauging finite subgroups}",
    eprint = "1712.09542",
    archivePrefix = "arXiv",
    primaryClass = "hep-th",
    reportNumber = "IPMU-17-0183",
    doi = "10.21468/SciPostPhys.8.1.015",
    journal = "SciPost Phys.",
    volume = "8",
    number = "1",
    pages = "015",
    year = "2020"
}

@article{Kong:2017etd,
    author = "Kong, Liang and Zheng, Hao",
    title = "{Gapless edges of 2d topological orders and enriched monoidal categories}",
    eprint = "1705.01087",
    archivePrefix = "arXiv",
    primaryClass = "cond-mat.str-el",
    doi = "10.1016/j.nuclphysb.2017.12.007",
    journal = "Nucl. Phys. B",
    volume = "927",
    pages = "140--165",
    year = "2018"
}

@article{Kong:2019byq,
    author = "Kong, Liang and Zheng, Hao",
    title = "{A mathematical theory of gapless edges of 2d topological orders. Part I}",
    eprint = "1905.04924",
    archivePrefix = "arXiv",
    primaryClass = "cond-mat.str-el",
    doi = "10.1007/JHEP02(2020)150",
    journal = "JHEP",
    volume = "02",
    pages = "150",
    year = "2020"
}

@article{Kong:2019cuu,
    author = "Kong, Liang and Zheng, Hao",
    title = "{A mathematical theory of gapless edges of 2d topological orders. Part II}",
    eprint = "1912.01760",
    archivePrefix = "arXiv",
    primaryClass = "cond-mat.str-el",
    doi = "10.1016/j.nuclphysb.2021.115384",
    journal = "Nucl. Phys. B",
    volume = "966",
    pages = "115384",
    year = "2021"
}

@article{Kong:2020iek,
    author = "Kong, Liang and Zheng, Hao",
    title = "{Categories of quantum liquids I}",
    eprint = "2011.02859",
    archivePrefix = "arXiv",
    primaryClass = "hep-th",
    doi = "10.1007/JHEP08(2022)070",
    journal = "JHEP",
    volume = "08",
    pages = "070",
    year = "2022"
}

@article{Johnson-Freyd:2020usu,
    author = "Johnson-Freyd, Theo",
    title = "{On the Classification of Topological Orders}",
    eprint = "2003.06663",
    archivePrefix = "arXiv",
    primaryClass = "math.CT",
    doi = "10.1007/s00220-022-04380-3",
    journal = "Commun. Math. Phys.",
    volume = "393",
    number = "2",
    pages = "989--1033",
    year = "2022"
}

@article{ChenKapustinTurzilloYou2019,
  author        = {Chen, Yu-An and Kapustin, Anton and Turzillo, Alex and You, Minyoung},
  title         = {Free and Interacting Short-Range Entangled Phases of Fermions: Beyond the Ten-Fold Way},
  journal       = {Physical Review B},
  volume        = {100},
  pages         = {195128},
  year          = {2019},
  doi           = {10.1103/PhysRevB.100.195128},
  eprint        = {1809.04958},
  archivePrefix = {arXiv},
  primaryClass  = {cond-mat.str-el}
}

@article{Xu:2026centers,
    author        = {Xu, Hao},
    title         = {{Centers of Algebras in Monoidal 2-Categories}},
    eprint        = {2607.05228},
    archivePrefix = {arXiv},
    primaryClass  = {math.QA},
    month         = {7},
    year          = {2026}
}

@article{ChenKapustinRadicevic2018,
  author        = {Chen, Yu-An and Kapustin, Anton and Radicevic, Djordje},
  title         = {Exact Bosonization in Two Spatial Dimensions and a New Class of Lattice Gauge Theories},
  journal       = {Annals of Physics},
  volume        = {393},
  pages         = {234--253},
  year          = {2018},
  doi           = {10.1016/j.aop.2018.03.024},
  eprint        = {1711.00515},
  archivePrefix = {arXiv},
  primaryClass  = {cond-mat.str-el}
}

@article{Debray:2025kfg,
    author = "Debray, Arun and Ye, Weicheng and Yu, Matthew",
    title = "{How to Build Anomalous (3+1)d Topological Quantum Field Theories}",
    eprint = "2510.24834",
    archivePrefix = "arXiv",
    primaryClass = "math-ph",
    month = "10",
    year = "2025"
}

@article{Davighi:2023luh,
    author = "Davighi, Joe and Lohitsiri, Nakarin and Debray, Arun",
    title = "{Toric 2-group anomalies via cobordism}",
    eprint = "2302.12853",
    archivePrefix = "arXiv",
    primaryClass = "hep-th",
    doi = "10.1007/JHEP07(2023)019",
    journal = "JHEP",
    volume = "07",
    pages = "019",
    year = "2023"
}

@article{Gaiotto:2019xmp,
    author = "Gaiotto, Davide and Johnson-Freyd, Theo",
    title = "{Condensations in higher categories}",
    eprint = "1905.09566",
    archivePrefix = "arXiv",
    primaryClass = "math.CT",
    month = "5",
    year = "2019"
}

@article{Liu:2024znj,
    author = "Liu, Ruizhi and Luo, Ran and Wang, Yi-Nan",
    title = "{Higher-Matter and Landau-Ginzburg Theory of Higher-Group Symmetries}",
    eprint = "2406.03974",
    archivePrefix = "arXiv",
    primaryClass = "hep-th",
    doi = "10.21468/SciPostPhys.18.2.052",
    journal = "SciPost Phys.",
    volume = "18",
    pages = "052",
    year = "2025"
}

@article{Bartsch:2023pzl,
    author = "Bartsch, Thomas and Bullimore, Mathew and Grigoletto, Andrea",
    title = "{Higher representations for extended operators}",
    eprint = "2304.03789",
    archivePrefix = "arXiv",
    primaryClass = "hep-th",
    month = "4",
    year = "2023"
}

@article{Kapustin:2013uxa,
    author = "Kapustin, Anton and Thorngren, Ryan",
    title = "{Higher Symmetry and Gapped Phases of Gauge Theories}",
    eprint = "1309.4721",
    archivePrefix = "arXiv",
    primaryClass = "hep-th",
    doi = "10.1007/978-3-319-59939-7_5",
    journal = "Prog. Math.",
    volume = "324",
    pages = "177--202",
    year = "2017"
}

@article{Zhang:2023ynd,
    author = "Zhang, Zhi-Feng and Wang, Qing-Rui and Ye, Peng",
    title = "{Continuum field theory of three-dimensional topological orders with emergent fermions and braiding statistics}",
    eprint = "2307.09983",
    archivePrefix = "arXiv",
    primaryClass = "cond-mat.str-el",
    doi = "10.1103/PhysRevResearch.5.043111",
    journal = "Phys. Rev. Res.",
    volume = "5",
    number = "4",
    pages = "043111",
    year = "2023"
}

@article{Zhang:2022rbg,
    author = "Zhang, Zhi-Feng and Wang, Qing-Rui and Ye, Peng",
    title = "{Non-Abelian fusion, shrinking, and quantum dimensions of Abelian gauge fluxes}",
    eprint = "2208.09228",
    archivePrefix = "arXiv",
    primaryClass = "cond-mat.str-el",
    doi = "10.1103/PhysRevB.107.165117",
    journal = "Phys. Rev. B",
    volume = "107",
    number = "16",
    pages = "165117",
    year = "2023"
}

@article{Decoppet:2024htz,
    author = "D{\'e}coppet, Thibault D. and Huston, Peter and Johnson-Freyd, Theo and Nikshych, Dmitri and Penneys, David and Plavnik, Julia and Reutter, David and Yu, Matthew",
    title = "{The Classification of Fusion 2-Categories}",
    eprint = "2411.05907",
    archivePrefix = "arXiv",
    primaryClass = "math.CT",
    month = "11",
    year = "2024"
}

@article{Putrov:2016qdo,
    author = "Putrov, Pavel and Wang, Juven and Yau, Shing-Tung",
    title = "{Braiding Statistics and Link Invariants of Bosonic/Fermionic Topological Quantum Matter in 2+1 and 3+1 dimensions}",
    eprint = "1612.09298",
    archivePrefix = "arXiv",
    primaryClass = "cond-mat.str-el",
    doi = "10.1016/j.aop.2017.06.019",
    journal = "Annals Phys.",
    volume = "384",
    pages = "254--287",
    year = "2017"
}

@article{Wang:2017loc,
    author = "Wang, Juven and Wen, Xiao-Gang and Witten, Edward",
    title = "{Symmetric Gapped Interfaces of SPT and SET States: Systematic Constructions}",
    eprint = "1705.06728",
    archivePrefix = "arXiv",
    primaryClass = "cond-mat.str-el",
    doi = "10.1103/PhysRevX.8.031048",
    journal = "Phys. Rev. X",
    volume = "8",
    number = "3",
    pages = "031048",
    year = "2018"
}

@phdthesis{clementintegral,
  title="Integral cohomology of finite Postnikov towers",
  year="2002",
  author="Cl{\'e}ment, A",
  school="Ph. D. Thesis, University of Lausanne, Switzerland",
  url="https://github.com/aclemen1/integral-cohomology-of-finite-postnikov-towers/blob/master/main.pdf"
}

@book{singer2006steenrod,
  title={Steenrod squares in spectral sequences},
  author={Singer, William M},
  series={Mathematical Surveys and Monographs},
  volume={129},
  year={2006},
  publisher={American Mathematical Society},
  address={Providence, RI}
}

@article{beaudry2018guide,
  title={A guide for computing stable homotopy groups},
  author={Beaudry, Agn{\`e}s and Campbell, Jonathan A},
  journal={arXiv preprint arXiv:1801.07530},
  year={2018}
}

@book{barnes2020foundations,
  title={Foundations of stable homotopy theory},
  author={Barnes, David and Roitzheim, Constanze},
  volume={185},
  year={2020},
  publisher={Cambridge University Press}
}

@book{adams1974stable,
  title={Stable homotopy and generalised homology},
  author={Adams, John Frank},
  year={1974},
  publisher={University of Chicago press}
}

@misc{hopkins2012spectra,
    author = "Hopkins, Michael",
    title = "{Spectra and stable homotopy theory}",
    howpublished = "\url{https://math.uchicago.edu/~amathew/256y.pdf}",
    year = "2012"
}

@article{ABP1967,
 ISSN = {0003486X, 19398980},
 URL = {http://www.jstor.org/stable/1970690},
 author = {D. W. Anderson and E. H. Brown and F. P. Peterson},
 journal = {Annals of Mathematics},
 number = {2},
 pages = {271--298},
 publisher = {[Annals of Mathematics, Trustees of Princeton University on Behalf of the Annals of Mathematics, Mathematics Department, Princeton University]},
 title = {The Structure of the Spin Cobordism Ring},
 urldate = {2026-06-15},
 volume = {86},
 year = {1967}
}

@book{McCleary_2000, place={Cambridge}, edition={2}, series={Cambridge Studies in Advanced Mathematics}, title={A User’s Guide to Spectral Sequences}, publisher={Cambridge University Press}, author={McCleary, John}, year={2000}, collection={Cambridge Studies in Advanced Mathematics}}

@article{Bason:2026njw,
    author = "Bason, Davide and Cui, Wei and Ruggeri, Lorenzo",
    title = "{Half-Spacetime Gauging of 2-Group Symmetry in 3d}",
    eprint = "2605.06287",
    archivePrefix = "arXiv",
    primaryClass = "hep-th",
    month = "5",
    year = "2026"
}

@article{Gaiotto:2014kfa,
    author = "Gaiotto, Davide and Kapustin, Anton and Seiberg, Nathan and Willett, Brian",
    title = "{Generalized Global Symmetries}",
    eprint = "1412.5148",
    archivePrefix = "arXiv",
    primaryClass = "hep-th",
    doi = "10.1007/JHEP02(2015)172",
    journal = "JHEP",
    volume = "02",
    pages = "172",
    year = "2015"
}

@article{Baez0801,
      title={The Classifying Space of a Topological 2-Group}, 
      author={John C. Baez and Danny Stevenson},
      year={2009},
      eprint={0801.3843},
      archivePrefix={arXiv},
      primaryClass={math.AT}
}

@article{Lee:2021crt,
    author = "Lee, Yasunori and Ohmori, Kantaro and Tachikawa, Yuji",
    title = "{Matching higher symmetries across Intriligator-Seiberg duality}",
    eprint = "2108.05369",
    archivePrefix = "arXiv",
    primaryClass = "hep-th",
    doi = "10.1007/JHEP10(2021)114",
    journal = "JHEP",
    volume = "10",
    pages = "114",
    year = "2021"
}

@article{Gukov:2020btk,
    author = "Gukov, Sergei and Hsin, Po-Shen and Pei, Du",
    title = "{Generalized global symmetries of $T[M]$ theories. Part I}",
    eprint = "2010.15890",
    archivePrefix = "arXiv",
    primaryClass = "hep-th",
    reportNumber = "CALT-TH-2020-045",
    doi = "10.1007/JHEP04(2021)232",
    journal = "JHEP",
    volume = "04",
    pages = "232",
    year = "2021"
}

@article{DelZotto:2020sop,
    author = "Del Zotto, Michele and Ohmori, Kantaro",
    title = "{2-Group Symmetries of 6D Little String Theories and T-Duality}",
    eprint = "2009.03489",
    archivePrefix = "arXiv",
    primaryClass = "hep-th",
    doi = "10.1007/s00023-021-01018-3",
    journal = "Annales Henri Poincare",
    volume = "22",
    number = "7",
    pages = "2451--2474",
    year = "2021"
}

@article{Bhardwaj:2021wif,
    author = "Bhardwaj, Lakshya",
    title = "{2-Group symmetries in class S}",
    eprint = "2107.06816",
    archivePrefix = "arXiv",
    primaryClass = "hep-th",
    doi = "10.21468/SciPostPhys.12.5.152",
    journal = "SciPost Phys.",
    volume = "12",
    number = "5",
    pages = "152",
    year = "2022"
}

@article{Apruzzi:2021nmk,
    author = "Apruzzi, Fabio and Bonetti, Federico and Etxebarria, I\~naki Garc\'\i{}a and Hosseini, Saghar S. and Schafer-Nameki, Sakura",
    title = "{Symmetry TFTs from String Theory}",
    eprint = "2112.02092",
    archivePrefix = "arXiv",
    primaryClass = "hep-th",
    doi = "10.1007/s00220-023-04737-2",
    journal = "Commun. Math. Phys.",
    volume = "402",
    number = "1",
    pages = "895--949",
    month = "12",
    year = "2023"
}

@article{Apruzzi:2021mlh,
    author = "Apruzzi, Fabio and Bhardwaj, Lakshya and Gould, Dewi S. W. and Schafer-Nameki, Sakura",
    title = "{2-Group symmetries and their classification in 6d}",
    eprint = "2110.14647",
    archivePrefix = "arXiv",
    primaryClass = "hep-th",
    doi = "10.21468/SciPostPhys.12.3.098",
    journal = "SciPost Phys.",
    volume = "12",
    number = "3",
    pages = "098",
    year = "2022"
}

@article{Hidaka:2020iaz,
    author = "Hidaka, Yoshimasa and Nitta, Muneto and Yokokura, Ryo",
    title = "{Higher-form symmetries and 3-group in axion electrodynamics}",
    eprint = "2006.12532",
    archivePrefix = "arXiv",
    primaryClass = "hep-th",
    reportNumber = "KEK-TH-2232, J-PARC-TH-0222, RIKEN-iTHEMS-Report-20",
    doi = "10.1016/j.physletb.2020.135672",
    journal = "Phys. Lett. B",
    volume = "808",
    pages = "135672",
    year = "2020"
}

@Article{Hidaka:2020izy,
  author        = {Hidaka, Yoshimasa and Nitta, Muneto and Yokokura, Ryo},
  journal       = {JHEP},
  title         = {{Global 3-group symmetry and 't Hooft anomalies in axion electrodynamics}},
  year          = {2021},
  pages         = {173},
  volume        = {01},
  archiveprefix = {arXiv},
  doi           = {10.1007/JHEP01(2021)173},
  eprint        = {2009.14368},
  primaryclass  = {hep-th},
  reportnumber  = {KEK-TH-2254, J-PARC-TH-0225, RIKEN-iTHEMS-Report-20},
}

@article{Benini:2018reh,
    author = "Benini, Francesco and C\'ordova, Clay and Hsin, Po-Shen",
    title = "{On 2-Group Global Symmetries and their Anomalies}",
    eprint = "1803.09336",
    archivePrefix = "arXiv",
    primaryClass = "hep-th",
    reportNumber = "SISSA 10/2018/FISI, SISSA-10-2018-FISI",
    doi = "10.1007/JHEP03(2019)118",
    journal = "JHEP",
    volume = "03",
    pages = "118",
    year = "2019"
}

@article{Cordova:2018cvg,
    author = "C\'ordova, Clay and Dumitrescu, Thomas T. and Intriligator, Kenneth",
    title = "{Exploring 2-Group Global Symmetries}",
    eprint = "1802.04790",
    archivePrefix = "arXiv",
    primaryClass = "hep-th",
    doi = "10.1007/JHEP02(2019)184",
    journal = "JHEP",
    volume = "02",
    pages = "184",
    year = "2019"
}

@book{brown2012cohomology,
  title={Cohomology of groups},
  author={Brown, Kenneth S},
  volume={87},
  year={2012},
  publisher={Springer Science \& Business Media},
isbn = {978-0-387-90688-1},
doi = {10.1007/978-1-4684-9327-6}
}

@article{Apruzzi:2022dlm,
    author = "Apruzzi, Fabio",
    title = "{Higher Form Symmetries TFT in 6d}",
    eprint = "2203.10063",
    archivePrefix = "arXiv",
    primaryClass = "hep-th",
    doi = "10.1007/JHEP11(2022)050",
    journal = "JHEP",
    volume = "11",
    pages = "050",
    month = "3",
    year = "2022"
}

@article{Bhardwaj:2022lsg,
   title={Universal Non-Invertible Symmetries},
   volume={70},
   ISSN={1521-3978},
   url={http://dx.doi.org/10.1002/prop.202200143},
   DOI={10.1002/prop.202200143},
   number={11},
   pages={2200143},
   journal={Fortschritte der Physik},
   publisher={Wiley},
   author={Bhardwaj, Lakshya and Schäfer-Nameki, Sakura and Wu, Jingxiang},
   year={2022},
   month=oct }

@article{Kapustin:2013qsa,
    author = "Kapustin, Anton and Thorngren, Ryan",
    title = "{Topological Field Theory on a Lattice, Discrete Theta-Angles and Confinement}",
    eprint = "1308.2926",
    archivePrefix = "arXiv",
    primaryClass = "hep-th",
    doi = "10.4310/ATMP.2014.v18.n5.a4",
    journal = "Adv. Theor. Math. Phys.",
    volume = "18",
    number = "5",
    pages = "1233--1247",
    year = "2014"
}

@article{Gukov:2013zka,
    author = "Gukov, Sergei and Kapustin, Anton",
    title = "{Topological Quantum Field Theory, Nonlocal Operators, and Gapped Phases of Gauge Theories}",
    eprint = "1307.4793",
    archivePrefix = "arXiv",
    primaryClass = "hep-th",
    month = "7",
    year = "2013"
}

@article{Yetter:1993dh,
    author = "Yetter, D. N.",
    title = "{TQFT's from homotopy 2 types}",
    doi = "10.1142/S0218216593000076",
    journal = "J. Knot Theor. Ramifications",
    volume = "2",
    pages = "113--123",
    year = "1993"
}

@article{elgueta2007representation,
title = {Representation theory of 2-groups on Kapranov and Voevodsky's 2-vector spaces},
journal = {Advances in Mathematics},
volume = {213},
number = {1},
pages = {53-92},
year = {2007},
issn = {0001-8708},
doi = {https://doi.org/10.1016/j.aim.2006.11.010},
url = {https://www.sciencedirect.com/science/article/pii/S0001870806003835},
author = {Josep Elgueta}
}

@article{Baez:2010ya,
    author = "Baez, John C. and Huerta, John",
    title = "{An Invitation to Higher Gauge Theory}",
    eprint = "1003.4485",
    archivePrefix = "arXiv",
    primaryClass = "hep-th",
    doi = "10.1007/s10714-010-1070-9",
    journal = "Gen. Rel. Grav.",
    volume = "43",
    pages = "2335--2392",
    year = "2011"
}

@article{Schafer-Nameki:2023jdn,
    author = "Schafer-Nameki, Sakura",
    title = "{ICTP lectures on (non-)invertible generalized symmetries}",
    eprint = "2305.18296",
    archivePrefix = "arXiv",
    primaryClass = "hep-th",
    doi = "10.1016/j.physrep.2024.01.007",
    journal = "Phys. Rept.",
    volume = "1063",
    pages = "1--55",
    year = "2024"
}

@article{Brennan:2023mmt,
    author = "Brennan, T. Daniel and Hong, Sungwoo",
    title = "{Introduction to Generalized Global Symmetries in QFT and Particle Physics}",
    eprint = "2306.00912",
    archivePrefix = "arXiv",
    primaryClass = "hep-ph",
    month = "6",
    year = "2023"
}

@article{Bhardwaj:2023kri,
    author = "Bhardwaj, Lakshya and Bottini, Lea E. and Fraser-Taliente, Ludovic and Gladden, Liam and Gould, Dewi S. W. and Platschorre, Arthur and Tillim, Hannah",
    title = "{Lectures on generalized symmetries}",
    eprint = "2307.07547",
    archivePrefix = "arXiv",
    primaryClass = "hep-th",
    doi = "10.1016/j.physrep.2023.11.002",
    journal = "Phys. Rept.",
    volume = "1051",
    pages = "1--87",
    year = "2024"
}

@article{Luo:2023ive,
    author = "Luo, Ran and Wang, Qing-Rui and Wang, Yi-Nan",
    title = "{Lecture notes on generalized symmetries and applications}",
    eprint = "2307.09215",
    archivePrefix = "arXiv",
    primaryClass = "hep-th",
    doi = "10.1016/j.physrep.2024.02.002",
    journal = "Phys. Rept.",
    volume = "1065",
    pages = "1--43",
    year = "2024"
}

@article{Kong:2020cie,
    author = "Kong, Liang and Lan, Tian and Wen, Xiao-Gang and Zhang, Zhi-Hao and Zheng, Hao",
    title = "{Algebraic higher symmetry and categorical symmetry -- a holographic and entanglement view of symmetry}",
    eprint = "2005.14178",
    archivePrefix = "arXiv",
    primaryClass = "cond-mat.str-el",
    doi = "10.1103/PhysRevResearch.2.043086",
    journal = "Phys. Rev. Res.",
    volume = "2",
    number = "4",
    pages = "043086",
    year = "2020"
}

@article{Brennan:2020ehu,
    author = "Brennan, T. Daniel and Cordova, Clay",
    title = "{Axions, higher-groups, and emergent symmetry}",
    eprint = "2011.09600",
    archivePrefix = "arXiv",
    primaryClass = "hep-th",
    doi = "10.1007/JHEP02(2022)145",
    journal = "JHEP",
    volume = "02",
    pages = "145",
    year = "2022"
}

@article{Brauner:2020rtz,
    author = "Brauner, Tom\'a\v{s}",
    title = "{Field theories with higher-group symmetry from composite currents}",
    eprint = "2012.00051",
    archivePrefix = "arXiv",
    primaryClass = "hep-th",
    doi = "10.1007/JHEP04(2021)045",
    journal = "JHEP",
    volume = "04",
    pages = "045",
    year = "2021"
}

@article{Hsin:2021qiy,
    author = "Hsin, Po-Shen and Ji, Wenjie and Jian, Chao-Ming",
    title = "{Exotic invertible phases with higher-group symmetries}",
    eprint = "2105.09454",
    archivePrefix = "arXiv",
    primaryClass = "cond-mat.str-el",
    reportNumber = "CALT-TH-2021-021",
    doi = "10.21468/SciPostPhys.12.2.052",
    journal = "SciPost Phys.",
    volume = "12",
    number = "2",
    pages = "052",
    year = "2022"
}

@article{Freed:2014iua,
    author = "Freed, Daniel S.",
    editor = "Donagi, Ron and Douglas, Michael R. and Kamenova, Ljudmila and Rocek, Martin",
    title = "{Anomalies and invertible field theories}",
    eprint = "1404.7224",
    archivePrefix = "arXiv",
    primaryClass = "hep-th",
    doi = "10.1090/pspum/088/01462",
    journal = "Proc. Symp. Pure Math.",
    volume = "88",
    pages = "25--46",
    year = "2014"
}

@article{Freed:2016rqq,
    author = "Freed, Daniel S. and Hopkins, Michael J.",
    title = "{Reflection positivity and invertible topological phases}",
    eprint = "1604.06527",
    archivePrefix = "arXiv",
    primaryClass = "hep-th",
    doi = "10.2140/gt.2021.25.1165",
    journal = "Geom. Topol.",
    volume = "25",
    pages = "1165--1330",
    year = "2021"
}

@article{Cordova:2022qtz,
    author = "Cordova, Clay and Koren, Seth",
    title = "{Higher Flavor Symmetries in the Standard Model}",
    eprint = "2212.13193",
    archivePrefix = "arXiv",
    primaryClass = "hep-ph",
    doi = "10.1002/andp.202300031",
    journal = "Annalen Phys.",
    volume = "535",
    number = "8",
    pages = "2300031",
    year = "2023"
}

@article{Barkeshli:2023bta,
   title={Higher-group symmetry of (3+1)D fermionic $\mathbb{Z}_2$ gauge theory: Logical CCZ, CS, and T gates from higher symmetry},
   volume={16},
   ISSN={2542-4653},
   url={http://dx.doi.org/10.21468/SciPostPhys.16.5.122},
   DOI={10.21468/SciPostPhys.16.5.122},
   number={5},
   pages={122},
   journal={SciPost Physics},
   publisher={Stichting SciPost},
   author={Barkeshli, Maissam and Hsin, Po-Shen and Kobayashi, Ryohei},
   year={2024},
   month=may }

@article{Sharpe:2015mja,
    author = "Sharpe, Eric",
    title = "{Notes on generalized global symmetries in QFT}",
    eprint = "1508.04770",
    archivePrefix = "arXiv",
    primaryClass = "hep-th",
    doi = "10.1002/prop.201500048",
    journal = "Fortsch. Phys.",
    volume = "63",
    pages = "659--682",
    year = "2015"
}

@article{Bhardwaj:2023wzd,
   title={Generalized charges, part I: Invertible symmetries and higher representations},
   volume={16},
   ISSN={2542-4653},
   url={http://dx.doi.org/10.21468/SciPostPhys.16.4.093},
   DOI={10.21468/SciPostPhys.16.4.093},
   number={4},
   pages={093},
   journal={SciPost Physics},
   publisher={Stichting SciPost},
   author={Bhardwaj, Lakshya and Schäfer-Nameki, Sakura},
   year={2024},
   month=apr }

@article{Bhardwaj:2023ayw,
    author = "Bhardwaj, Lakshya and Schafer-Nameki, Sakura",
    title = "{Generalized Charges, Part II: Non-Invertible Symmetries and the Symmetry TFT}",
    eprint = "2305.17159",
    archivePrefix = "arXiv",
    primaryClass = "hep-th",
    doi = "10.21468/SciPostPhys.19.4.098",
    journal = "SciPost Phys.",
    volume = "19",
    number = "4",
    pages = "098",
    month = "5",
    year = "2025"
}

@article{Armas:2024caa,
    author = "Armas, Jay and Batzios, Giorgos and Jain, Akash",
    title = "{Higher-group global symmetry and the bosonic M5 brane}",
    eprint = "2402.19458",
    archivePrefix = "arXiv",
    primaryClass = "hep-th",
    doi = "10.1007/JHEP08(2024)003",
    journal = "JHEP",
    volume = "08",
    pages = "003",
    year = "2024"
}

@article{Wan:2018djl,
    author = "Wan, Zheyan and Wang, Juven",
    title = "{Adjoint QCD$_4$, Deconfined Critical Phenomena, Symmetry-Enriched Topological Quantum Field Theory, and Higher Symmetry-Extension}",
    eprint = "1812.11955",
    archivePrefix = "arXiv",
    primaryClass = "hep-th",
    doi = "10.1103/PhysRevD.99.065013",
    journal = "Phys. Rev. D",
    volume = "99",
    number = "6",
    pages = "065013",
    year = "2019"
}

@article{Moradi:2023dan,
    author = {Moradi, Heidar and Aksoy, \"Omer M. and Bardarson, Jens H. and Tiwari, Apoorv},
    title = "{Symmetry fractionalization, mixed-anomalies and dualities in quantum spin models with generalized symmetries}",
    eprint = "2307.01266",
    archivePrefix = "arXiv",
    primaryClass = "cond-mat.str-el",
    month = "7",
    year = "2023"
}

@article{Cvetic:2023pgm,
    author = {Cveti\v{c}, Mirjam and Heckman, Jonathan J. and H\"ubner, Max and Torres, Ethan},
    title = "{Generalized symmetries, gravity, and the swampland}",
    eprint = "2307.13027",
    archivePrefix = "arXiv",
    primaryClass = "hep-th",
    doi = "10.1103/PhysRevD.109.026012",
    journal = "Phys. Rev. D",
    volume = "109",
    number = "2",
    pages = "026012",
    year = "2024"
}

@article{Iqbal:2020lrt,
    author = "Iqbal, Nabil and Poovuttikul, Napat",
    title = "{2-group global symmetries, hydrodynamics and holography}",
    eprint = "2010.00320",
    archivePrefix = "arXiv",
    primaryClass = "hep-th",
    doi = "10.21468/SciPostPhys.15.2.063",
    journal = "SciPost Phys.",
    volume = "15",
    number = "2",
    pages = "063",
    year = "2023"
}

@article{Delcamp:2018wlb,
    author = "Delcamp, Clement and Tiwari, Apoorv",
    title = "{From gauge to higher gauge models of topological phases}",
    eprint = "1802.10104",
    archivePrefix = "arXiv",
    primaryClass = "cond-mat.str-el",
    doi = "10.1007/JHEP10(2018)049",
    journal = "JHEP",
    volume = "10",
    pages = "049",
    year = "2018"
}

@article{Barkeshli:2022wuz,
    author = "Barkeshli, Maissam and Chen, Yu-An and Huang, Sheng-Jie and Kobayashi, Ryohei and Tantivasadakarn, Nathanan and Zhu, Guanyu",
    title = "{Codimension-2 defects and higher symmetries in (3+1)D topological phases}",
    eprint = "2208.07367",
    archivePrefix = "arXiv",
    primaryClass = "cond-mat.str-el",
    doi = "10.21468/SciPostPhys.14.4.065",
    journal = "SciPost Phys.",
    volume = "14",
    number = "4",
    pages = "065",
    year = "2023"
}

@article{Cvetic:2022imb,
    author = {Cveti\v{c}, Mirjam and Heckman, Jonathan J. and H\"ubner, Max and Torres, Ethan},
    title = "{0-form, 1-form, and 2-group symmetries via cutting and gluing of orbifolds}",
    eprint = "2203.10102",
    archivePrefix = "arXiv",
    primaryClass = "hep-th",
    reportNumber = "UPR-1317-T, CERN-TH-2022-053",
    doi = "10.1103/PhysRevD.106.106003",
    journal = "Phys. Rev. D",
    volume = "106",
    number = "10",
    pages = "106003",
    year = "2022"
}

@article{Bhardwaj:2020phs,
    author = {Bhardwaj, Lakshya and Sch\"afer-Nameki, Sakura},
    title = "{Higher-form symmetries of 6d and 5d theories}",
    eprint = "2008.09600",
    archivePrefix = "arXiv",
    primaryClass = "hep-th",
    doi = "10.1007/JHEP02(2021)159",
    journal = "JHEP",
    volume = "02",
    pages = "159",
    year = "2021"
}

@article{Apruzzi:2021vcu,
    author = "Apruzzi, Fabio and Schafer-Nameki, Sakura and Bhardwaj, Lakshya and Oh, Jihwan",
    title = "{The Global Form of Flavor Symmetries and 2-Group Symmetries in 5d SCFTs}",
    eprint = "2105.08724",
    archivePrefix = "arXiv",
    primaryClass = "hep-th",
    doi = "10.21468/SciPostPhys.13.2.024",
    journal = "SciPost Phys.",
    volume = "13",
    number = "2",
    pages = "024",
    year = "2022"
}

@article{DelZotto:2022fnw,
    author = "Del Zotto, Michele and Heckman, Jonathan J. and Meynet, Shani Nadir and Moscrop, Robert and Zhang, Hao Y.",
    title = "{Higher symmetries of 5D orbifold SCFTs}",
    eprint = "2201.08372",
    archivePrefix = "arXiv",
    primaryClass = "hep-th",
    doi = "10.1103/PhysRevD.106.046010",
    journal = "Phys. Rev. D",
    volume = "106",
    number = "4",
    pages = "046010",
    year = "2022"
}

@article{Hsin:2019fhf,
    author = "Hsin, Po-Shen and Turzillo, Alex",
    title = "{Symmetry-enriched quantum spin liquids in (3 + 1)$d$}",
    eprint = "1904.11550",
    archivePrefix = "arXiv",
    primaryClass = "cond-mat.str-el",
    reportNumber = "CALT-TH-2019-014",
    doi = "10.1007/JHEP09(2020)022",
    journal = "JHEP",
    volume = "09",
    pages = "022",
    year = "2020"
}

@article{Bartsch:2023wvv,
    author = "Bartsch, Thomas and Bullimore, Mathew and Grigoletto, Andrea",
    title = "{Representation theory for categorical symmetries}",
    eprint = "2305.17165",
    archivePrefix = "arXiv",
    primaryClass = "hep-th",
    month = "5",
    year = "2023"
}

@article{Kang:2023uvm,
    author = "Kang, Monica Jinwoo and Kang, Sungkyung",
    title = "{Central extensions of higher groups: Green-Schwarz mechanism and 2-connections}",
    eprint = "2311.14666",
    archivePrefix = "arXiv",
    primaryClass = "hep-th",
    reportNumber = "CALT-TH-2023-048",
    month = "11",
    year = "2023"
}

@article{Pace:2023kyi,
    author = "Pace, Salvatore D.",
    title = "{Emergent generalized symmetries in ordered phases and applications to quantum disordering}",
    eprint = "2308.05730",
    archivePrefix = "arXiv",
    primaryClass = "cond-mat.str-el",
    doi = "10.21468/SciPostPhys.17.3.080",
    journal = "SciPost Phys.",
    volume = "17",
    number = "3",
    pages = "080",
    year = "2024"
}

@article{Debray:2023rlx,
    author = "Debray, Arun",
    title = "{Bordism for the 2-group symmetries of the heterotic and CHL strings}",
    eprint = "2304.14764",
    archivePrefix = "arXiv",
    primaryClass = "math.AT",
    doi = "10.1090/conm/802/16079",
    journal = "Contemp. Math.",
    volume = "802",
    pages = "227--98",
    year = "2024"
}

@article{Bhardwaj:2022scy,
    author = "Bhardwaj, Lakshya and Gould, Dewi S. W.",
    title = "{Disconnected 0-form and 2-group symmetries}",
    eprint = "2206.01287",
    archivePrefix = "arXiv",
    primaryClass = "hep-th",
    doi = "10.1007/JHEP07(2023)098",
    journal = "JHEP",
    volume = "07",
    pages = "098",
    year = "2023"
}

@article{DelZotto:2022joo,
    author = "Del Zotto, Michele and Garc\'\i{}a Etxebarria, I\~naki and Schafer-Nameki, Sakura",
    title = "{2-Group Symmetries and M-Theory}",
    eprint = "2203.10097",
    archivePrefix = "arXiv",
    primaryClass = "hep-th",
    doi = "10.21468/SciPostPhys.13.5.105",
    journal = "SciPost Phys.",
    volume = "13",
    pages = "105",
    year = "2022"
}

@article{Yu:2020twi,
    author = "Yu, Matthew",
    title = "{Symmetries and anomalies of (1+1)d theories: 2-groups and symmetry fractionalization}",
    eprint = "2010.01136",
    archivePrefix = "arXiv",
    primaryClass = "hep-th",
    doi = "10.1007/JHEP08(2021)061",
    journal = "JHEP",
    volume = "08",
    pages = "061",
    year = "2021"
}

@article{Cordova:2020tij,
    author = "Cordova, Clay and Dumitrescu, Thomas T. and Intriligator, Kenneth",
    title = "{2-Group Global Symmetries and Anomalies in Six-Dimensional Quantum Field Theories}",
    eprint = "2009.00138",
    archivePrefix = "arXiv",
    primaryClass = "hep-th",
    doi = "10.1007/JHEP04(2021)252",
    journal = "JHEP",
    volume = "04",
    pages = "252",
    year = "2021"
}

@article{DeWolfe:2020uzb,
    author = "DeWolfe, Oliver and Higginbotham, Kenneth",
    title = "{Generalized symmetries and 2-groups via electromagnetic duality in $AdS/CFT$}",
    eprint = "2010.06594",
    archivePrefix = "arXiv",
    primaryClass = "hep-th",
    doi = "10.1103/PhysRevD.103.026011",
    journal = "Phys. Rev. D",
    volume = "103",
    number = "2",
    pages = "026011",
    year = "2021"
}

@article{Apruzzi:2024htg,
    author = "Apruzzi, Fabio and Bedogna, Francesco and Dondi, Nicola",
    title = "{SymTh for non-finite symmetries}",
    eprint = "2402.14813",
    archivePrefix = "arXiv",
    primaryClass = "hep-th",
    doi = "10.1007/JHEP04(2026)153",
    journal = "JHEP",
    volume = "04",
    pages = "153",
    month = "2",
    year = "2026"
}

@article{Huang:2024wdr,
    author = "Huang, Mo and Xu, Hao and Zhang, Zhi-Hao",
    title = "{The 2-character theory for finite 2-groups}",
    eprint = "2404.01162",
    archivePrefix = "arXiv",
    primaryClass = "math.RT",
    month = "4",
    year = "2024"
}

@article{Ambrosino:2024ggh,
    author = "Ambrosino, Federico and Luo, Ran and Wang, Yi-Nan and Zhang, Yi",
    title = "{Understanding fermionic generalized symmetries}",
    eprint = "2404.12301",
    archivePrefix = "arXiv",
    primaryClass = "hep-th",
    reportNumber = "DESY-24-055",
    doi = "10.1103/PhysRevD.110.105020",
    journal = "Phys. Rev. D",
    volume = "110",
    number = "10",
    pages = "105020",
    year = "2024"
}

@article{Chen:2020msl,
    author = "Chen, Yu-An and Ellison, Tyler D. and Tantivasadakarn, Nathanan",
    title = "{Disentangling supercohomology symmetry-protected topological phases in three spatial dimensions}",
    eprint = "2008.05652",
    archivePrefix = "arXiv",
    primaryClass = "cond-mat.str-el",
    doi = "10.1103/PhysRevResearch.3.013056",
    journal = "Phys. Rev. Res.",
    volume = "3",
    number = "1",
    pages = "013056",
    year = "2021"
}

@article{Costa:2024wks,
    author = "Costa, Davi and others",
    title = "{Simons lectures on categorical symmetries}",
    eprint = "2411.09082",
    archivePrefix = "arXiv",
    primaryClass = "math-ph",
    year = "2024"
}

@article{Gomes:2023ahz,
    author = "Gomes, Pedro R. S.",
    title = "{An introduction to higher-form symmetries}",
    eprint = "2303.01817",
    archivePrefix = "arXiv",
    primaryClass = "hep-th",
    doi = "10.21468/SciPostPhysLectNotes.74",
    journal = "SciPost Phys. Lect. Notes",
    volume = "74",
    pages = "1",
    year = "2023"
}

@inproceedings{Iqbal:2024pee,
    author = "Iqbal, Nabil",
    title = "{Jena lectures on generalized global symmetries: principles and applications}",
    eprint = "2407.20815",
    archivePrefix = "arXiv",
    primaryClass = "hep-th",
    year = "2024"
}

@article{Kaidi:2026urc,
    author = "Kaidi, Justin",
    title = "{Introduction to generalized symmetries}",
    eprint = "2603.08798",
    archivePrefix = "arXiv",
    primaryClass = "hep-th",
    month = "3",
    year = "2026"
}

@article{Bhardwaj:2025jtf,
    author = "Bhardwaj, Lakshya and Gai, Yuhan and Huang, Sheng-Jie and Inamura, Kansei and Schafer-Nameki, Sakura and Tiwari, Apoorv and Warman, Alison",
    title = "{Gapless Phases in (2+1)d with Non-Invertible Symmetries}",
    eprint = "2503.12699",
    archivePrefix = "arXiv",
    primaryClass = "cond-mat.str-el",
    month = "3",
    year = "2025"
}

@article{Bhardwaj:2025piv,
    author = "Bhardwaj, Lakshya and Schafer-Nameki, Sakura and Tiwari, Apoorv and Warman, Alison",
    title = "{Gapped Phases in (2+1)d with Non-Invertible Symmetries: Part II}",
    eprint = "2502.20440",
    archivePrefix = "arXiv",
    primaryClass = "hep-th",
    month = "2",
    year = "2025"
}

@article{Bhardwaj:2024qiv,
    author = "Bhardwaj, Lakshya and Pajer, Daniel and Schafer-Nameki, Sakura and Tiwari, Apoorv and Warman, Alison and Wu, Jingxiang",
    title = "{Gapped phases in (2+1)d with non-invertible symmetries: Part I}",
    eprint = "2408.05266",
    archivePrefix = "arXiv",
    primaryClass = "hep-th",
    doi = "10.21468/SciPostPhys.19.2.056",
    journal = "SciPost Phys.",
    volume = "19",
    number = "2",
    pages = "056",
    year = "2025"
}

@misc{Chen2024,
      title={Topological defects of 2+1D systems from line excitations in 3+1D bulk}, 
      author={Wenjie Ji and Xie Chen},
      year={2024},
      eprint={2407.02488},
      archivePrefix={arXiv},
      primaryClass={cond-mat.str-el},
      url={https://arxiv.org/abs/2407.02488}, 
}

@Article{Xu2023,
	title={{Boundary states of three dimensional topological order and the deconfined quantum critical point}},
	author={Wenjie Ji and Nathanan Tantivasadakarn and Cenke Xu},
	journal={SciPost Phys.},
	volume={15},
	pages={231},
	year={2023},
	publisher={SciPost},
	doi={10.21468/SciPostPhys.15.6.231},
	url={https://scipost.org/10.21468/SciPostPhys.15.6.231},
}

@article{Kaidi:2022cpf,
    author = "Kaidi, Justin and Ohmori, Kantaro and Zheng, Yunqin",
    title = "{Symmetry TFTs for Non-invertible Defects}",
    eprint = "2209.11062",
    archivePrefix = "arXiv",
    primaryClass = "hep-th",
    doi = "10.1007/s00220-023-04859-7",
    journal = "Commun. Math. Phys.",
    volume = "404",
    number = "2",
    pages = "1021--1124",
    year = "2023"
}

@article{Witten:1998wy,
    author = "Witten, Edward",
    title = "{AdS / CFT correspondence and topological field theory}",
    eprint = "hep-th/9812012",
    archivePrefix = "arXiv",
    reportNumber = "IASSNS-HEP-98-96",
    doi = "10.1088/1126-6708/1998/12/012",
    journal = "JHEP",
    volume = "12",
    pages = "012",
    year = "1998"
}

@article{Gaiotto:2020iye,
    author = "Gaiotto, Davide and Kulp, Justin",
    title = "{Orbifold groupoids}",
    eprint = "2008.05960",
    archivePrefix = "arXiv",
    primaryClass = "hep-th",
    doi = "10.1007/JHEP02(2021)132",
    journal = "JHEP",
    volume = "02",
    pages = "132",
    year = "2021"
}

@article{Moradi:2022lqp,
    author = "Moradi, Heidar and Moosavian, Seyed Faroogh and Tiwari, Apoorv",
    title = "{Topological holography: Towards a unification of Landau and beyond-Landau physics}",
    eprint = "2207.10712",
    archivePrefix = "arXiv",
    primaryClass = "cond-mat.str-el",
    doi = "10.21468/SciPostPhysCore.6.4.066",
    journal = "SciPost Phys. Core",
    volume = "6",
    pages = "066",
    year = "2023"
}

@article{Freed:2022qnc,
    author = "Freed, Daniel S. and Moore, Gregory W. and Teleman, Constantin",
    title = "{Topological symmetry in quantum field theory}",
    eprint = "2209.07471",
    archivePrefix = "arXiv",
    primaryClass = "hep-th",
    month = "9",
    year = "2022"
}

@article{Kaidi:2023maf,
    author = "Kaidi, Justin and Nardoni, Emily and Zafrir, Gabi and Zheng, Yunqin",
    title = "{Symmetry TFTs and anomalies of non-invertible symmetries}",
    eprint = "2301.07112",
    archivePrefix = "arXiv",
    primaryClass = "hep-th",
    doi = "10.1007/JHEP10(2023)053",
    journal = "JHEP",
    volume = "10",
    pages = "053",
    year = "2023"
}

@article{Bhardwaj:2023fca,
    author = "Bhardwaj, Lakshya and Bottini, Lea E. and Pajer, Daniel and Schafer-Nameki, Sakura",
    title = "{Categorical Landau Paradigm for Gapped Phases}",
    eprint = "2310.03786",
    archivePrefix = "arXiv",
    primaryClass = "cond-mat.str-el",
    doi = "10.1103/PhysRevLett.133.161601",
    journal = "Phys. Rev. Lett.",
    volume = "133",
    number = "16",
    pages = "161601",
    year = "2024"
}

@article{Xu:2022rtj,
    author = "Xu, Rongge and Zhang, Zhi-Hao",
    title = "{Categorical descriptions of one-dimensional gapped phases with Abelian onsite symmetries}",
    eprint = "2205.09656",
    archivePrefix = "arXiv",
    primaryClass = "cond-mat.str-el",
    doi = "10.1103/PhysRevB.110.155106",
    journal = "Phys. Rev. B",
    volume = "110",
    number = "15",
    pages = "155106",
    year = "2024"
}

@article{Bhardwaj:2023idu,
    author = {Bhardwaj, Lakshya and Bottini, Lea E. and Pajer, Daniel and Sch\"afer-Nameki, Sakura},
    title = "{Gapped Phases with Non-Invertible Symmetries: (1+1)d}",
    eprint = "2310.03784",
    archivePrefix = "arXiv",
    primaryClass = "hep-th",
    doi = "10.21468/SciPostPhys.18.1.032",
    journal = "SciPost Phys.",
    volume = "18",
    number = "1",
    pages = "032",
    month = "10",
    year = "2025"
}

@article{Hai:2023osv,
    author = "Hai, Yong-Ju and Zhang, Ze and Zheng, Hao and Kong, Liang and Wu, Jiansheng and Yu, Dapeng",
    title = "{Uniquely identifying topological order based on boundary-bulk duality and anyon condensation}",
    doi = "10.1093/nsr/nwac264",
    journal = "Natl. Sci. Rev.",
    volume = "10",
    number = "3",
    pages = "nwac264",
    year = "2023"
}

@article{Bhardwaj:2024qrf,
    author = "Bhardwaj, Lakshya and Pajer, Daniel and Schafer-Nameki, Sakura and Warman, Alison",
    title = "{Hasse Diagrams for Gapless SPT and SSB Phases with Non-Invertible Symmetries}",
    eprint = "2403.00905",
    archivePrefix = "arXiv",
    primaryClass = "cond-mat.str-el",
    month = "3",
    year = "2024"
}

@article{Bhardwaj:2023bbf,
    author = "Bhardwaj, Lakshya and Bottini, Lea E. and Pajer, Daniel and Schafer-Nameki, Sakura",
    title = "{The Club Sandwich: Gapless Phases and Phase Transitions with Non-Invertible Symmetries}",
    eprint = "2312.17322",
    archivePrefix = "arXiv",
    primaryClass = "hep-th",
    doi = "10.21468/SciPostPhys.18.5.156",
    journal = "SciPost Phys.",
    volume = "18",
    number = "5",
    pages = "156",
    month = "12",
    year = "2025"
}

@article{Ji:2019jhk,
    author = "Ji, Wenjie and Wen, Xiao-Gang",
    title = "{Categorical symmetry and noninvertible anomaly in symmetry-breaking and topological phase transitions}",
    eprint = "1912.13492",
    archivePrefix = "arXiv",
    primaryClass = "cond-mat.str-el",
    doi = "10.1103/PhysRevResearch.2.033417",
    journal = "Phys. Rev. Res.",
    volume = "2",
    number = "3",
    pages = "033417",
    year = "2020"
}

@article{Antinucci:2024ltv,
    author = "Antinucci, Andrea and Copetti, Christian and Schafer-Nameki, Sakura",
    title = "{SymTFT for (3+1)d Gapless SPTs and Obstructions to Confinement}",
    eprint = "2408.05585",
    archivePrefix = "arXiv",
    primaryClass = "hep-th",
    doi = "10.21468/SciPostPhys.18.3.114",
    journal = "SciPost Phys.",
    volume = "18",
    number = "3",
    pages = "114",
    month = "8",
    year = "2025"
}

@article{Gagliano:2024off,
    author = "Gagliano, Finn and Garc\'\i{}a Etxebarria, I\~naki",
    title = "{SymTFTs for $U(1)$ symmetries from descent}",
    eprint = "2411.15126",
    archivePrefix = "arXiv",
    primaryClass = "hep-th",
    month = "11",
    year = "2024"
}

@article{Bonetti:2024cjk,
    author = "Bonetti, Federico and Del Zotto, Michele and Minasian, Ruben",
    title = "{SymTFTs for Continuous non-Abelian Symmetries}",
    eprint = "2402.12347",
    archivePrefix = "arXiv",
    primaryClass = "hep-th",
    month = "2",
    year = "2024"
}

@article{Brennan:2024fgj,
    author = "Brennan, T. Daniel and Sun, Zhengdi",
    title = "{A SymTFT for continuous symmetries}",
    eprint = "2401.06128",
    archivePrefix = "arXiv",
    primaryClass = "hep-th",
    doi = "10.1007/JHEP12(2024)100",
    journal = "JHEP",
    volume = "12",
    pages = "100",
    year = "2024"
}

@article{Tian:2025ooo,
    author = "Jia, Qiang and Luo, Ran and Tian, Jiahua and Wang, Yi-Nan and Zhang, Yi",
    title = "{Symmetry Topological Field Theory for Flavor Symmetry}",
    eprint = "2503.04546",
    archivePrefix = "arXiv",
    primaryClass = "hep-th",
    month = "3",
    year = "2025"
}

@article{Chen:2024ulc,
    author = "Chen, Jin and Jia, Qiang",
    title = "{SymTFT Approach to 2D Orbifold Groupoids: `t Hooft Anomalies, Gauging, and Partition Functions}",
    eprint = "2411.18056",
    archivePrefix = "arXiv",
    primaryClass = "hep-th",
    reportNumber = "USTC-ICTS/PCFT-24-46",
    month = "11",
    year = "2024"
}

@article{Lan:2014uaa,
    author = "Lan, Tian and Wang, Juven C. and Wen, Xiao-Gang",
    title = "{Gapped Domain Walls, Gapped Boundaries and Topological Degeneracy}",
    eprint = "1408.6514",
    archivePrefix = "arXiv",
    primaryClass = "cond-mat.str-el",
    doi = "10.1103/PhysRevLett.114.076402",
    journal = "Phys. Rev. Lett.",
    volume = "114",
    number = "7",
    pages = "076402",
    year = "2015"
}

@article{Decoppet2022Drinfeld,
      title={Drinfeld Centers and Morita Equivalence Classes of Fusion 2-Categories},
      author={Thibault D. Décoppet},
      journal={Compositio Mathematica},
      volume={161},
      number={2},
      pages={305--340},
      year={2025},
      eprint={2211.04917},
      archivePrefix={arXiv},
      primaryClass={math.QA},
      url={https://arxiv.org/abs/2211.04917},
}

@article{Ostrik:2002ohv,
    author = "Ostrik, Victor",
    title = "{Module categories over the Drinfeld double of a finite group}",
    eprint = "math/0202130",
    archivePrefix = "arXiv",
    doi = "10.1155/S1073792803205079",
    journal = "Int. Math. Res. Not.",
    volume = "2003",
    number = "27",
    pages = "1507--1520",
    month = "2",
    year = "2003"
}

@article{DelZotto:2022ras,
    author = "Del Zotto, Michele and Garc\'\i{}a Etxebarria, I\~naki",
    title = "{Global structures from the infrared}",
    eprint = "2204.06495",
    archivePrefix = "arXiv",
    primaryClass = "hep-th",
    doi = "10.1007/JHEP11(2023)058",
    journal = "JHEP",
    volume = "11",
    pages = "058",
    year = "2023"
}

@article{vanBeest:2022fss,
    author = "van Beest, Marieke and Gould, Dewi S. W. and Schafer-Nameki, Sakura and Wang, Yi-Nan",
    title = "{Symmetry TFTs for 3d QFTs from M-theory}",
    eprint = "2210.03703",
    archivePrefix = "arXiv",
    primaryClass = "hep-th",
    doi = "10.1007/JHEP02(2023)226",
    journal = "JHEP",
    volume = "02",
    pages = "226",
    year = "2023"
}

@article{Apruzzi:2023uma,
    author = "Apruzzi, Fabio and Bonetti, Federico and Gould, Dewi S. W. and Schafer-Nameki, Sakura",
    title = "{Aspects of categorical symmetries from branes: SymTFTs and generalized charges}",
    eprint = "2306.16405",
    archivePrefix = "arXiv",
    primaryClass = "hep-th",
    doi = "10.21468/SciPostPhys.17.1.025",
    journal = "SciPost Phys.",
    volume = "17",
    number = "1",
    pages = "025",
    year = "2024"
}

@article{Baume:2023kkf,
    author = {Baume, Florent and Heckman, Jonathan J. and H\"ubner, Max and Torres, Ethan and Turner, Andrew P. and Yu, Xingyang},
    title = "{SymTrees and Multi-Sector QFTs}",
    eprint = "2310.12980",
    archivePrefix = "arXiv",
    primaryClass = "hep-th",
    reportNumber = "ZMP-HH/23-13, CERN-TH-2023-183",
    doi = "10.1103/PhysRevD.109.106013",
    journal = "Phys. Rev. D",
    volume = "109",
    number = "10",
    pages = "106013",
    year = "2024"
}

@article{Antinucci:2024zjp,
    author = "Antinucci, Andrea and Benini, Francesco",
    title = "{Anomalies and gauging of U(1) symmetries}",
    eprint = "2401.10165",
    archivePrefix = "arXiv",
    primaryClass = "hep-th",
    reportNumber = "SISSA 01/2024/FISI",
    doi = "10.1103/PhysRevB.111.024110",
    journal = "Phys. Rev. B",
    volume = "111",
    number = "2",
    pages = "024110",
    year = "2025"
}

@article{Antinucci:2024bcm,
    author = "Antinucci, Andrea and Benini, Francesco and Rizi, Giovanni",
    title = "{Holographic Duals of Symmetry Broken Phases}",
    eprint = "2408.01418",
    archivePrefix = "arXiv",
    primaryClass = "hep-th",
    reportNumber = "SISSA 15/2024/FISI",
    doi = "10.1002/prop.202400172",
    journal = "Fortsch. Phys.",
    volume = "72",
    number = "12",
    pages = "2400172",
    year = "2024"
}

@article{Kong:2013aya,
    author = "Kong, Liang",
    title = "{Anyon condensation and tensor categories}",
    eprint = "1307.8244",
    archivePrefix = "arXiv",
    primaryClass = "cond-mat.str-el",
    doi = "10.1016/j.nuclphysb.2014.07.003",
    journal = "Nucl. Phys. B",
    volume = "886",
    pages = "436--482",
    year = "2014"
}

@article{Zhao:2022yaw,
    author = "Zhao, Jiaheng and Lou, Jia-Qi and Zhang, Zhi-Hao and Hung, Ling-Yan and Kong, Liang and Tian, Yin",
    title = "{String condensations in $3+1D$ and Lagrangian algebras}",
    eprint = "2208.07865",
    archivePrefix = "arXiv",
    primaryClass = "cond-mat.str-el",
    doi = "10.4310/ATMP.2023.v27.n2.a5",
    journal = "Adv. Theor. Math. Phys.",
    volume = "27",
    number = "2",
    pages = "583--622",
    year = "2023"
}

@article{Kong:2024ykr,
    author = "Kong, Liang and Zhang, Zhi-Hao and Zhao, Jiaheng and Zheng, Hao",
    title = "{Higher condensation theory}",
    eprint = "2403.07813",
    archivePrefix = "arXiv",
    primaryClass = "cond-mat.str-el",
    month = "3",
    year = "2024"
}

@article{Bhardwaj:2024wlr,
    author = "Bhardwaj, Lakshya and Bottini, Lea E. and Schafer-Nameki, Sakura and Tiwari, Apoorv",
    title = "{Illustrating the categorical Landau paradigm in lattice models}",
    eprint = "2405.05302",
    archivePrefix = "arXiv",
    primaryClass = "cond-mat.str-el",
    doi = "10.1103/PhysRevB.111.054432",
    journal = "Phys. Rev. B",
    volume = "111",
    number = "5",
    pages = "054432",
    year = "2025"
}

@article{Argurio:2024oym,
    author = "Argurio, Riccardo and Benini, Francesco and Bertolini, Matteo and Galati, Giovanni and Niro, Pierluigi",
    title = "{On the symmetry TFT of Yang-Mills-Chern-Simons theory}",
    eprint = "2404.06601",
    archivePrefix = "arXiv",
    primaryClass = "hep-th",
    reportNumber = "SISSA 05/2024/FISI",
    doi = "10.1007/JHEP07(2024)130",
    journal = "JHEP",
    volume = "07",
    pages = "130",
    year = "2024"
}

@article{Bhardwaj:2024igy,
    author = "Bhardwaj, Lakshya and Copetti, Christian and Pajer, Daniel and Schafer-Nameki, Sakura",
    title = "{Boundary SymTFT}",
    eprint = "2409.02166",
    archivePrefix = "arXiv",
    primaryClass = "hep-th",
    doi = "10.21468/SciPostPhys.19.2.061",
    journal = "SciPost Phys.",
    volume = "19",
    number = "2",
    pages = "061",
    month = "9",
    year = "2025"
}

@article{Choi:2024tri,
    author = "Choi, Yichul and Rayhaun, Brandon C. and Zheng, Yunqin",
    title = "{Generalized Tube Algebras, Symmetry-Resolved Partition Functions, and Twisted Boundary States}",
    eprint = "2409.02159",
    archivePrefix = "arXiv",
    primaryClass = "hep-th",
    journal = "Commun. Math. Phys.",
    volume = "407",
    pages = "62",
    month = "9",
    year = "2026"
}

@article{Tian:2024dgl,
    author = "Tian, Jiahua and Wang, Yi-Nan",
    title = "{A Tale of Bulk and Branes: Symmetry TFT of 6D SCFTs from IIB/F-theory}",
    eprint = "2410.23076",
    archivePrefix = "arXiv",
    primaryClass = "hep-th",
    doi = "10.1007/JHEP03(2025)085",
    journal = "JHEP",
    volume = "03",
    pages = "085",
    month = "10",
    year = "2025"
}

@article{Najjar:2024vmm,
    author = "Najjar, Marwan and Santilli, Leonardo and Wang, Yi-Nan",
    title = "{(-1)-form symmetries from M-theory and SymTFTs}",
    eprint = "2411.19683",
    archivePrefix = "arXiv",
    primaryClass = "hep-th",
    reportNumber = "USTC-ICTS/PCFT-24-53",
    month = "11",
    year = "2024"
}

@article{Bonetti:2024etn,
    author = "Bonetti, Federico and Del Zotto, Michele and Minasian, Ruben",
    title = "{SymTFTs and Non-Invertible Symmetries of 6d (2,0) SCFTs of Type $D$ from M-theory}",
    eprint = "2412.07842",
    archivePrefix = "arXiv",
    primaryClass = "hep-th",
    month = "12",
    year = "2024"
}

@article{Wen:2024udn,
    author = "Wen, Rui and Ye, Weicheng and Potter, Andrew C.",
    title = "{Topological holography for fermions}",
    eprint = "2404.19004",
    archivePrefix = "arXiv",
    primaryClass = "cond-mat.str-el",
    month = "4",
    year = "2024"
}

@article{Kong:2015flk,
    author = "Kong, Liang and Wen, Xiao-Gang and Zheng, Hao",
    title = "{Boundary-bulk relation for topological orders as the functor mapping higher categories to their centers}",
    eprint = "1502.01690",
    archivePrefix = "arXiv",
    primaryClass = "cond-mat.str-el",
    month = "2",
    year = "2015"
}

@inproceedings{Cordova:2022ruw,
    author = "Cordova, Clay and Dumitrescu, Thomas T. and Intriligator, Kenneth and Shao, Shu-Heng",
    title = "{Snowmass White Paper: Generalized Symmetries in Quantum Field Theory and Beyond}",
    booktitle = "{Snowmass 2021}",
    eprint = "2205.09545",
    archivePrefix = "arXiv",
    primaryClass = "hep-th",
    month = "5",
    year = "2022"
}

@article{McGreevy:2022oyu,
    author = "McGreevy, John",
    title = "{Generalized Symmetries in Condensed Matter}",
    eprint = "2204.03045",
    archivePrefix = "arXiv",
    primaryClass = "cond-mat.str-el",
    doi = "10.1146/annurev-conmatphys-040721-021029",
    journal = "Ann. Rev. Condensed Matter Phys.",
    volume = "14",
    pages = "57--82",
    year = "2023"
}

@article{Shao:2023gho,
    author = "Shao, Shu-Heng",
    title = "{What's Done Cannot Be Undone: TASI Lectures on Non-Invertible Symmetries}",
    eprint = "2308.00747",
    archivePrefix = "arXiv",
    primaryClass = "hep-th",
    month = "8",
    year = "2023"
}

@article{Baez:2003yaq,
    author = "Baez, John C. and Lauda, Aaron D.",
    title = "{Higher-Dimensional Algebra V: 2-Groups}",
    eprint = "math/0307200",
    archivePrefix = "arXiv",
    primaryClass = "math.QA",
    journal = "Theory Appl. Categ.",
    volume = "12",
    pages = "423--491",
    month = "7",
    year = "2004"
}

@article{Baez:2004in,
    author = "Baez, John and Schreiber, Urs",
    title = "{Higher gauge theory: 2-connections on 2-bundles}",
    eprint = "hep-th/0412325",
    archivePrefix = "arXiv",
    primaryClass = "hep-th",
    month = "12",
    year = "2004"
}

@article{Bullivant:2016clk,
    author = "Bullivant, Alex and Cal{\c{c}}ada, Marcos and K{\'a}d{\'a}r, Zolt{\'a}n and Martin, Paul and Faria Martins, Jo{\~a}o",
    title = "{Topological phases from higher gauge symmetry in 3+1 dimensions}",
    eprint = "1606.06639",
    archivePrefix = "arXiv",
    primaryClass = "cond-mat.str-el",
    doi = "10.1103/PhysRevB.95.155118",
    journal = "Phys. Rev. B",
    volume = "95",
    number = "15",
    pages = "155118",
    year = "2017"
}

@article{Delcamp:2019fdp,
    author = "Delcamp, Clement and Tiwari, Apoorv",
    title = "{On 2-form gauge models of topological phases}",
    eprint = "1901.02249",
    archivePrefix = "arXiv",
    primaryClass = "hep-th",
    doi = "10.1007/JHEP05(2019)064",
    journal = "JHEP",
    volume = "05",
    pages = "064",
    year = "2019"
}

@article{Garcia-Etxebarria:2018ajm,
    author = "Garc{\'\i}a-Etxebarria, I{\~n}aki and Montero, Miguel",
    title = "{Dai-Freed anomalies in particle physics}",
    eprint = "1808.00009",
    archivePrefix = "arXiv",
    primaryClass = "hep-th",
    doi = "10.1007/JHEP08(2019)003",
    journal = "JHEP",
    volume = "08",
    pages = "003",
    year = "2019"
}

@article{Wan:2018bns,
    author = "Wan, Zheyan and Wang, Juven",
    title = "{Higher anomalies, higher symmetries, and cobordisms I: classification of higher-symmetry-protected topological states and their boundary fermionic/bosonic anomalies via a generalized cobordism theory}",
    eprint = "1812.11967",
    archivePrefix = "arXiv",
    primaryClass = "hep-th",
    doi = "10.4310/AMSA.2019.v4.n2.a2",
    journal = "Ann. Math. Sci. Appl.",
    volume = "4",
    number = "2",
    pages = "107--311",
    year = "2019"
}

@article{Wan:2019soo,
    author = "Wan, Zheyan and Wang, Juven and Zheng, Yunqin",
    title = "{Higher anomalies, higher symmetries, and cobordisms II: Lorentz symmetry extension and enriched bosonic / fermionic quantum gauge theory}",
    eprint = "1912.13504",
    archivePrefix = "arXiv",
    primaryClass = "hep-th",
    doi = "10.4310/AMSA.2020.v5.n2.a2",
    journal = "Ann. Math. Sci. Appl.",
    volume = "05",
    number = "2",
    pages = "171--257",
    year = "2020"
}

@article{Kitaev:2011dxc,
    author = "Kitaev, Alexei and Kong, Liang",
    title = "{Models for Gapped Boundaries and Domain Walls}",
    eprint = "1104.5047",
    archivePrefix = "arXiv",
    primaryClass = "cond-mat.str-el",
    doi = "10.1007/s00220-012-1500-5",
    journal = "Commun. Math. Phys.",
    volume = "313",
    number = "2",
    pages = "351--373",
    year = "2012"
}

@article{Cong:2017ffh,
    author = "Cong, Iris and Cheng, Meng and Wang, Zhenghan",
    title = "{Hamiltonian and Algebraic Theories of Gapped Boundaries in Topological Phases of Matter}",
    eprint = "1707.04564",
    archivePrefix = "arXiv",
    primaryClass = "cond-mat.str-el",
    doi = "10.1007/s00220-017-2960-4",
    journal = "Commun. Math. Phys.",
    volume = "355",
    number = "2",
    pages = "645--689",
    year = "2017"
}

@article{Bhardwaj:2017xup,
    author = "Bhardwaj, Lakshya and Tachikawa, Yuji",
    title = "{On finite symmetries and their gauging in two dimensions}",
    eprint = "1704.02330",
    archivePrefix = "arXiv",
    primaryClass = "hep-th",
    doi = "10.1007/JHEP03(2018)189",
    journal = "JHEP",
    volume = "03",
    pages = "189",
    year = "2018"
}

@article{Chang:2018iay,
    author = "Chang, Chi-Ming and Lin, Ying-Hsuan and Shao, Shu-Heng and Wang, Yifan and Yin, Xi",
    title = "{Topological Defect Lines and Renormalization Group Flows in Two Dimensions}",
    eprint = "1802.04445",
    archivePrefix = "arXiv",
    primaryClass = "hep-th",
    doi = "10.1007/JHEP01(2019)026",
    journal = "JHEP",
    volume = "01",
    pages = "026",
    year = "2019"
}

@article{Thorngren:2019iar,
    author = "Thorngren, Ryan and Wang, Yifan",
    title = "{Fusion category symmetry. Part I. Anomaly in-flow and gapped phases}",
    eprint = "1912.02817",
    archivePrefix = "arXiv",
    primaryClass = "hep-th",
    doi = "10.1007/JHEP04(2024)132",
    journal = "JHEP",
    volume = "04",
    pages = "132",
    year = "2024"
}

@article{Ji:2019eqo,
    author = "Ji, Wenjie and Wen, Xiao-Gang",
    title = "{Non-invertible anomalies and mapping-class-group transformation of anomalous partition functions}",
    eprint = "1905.13279",
    archivePrefix = "arXiv",
    primaryClass = "cond-mat.str-el",
    doi = "10.1103/PhysRevResearch.1.033054",
    journal = "Phys. Rev. Res.",
    volume = "1",
    number = "3",
    pages = "033054",
    year = "2019"
}

@article{Ji:2021esj,
    author = "Ji, Wenjie and Wen, Xiao-Gang",
    title = "{A unified view on symmetry, anomalous symmetry and non-invertible gravitational anomaly}",
    eprint = "2106.02069",
    archivePrefix = "arXiv",
    primaryClass = "cond-mat.str-el",
    month = "6",
    year = "2021"
}

@article{Lichtman:2020nuw,
    author = "Lichtman, Tsuf and Thorngren, Ryan and Lindner, Netanel H. and Stern, Ady and Berg, Erez",
    title = "{Bulk anyons as edge symmetries: Boundary phase diagrams of topologically ordered states}",
    eprint = "2003.04328",
    archivePrefix = "arXiv",
    primaryClass = "cond-mat.str-el",
    doi = "10.1103/PhysRevB.104.075141",
    journal = "Phys. Rev. B",
    volume = "104",
    number = "7",
    pages = "075141",
    year = "2021"
}

@article{Chatterjee:2022kxb,
    author = "Chatterjee, Arkya and Wen, Xiao-Gang",
    title = "{Symmetry as a shadow of topological order and a derivation of topological holographic principle}",
    eprint = "2203.03596",
    archivePrefix = "arXiv",
    primaryClass = "cond-mat.str-el",
    doi = "10.1103/PhysRevB.107.155136",
    journal = "Phys. Rev. B",
    volume = "107",
    number = "15",
    pages = "155136",
    year = "2023"
}

@article{Chatterjee:2022tyg,
    author = "Chatterjee, Arkya and Wen, Xiao-Gang",
    title = "{Holographic theory for continuous phase transitions: Emergence and symmetry protection of gaplessness}",
    eprint = "2205.06244",
    archivePrefix = "arXiv",
    primaryClass = "cond-mat.str-el",
    doi = "10.1103/PhysRevB.108.075105",
    journal = "Phys. Rev. B",
    volume = "108",
    number = "7",
    pages = "075105",
    year = "2023"
}

@article{Chatterjee:2022jll,
    author = "Chatterjee, Arkya and Ji, Wenjie and Wen, Xiao-Gang",
    title = "{Emergent generalized symmetry and maximal symmetry topological order}",
    eprint = "2212.14432",
    archivePrefix = "arXiv",
    primaryClass = "cond-mat.str-el",
    doi = "10.1103/jzfv-ygmr",
    journal = "Phys. Rev. B",
    volume = "112",
    number = "11",
    pages = "115142",
    year = "2025"
}

@article{Huang:2023pyk,
    author = "Huang, Sheng-Jie and Cheng, Meng",
    title = "{Topological holography, quantum criticality, and boundary states}",
    eprint = "2310.16878",
    archivePrefix = "arXiv",
    primaryClass = "cond-mat.str-el",
    doi = "10.21468/SciPostPhys.18.6.213",
    journal = "SciPost Phys.",
    volume = "18",
    number = "6",
    pages = "213",
    year = "2025"
}

@article{Wen:2022tkg,
    author = "Wen, Rui and Potter, Andrew C.",
    title = "{Bulk-boundary correspondence for intrinsically gapless symmetry-protected topological phases from group cohomology}",
    eprint = "2208.09001",
    archivePrefix = "arXiv",
    primaryClass = "cond-mat.str-el",
    doi = "10.1103/PhysRevB.107.245127",
    journal = "Phys. Rev. B",
    volume = "107",
    number = "24",
    pages = "245127",
    year = "2023"
}

@article{Wen:2023otf,
    author = "Wen, Rui and Potter, Andrew C.",
    title = "{Classification of 1+1D gapless symmetry protected phases via topological holography}",
    eprint = "2311.00050",
    archivePrefix = "arXiv",
    primaryClass = "cond-mat.str-el",
    doi = "10.1103/PhysRevB.111.115161",
    journal = "Phys. Rev. B",
    volume = "111",
    number = "11",
    pages = "115161",
    year = "2025"
}

@article{Luo:2022krz,
    author = "Luo, Zhu-Xi",
    title = "{Gapped boundaries of (3+1)-dimensional topological orders}",
    eprint = "2212.09779",
    archivePrefix = "arXiv",
    primaryClass = "cond-mat.str-el",
    doi = "10.1103/PhysRevB.107.125425",
    journal = "Phys. Rev. B",
    volume = "107",
    number = "12",
    pages = "125425",
    year = "2023"
}

@article{Kan:2023yhz,
    author = "Kan, Naoto and Morikawa, Okuto and Nagoya, Yuta and Wada, Hiroki",
    title = "{Higher-group structure in lattice Abelian gauge theory under instanton-sum modification}",
    eprint = "2302.13466",
    archivePrefix = "arXiv",
    primaryClass = "hep-th",
    doi = "10.1140/epjc/s10052-023-11616-6",
    journal = "Eur. Phys. J. C",
    volume = "83",
    number = "6",
    pages = "481",
    year = "2023"
}

@article{Nakajima:2024vgc,
    author = "Nakajima, Tatsuki and Nakamura, Kikyo and Sakai, Tadakatsu",
    title = "{Note on higher-group structure in 6d self-dual gauge theory}",
    eprint = "2406.10518",
    archivePrefix = "arXiv",
    primaryClass = "hep-th",
    doi = "10.1007/JHEP10(2024)093",
    journal = "JHEP",
    volume = "10",
    pages = "093",
    year = "2024"
}

@article{Bartsch:2024ech,
    author = "Bartsch, Thomas",
    title = "{On Unitary 2-Group Symmetries}",
    eprint = "2411.05067",
    archivePrefix = "arXiv",
    primaryClass = "math-ph",
    month = "11",
    year = "2024"
}

@article{Tian:2025yrj,
    author = "Tian, Jiahua and Wang, Xin",
    title = "{Higher Form and Higher Group Symmetries via Mirror Symmetry}",
    eprint = "2503.09967",
    archivePrefix = "arXiv",
    primaryClass = "hep-th",
    doi = "10.1007/JHEP10(2025)075",
    journal = "JHEP",
    volume = "10",
    pages = "075",
    year = "2025"
}

@article{Carqueville:2025kqs,
    author = "Carqueville, Nils and Haake, Benjamin",
    title = "{2-Group Symmetries of 3-dimensional Defect TQFTs and Their Gauging}",
    eprint = "2506.08178",
    archivePrefix = "arXiv",
    primaryClass = "math.QA",
    month = "6",
    year = "2025"
}

@article{Huang:2020pki,
    author = "Huang, Mo",
    title = "{Finite 2-group gauge theory and its 3+1D lattice realization}",
    eprint = "2508.04693",
    archivePrefix = "arXiv",
    primaryClass = "math-ph",
    doi = "10.1007/JHEP03(2026)133",
    journal = "JHEP",
    volume = "03",
    pages = "133",
    year = "2026"
}

@article{Caro-Perez:2026esp,
    author = "Caro-P{\'e}rez, F. and Garcia del Moral, M. P. and Restuccia, A.",
    title = "{2-Group global symmetry in the compactified M2-brane}",
    eprint = "2607.07924",
    archivePrefix = "arXiv",
    primaryClass = "hep-th",
    month = "7",
    year = "2026"
}

@article{Putrov:2025xmw,
    author = "Putrov, Pavel and Radhakrishnan, Rajath",
    title = "{Braidings on topological operators, anomaly of higher-form symmetries and the SymTFT}",
    eprint = "2503.13633",
    archivePrefix = "arXiv",
    primaryClass = "hep-th",
    month = "3",
    year = "2025"
}

@article{DelZotto:2025yoy,
    author = "Del Zotto, Michele and Hasan, Azeem and Riedel G{\aa}rding, Elias",
    title = "{SymTFT, Protected Gaplessness, and Spontaneous Breaking of Non-invertible Symmetries}",
    eprint = "2504.18501",
    archivePrefix = "arXiv",
    primaryClass = "hep-th",
    month = "4",
    year = "2025"
}

@article{Hung:2025gcp,
    author = "Hung, Ling-Yan and Ji, Kaixin and Shen, Ce and Wan, Yidun and Zhao, Yu",
    title = "{A 2D-CFT Factory: Critical Lattice Models from Competing Anyon Condensation Processes in SymTO/SymTFT}",
    eprint = "2506.05324",
    archivePrefix = "arXiv",
    primaryClass = "cond-mat.str-el",
    month = "6",
    year = "2025"
}

@article{Robbins:2025puq,
    author = "Robbins, Daniel and Roy, Subham",
    title = "{SymTFT actions, Condensable algebras and Categorical anomaly resolutions}",
    eprint = "2509.05408",
    archivePrefix = "arXiv",
    primaryClass = "hep-th",
    doi = "10.1007/JHEP06(2026)058",
    journal = "JHEP",
    volume = "06",
    pages = "058",
    year = "2026"
}

@article{Bonetti:2025dvm,
    author = "Bonetti, Federico and Del Zotto, Michele and Minasian, Ruben",
    title = "{SymTFT for Continuous Symmetries: Non-linear Realizations and Spontaneous Breaking}",
    eprint = "2509.10343",
    archivePrefix = "arXiv",
    primaryClass = "hep-th",
    doi = "10.1007/JHEP05(2026)048",
    journal = "JHEP",
    volume = "05",
    pages = "048",
    year = "2026"
}

@article{Jia:2026tfh,
    author = "Jia, Qiang and Zhang, Yi",
    title = "{Flat Gauging of Continuous (Non-invertible) Symmetries and Non-compact BF SymTFT for Compact Boson}",
    eprint = "2606.15732",
    archivePrefix = "arXiv",
    primaryClass = "hep-th",
    month = "6",
    year = "2026"
}

@article{Lootens:2024gfp,
    author = "Lootens, Laurens and Delcamp, Clement and Verstraete, Frank",
    title = "{Entanglement and the density matrix renormalization group in the generalized Landau paradigm}",
    eprint = "2408.06334",
    archivePrefix = "arXiv",
    primaryClass = "quant-ph",
    doi = "10.1038/s41567-025-02961-2",
    journal = "Nature Phys.",
    volume = "21",
    number = "10",
    pages = "1657--1663",
    year = "2025"
}

@article{Chen:2025uno,
    author = "Chen, Xie",
    title = "{Essay: Generalized Landau Paradigm for Quantum Phases and Phase Transitions}",
    eprint = "2511.19793",
    archivePrefix = "arXiv",
    primaryClass = "hep-th",
    doi = "10.1103/tmvy-vsqd",
    journal = "Phys. Rev. Lett.",
    volume = "135",
    number = "25",
    pages = "250001",
    year = "2025"
}

@article{Ebisu:2026rnu,
    author = "Ebisu, Hiromi and Han, Bo and Cao, Weiguang",
    title = "{Understanding deconfined quantum critical points from crystalline categorical Landau paradigm}",
    eprint = "2606.05856",
    archivePrefix = "arXiv",
    primaryClass = "cond-mat.str-el",
    month = "6",
    year = "2026"
}

@article{Zhang:2026uhx,
    author = "Zhang, Zhi-Feng and Huang, Yizhou and Wang, Qing-Rui and Ye, Peng",
    title = "{Non-invertible symmetries and mixed anomalies from conserved current construction in (3+1)D twisted BF topological quantum field theories}",
    eprint = "2601.01523",
    archivePrefix = "arXiv",
    primaryClass = "cond-mat.str-el",
    month = "1",
    year = "2026"
}

@misc{HatcherSpectralSequences,
    author = "Hatcher, Allen",
    title = "{Spectral Sequences: Preliminary Chapter 5 for Algebraic Topology}",
    howpublished = "\url{https://pi.math.cornell.edu/~hatcher/AT/ATch5.pdf}",
    year = "2004"
}

@book{MilnorStasheff1974,
    author = "Milnor, John W. and Stasheff, James D.",
    title = "{Characteristic Classes}",
    series = "Annals of Mathematics Studies",
    volume = "76",
    publisher = "Princeton University Press",
    address = "Princeton, NJ",
    year = "1974"
}

@article{Troue,
    author = "Trou{\'e}, Jacques",
    title = "{The orders of the {P}ostnikov invariants of the {T}hom spectrum {$MSO$}}",
    journal = "Illinois J. Math.",
    volume = "10",
    pages = "592--604",
    year = "1966"
}

@article{SaitoTachikawaZhang2026,
    author = "Saito, Shota and Tachikawa, Yuji and Zhang, Yi",
    title = "{Bosonic SPT and invertible phases and its relation to Steenrod's problem}",
    eprint = "2607.14662",
    archivePrefix = "arXiv",
    primaryClass = "hep-th",
    month = "7",
    year = "2026"
}

@article{Lee:2020ewl,
    author = "Lee, Yasunori and Tachikawa, Yuji",
    title = "{Some comments on 6D global gauge anomalies}",
    eprint = "2012.11622",
    archivePrefix = "arXiv",
    primaryClass = "hep-th",
    doi = "10.1093/ptep/ptab015",
    journal = "PTEP",
    volume = "2021",
    number = "8",
    pages = "08B103",
    year = "2021"
}

@article{Gu:2012ib,
    author = "Gu, Zheng-Cheng and Wen, Xiao-Gang",
    title = "{Symmetry-protected topological orders for interacting fermions: Fermionic topological nonlinear $\sigma$ models and a special group supercohomology theory}",
    eprint = "1201.2648",
    archivePrefix = "arXiv",
    primaryClass = "cond-mat.str-el",
    doi = "10.1103/PhysRevB.90.115141",
    journal = "Phys. Rev. B",
    volume = "90",
    number = "11",
    pages = "115141",
    year = "2014"
}

@article{Delmastro:2022pfo,
    author = "Delmastro, Diego G. and Gomis, Jaume and Hsin, Po-Shen and Komargodski, Zohar",
    title = "{Anomalies and symmetry fractionalization}",
    eprint = "2206.15118",
    archivePrefix = "arXiv",
    primaryClass = "hep-th",
    doi = "10.21468/SciPostPhys.15.3.079",
    journal = "SciPost Phys.",
    volume = "15",
    pages = "079",
    year = "2023"
}

@article{Jia:2026jhj,
    author = "Jia, Weizhen and Wang, Yi-Nan and Zhang, Yi",
    title = "{On Quantum Aspects of 1-Form Symmetries II: Bordism, Invertible Phases, and Anomalies}",
    eprint = "2606.07056",
    archivePrefix = "arXiv",
    primaryClass = "hep-th",
    month = "6",
    year = "2026"
}

@article{Jia:2026vcr,
    author = "Jia, Qiang and Luo, Ran and Tian, Jiahua and Wang, Yi-Nan and Zhang, Yi",
    title = "{Categorical Symmetries via Operator Algebras}",
    eprint = "2604.25821",
    archivePrefix = "arXiv",
    primaryClass = "hep-th",
    month = "4",
    year = "2026"
}

@article{Jia:2025vrj,
    author = "Jia, Qiang and Luo, Ran and Tian, Jiahua and Wang, Yi-Nan and Zhang, Yi",
    title = "{Categorical continuous symmetry}",
    eprint = "2509.13170",
    archivePrefix = "arXiv",
    primaryClass = "hep-th",
    doi = "10.1007/JHEP06(2026)031",
    journal = "JHEP",
    volume = "06",
    pages = "031",
    year = "2026"
}

@article{Luo:2025phx,
    author = "Luo, Ran and Wang, Yi-Nan and Bi, Zhen",
    title = "{Topological Holography for Mixed-State Phases and Phase Transitions}",
    eprint = "2507.06218",
    archivePrefix = "arXiv",
    primaryClass = "cond-mat.str-el",
    doi = "10.1103/9kmh-gjf8",
    journal = "PRX Quantum",
    volume = "6",
    number = "4",
    pages = "040358",
    year = "2025"
}

@article{Schafer-Nameki:2025fiy,
    author = "Schafer-Nameki, Sakura and Tiwari, Apoorv and Warman, Alison and Zhang, Carolyn",
    title = "{SymTFT Approach for Mixed States with Non-Invertible Symmetries}",
    eprint = "2507.05350",
    archivePrefix = "arXiv",
    primaryClass = "quant-ph",
    month = "7",
    year = "2025"
}

@article{Qi:2025tal,
    author = "Qi, Marvin and Sohal, Ramanjit and Chen, Xie and Stephen, David T. and Prem, Abhinav",
    title = "{The Symmetry Taco: Equivalences between Gapped, Gapless, and Mixed-State SPTs}",
    eprint = "2507.05335",
    archivePrefix = "arXiv",
    primaryClass = "cond-mat.str-el",
    month = "7",
    year = "2025"
}

@article{Chen-Feng,

author="Yitao Feng and Yu-An Chen",
title = "{Defining 2-group symmetry actions and anomalies on lattices, to Appear}"


}

@article{Wen:2011np,
    author = "Wen, Xiao-Gang",
    title = "{Symmetry protected topological phases in non-interacting fermion systems}",
    eprint = "1111.6341",
    archivePrefix = "arXiv",
    primaryClass = "cond-mat.str-el",
    doi = "10.1103/PhysRevB.85.085103",
    journal = "Phys. Rev. B",
    volume = "85",
    pages = "085103",
    year = "2012"
}

@article{Chan:2017eov,
    author = "Chan, AtMa P. O. and Ye, Peng and Ryu, Shinsei",
    title = "{Braiding with Borromean Rings in (3+1)-Dimensional Spacetime}",
    eprint = "1703.01926",
    archivePrefix = "arXiv",
    primaryClass = "cond-mat.str-el",
    doi = "10.1103/PhysRevLett.121.061601",
    journal = "Phys. Rev. Lett.",
    volume = "121",
    number = "6",
    pages = "061601",
    year = "2018"
}

@article{Zhang:2020kgc,
    author = "Zhang, Zhi-Feng and Ye, Peng",
    title = "{Compatible braidings with Hopf links, multiloop, and Borromean rings in $(3+1)$-dimensional spacetime}",
    eprint = "2012.13761",
    archivePrefix = "arXiv",
    primaryClass = "cond-mat.str-el",
    doi = "10.1103/PhysRevResearch.3.023132",
    journal = "Phys. Rev. Res.",
    volume = "3",
    number = "2",
    pages = "023132",
    year = "2021"
}

@article{Wang:2018iwz,
    author = "Wang, Qing-Rui and Cheng, Meng and Wang, Chenjie and Gu, Zheng-Cheng",
    title = "{Topological Quantum Field Theory for Abelian Topological Phases and Loop Braiding Statistics in $(3+1)$-Dimensions}",
    eprint = "1810.13428",
    archivePrefix = "arXiv",
    primaryClass = "cond-mat.str-el",
    doi = "10.1103/PhysRevB.99.235137",
    journal = "Phys. Rev. B",
    volume = "99",
    number = "23",
    pages = "235137",
    year = "2019"
}

@article{Teichner1993,
  author  = {Teichner, Peter},
  title   = {On the signature of four-manifolds with universal covering spin},
  journal = {Mathematische Annalen},
  volume  = {295},
  number  = {4},
  pages   = {745--759},
  year    = {1993},
  doi     = {10.1007/BF01444915}
}

\end{document}